\documentclass[3p,times]{elsarticle}

\usepackage{array}
\usepackage{color} 
\usepackage{tabularx}
\usepackage{graphicx} 
\usepackage{amsmath}
\usepackage{amssymb}
\usepackage{amsfonts}	
\usepackage{moreverb}
\usepackage{dsfont}
\usepackage{tipa}
\usepackage{upgreek}
\usepackage{bm}
\usepackage{multirow}
\usepackage{soul}
\usepackage{ textcomp }

\newcommand\BibTeX{{\rmfamily B\kern-.05em \textsc{i\kern-.025em b}\kern-.08em
T\kern-.1667em\lower.7ex\hbox{E}\kern-.125emX}}

\usepackage{hyperref} 
\hypersetup{
    colorlinks=true,                          
    linkcolor=blue, 
    citecolor=red, 
    urlcolor=blue  } 

\newcommand{\pd}{\mathcal{\partial}}
\newcommand{\xx}{{\boldsymbol{x}}}
\newcommand{\yy}{{\boldsymbol{y}}}

\newcommand{\x}{\mbf{x}}

\newcommand{\mbf}[1]{\mathbf{#1}}			%

\newcommand{\Q}{\mathbf{Q}}
\renewcommand{\S}{\mathbf{S}}
\renewcommand{\u}{\mathbf{u}}
\newcommand{\w}{\mathbf{w}}
\newcommand{\m}{\mathbf{m}}
\newcommand{\q}{\mathbf{q}}
\newcommand{\F}{\mathbf{F}}
\newcommand{\f}{\mathbf{f}}
\newcommand{\g}{\mathbf{g}}
\newcommand{\h}{\mathbf{h}}

\renewcommand{\v}{\mathbf{v}}

\newcommand{\rmd}{{\rm d}}

\newcommand{\G}{\mathbf{G}}
\newcommand{\A}{\AAA}

\newcommand{\halb}{\frac{1}{2}}

\newcommand{\bdm}{\begin{displaymath}}
\newcommand{\edm}{\end{displaymath}}

\newcommand{\bea}{\begin{eqnarray} }
\newcommand{\eea}{\end{eqnarray} }

\newcommand{\dev}{\textnormal{dev}} 

\newcommand{\AAA}{{\boldsymbol{A}}}
\newcommand{\aaa}{{\boldsymbol{a}}}
\newcommand{\DD}{{\mathbf{D}}}
\newcommand{\HH}{{\mathbf{H}}}

\newcommand{\GG}{{\mathbf{G}}}

\newcommand{\ee}{{\mathbf{e}}}
\newcommand{\bb}{{\mathbf{b}}}
\newcommand{\hh}{{\mathbf{h}}}
\newcommand{\dd}{{\mathbf{d}}}
\newcommand{\vv}{{\mathbf{v}}}
\newcommand{\uu}{{\mathbf{u}}}
\newcommand{\mcE}{{\mathcal{E}}}

\newcommand{\EE}{{\mathbf{E}}}
\newcommand{\BB}{{\mathbf{B}}}

\newcommand{\FF}{{\boldsymbol{F}}}
\newcommand{\II}{{\mathbf{I}}}
\newcommand{\JJ}{{\mathbf{J}}}

\newcommand{\pp}{{\mathbf{p}}}
\newcommand{\qq}{{\mathbf{q}}}

\newcommand{\Id}{{\mathbf{I}}}
\newcommand{\tr}{\textnormal{tr}}
\newcommand{\BS}{{\boldsymbol{\sigma}}}

\newfont{\numerikEleven}{ecrm1000}
\newfont{\numerikTen}{cmss10}
\newfont{\numerikNine}{cmss9}
\newfont{\numerikEight}{cmss8}

\journal{Journal of Computational Physics}

\begin{document} 
\begin{frontmatter}
\title{High order ADER schemes for a unified first order hyperbolic formulation of Newtonian continuum mechanics coupled with electro-dynamics} 
\author[UniTN]{Michael Dumbser$^{*}$}
\ead{michael.dumbser@unitn.it}
\cortext[cor1]{Corresponding author}

\author[IMT,NSC]{Ilya Peshkov}
\ead{peshkov@math.nsc.ru}

\author[NSC,NSU]{Evgeniy Romenski}
\ead{evrom@math.nsc.ru}

\author[UniTN]{Olindo Zanotti}
\ead{olindo.zanotti@unitn.it}


\address[UniTN]{Department of Civil, Environmental and Mechanical Engineering, 
University of Trento, Via Mesiano 77, 38123 Trento, Italy.} 
\address[IMT]{{Institut de Math\'{e}matiques de Toulouse, Universit\'{e} Toulouse III, F-31062 Toulouse, France.}}
\address[NSC]{{Sobolev Institute of Mathematics, 4 Acad. Koptyug Avenue, 630090 Novosibirsk, Russia}}
\address[NSU]{{Novosibirsk State University, 2 Pirogova Str., 630090 Novosibirsk, Russia}}

\begin{abstract} \color[rgb]{0,0,0}
In this paper, we propose a new unified \textit{first order hyperbolic} model of Newtonian continuum mechanics coupled with electro-dynamics. 
The model is able to describe the behavior of moving elasto-plastic dielectric solids as well as viscous and inviscid fluids in the presence of electro-magnetic 
fields. 
It is actually a very peculiar feature of the proposed PDE system that viscous fluids are treated just as a special case of elasto-plastic solids. This is 
achieved by introducing a \textit{strain relaxation} mechanism in the evolution equations of the distortion matrix $\AAA$, which in the case of
purely elastic solids maps the current configuration to the reference configuration. The model also contains a hyperbolic formulation of heat conduction 
as well as a dissipative source term in the evolution equations for the electric field given by Ohm's law. 
Via formal asymptotic analysis we show that in the stiff limit, the governing first order hyperbolic PDE system with relaxation source terms tends 
\textit{asymptotically} to the well-known viscous and resistive magnetohydrodynamics (MHD) equations. Furthermore, a rigorous derivation of the model from 
variational principles is presented, together with the transformation of the Euler-Lagrange differential equations associated with the underlying variational 
problem from Lagrangian coordinates to Eulerian coordinates in a fixed laboratory frame. 
The present paper hence extends the unified first order hyperbolic model of Newtonian continuum mechanics recently proposed in \cite{PeshRom2014,DPRZ2016} 
to the more general case where the continuum is coupled with electro-magnetic fields. The governing PDE system is \textit{symmetric hyperbolic} and 
satisfies the first and second principle of thermodynamics, hence it belongs to the so-called class of symmetric hyperbolic thermodynamically compatible systems (HTC),  
which have been studied for the first time by Godunov in 1961 \cite{God1961} and later in a series of papers by Godunov and Romenski \cite{GodRom1972,GodRom1995,Rom1998}. 
An important feature of the proposed model is that the propagation speeds of all physical processes, including dissipative processes, are \textit{finite}. 
The model is discretized using high order accurate ADER discontinuous Galerkin (DG) finite element schemes with \textit{a posteriori} subcell finite volume limiter 
and using high order ADER-WENO finite volume schemes. 
We show numerical test problems that explore a rather large parameter space of the model ranging from ideal MHD, viscous and resistive MHD over pure electro-dynamics to 
moving dielectric elastic solids in a magnetic field. 
\end{abstract}

\begin{keyword}
 symmetric hyperbolic thermodynamically compatible systems (HTC) \sep 
 unified first order hyperbolic model of continuum mechanics 
 (fluid mechanics, solid mechanics, electro-dynamics) \sep 
 finite signal speeds of all physical processes \sep 
 arbitrary high-order ADER Discontinuous Galerkin schemes \sep 
 path-conservative methods and stiff source terms \sep 
 nonlinear hyperelasticity   
%
\end{keyword}
\end{frontmatter}


%
\section{Introduction} \label{sec:introduction}

In this paper, we propose a new unified \textit{first order hyperbolic} model of Newtonian continuum mechanics coupled with electro-dynamics. The model is the extension of our previous results~\cite{DPRZ2016}, hereafter Paper~I, on a unified formulation of continuum mechanics towards the coupling of the time evolution equations for the matter with the electric and magnetic fields. 
The problem of determining the force acting on a medium in an electromagnetic field, as well as the related problem of determining the energy-momentum tensor of an electromagnetic field in a medium, has been discussed in the literature over the years since the work by Minkowski~\cite{Minkowski1908} and Abraham~\cite{Abraham1909}. However, to the best of our knowledge, a universally accepted solution to this problem has been absent to date~\cite{Landau1984electrodynamics,MakarovRukhadze,GinzbUgarov,eringen2012electrodynamics,DOROVSKY199491}.

In this respect, our work can be broadly considered as a contribution to the modeling of electrodynamics of moving continuous media. We   do not claim to give an ultimate solution to the problem, but rather to show that, within our formalism, all  the equations can be obtained in a consistent way with rather good mathematical properties (symmetric hyperbolicity, first order PDEs, well posedeness of the initial value problem, finite speeds of perturbation propagation even for dissipative processes in the diffusive regime) and that the corresponding physical effects are correctly described. By an extensive comparison with the numerical and analytical solutions to the well established models as the Maxwell equations, ideal MHD equations and viscous resistive MHD (VRMHD) equations, we demonstrate that the proposed nonlinear hyperbolic dissipative model is able to describe dielectrics ($ \eta\rightarrow \infty $), ideal conductors ($ \eta\rightarrow0 $), and resistive conductors ($ 0<\eta<\infty $) as particular cases, where $ \eta $ is the resistivity. Thus, the applicability range of the proposed model is larger than those for ideal and resistive MHD models, because the electric and magnetic fields are genuinely independent and are governed by their own time evolution equations as in the Maxwell equations. 

In Paper~I and~\cite{PeshRom2014}, we provided a unified first-order hyperbolic formulation of the equations of
continuum mechanics, showing for the first time that the dynamics of fluids 
and solids can be cast in a \textit{single mathematical framework}. This 
becomes possible due to the use of a characteristic strain dissipation time $ 
\tau $, which is the characteristic time for continuum particle 
\textit{rearrangements}. 
By its definition, the characteristic time $ \tau $, as opposed to the 
viscosity coefficient, is 
applicable to the dynamics of both fluids and solids (see the discussions 
in~\cite{PeshRom2014} and Paper~I) and is a continuum interpretation of the 
seminal idea of the so-called \textit{particle settled life time} (PSI) of 
Frenkel~\cite{Frenkel1955}, who applied it to describe the fluidity  of 
liquids, see also 
\cite{brazhkin2012two,bolmatov2013thermodynamic,bolmatov2015revealing} and 
references therein. In addition, the definition of $ \tau $ assumes the 
continuum particles to have a 
\textit{finite scale} and thus to be deformable as opposed to the 
\textit{scaleless} \textit{mathematical points} in classical continuum 
mechanics. We note that the model studied in~\cite{PeshRom2014} and Paper~I was 
used by several 
authors, 
e.g.~\cite{Rom1989,Resnyansky1995,Resnyansky2002709,GavrFavr2008,BartonRom2010,Pesh2010,favrie2009solid,resnyansky2011constitutive,Barton2013,Ndanou2014,Pesh2015}
 to cite just a few, in the 
solid dynamics context since its 
original invention 
in 1970th by Godunov and Romenski~\cite{GodunovRomenski72,God1978} but the 
recognition that the same model is also applicable to the dynamics of viscous fluids 
and its extensive validation in the fluid dynamics context was made only 
recently in~\cite{PeshRom2014} and Paper~I.

What concerns a mathematical guide to derive time evolution 
equations, as in~\cite{PeshRom2014} and Paper~I, we follow the so-called 
formalism  of \textit{\textbf{H}yperbolic \textbf{T}hermodynamically 
\textbf{C}ompatible systems of conservation laws}, or simply HTC formalism 
here. This formalism is described in Section~\ref{sec.HTC.Lagr}. 
We recall that \textit{hyperbolicity} naturally  accounts for the two most 
relevant features of fundamental physical systems, namely the unique and 
continuous dependence of the solution on the initial data,  and the  finite 
velocity for perturbation propagation~(causality).

At this point we stress that the HTC theory is \textit{radically different} from classical  Maxwell-Cattaneo hyperbolic relaxation models \cite{Cattaneo1948,Jou1996,Muller1998} typically used in extended irreversible thermodynamics (EIT), since the propagation speeds of all physical processes
remain \textit{finite}, even in the stiff relaxation limit (parabolic diffusion limit), see \cite{PeshRom2014} for a more detailed discussion. The differences between the approaches become also apparent if one takes a look at the physical meaning of the state variables used in both approaches. We recall that in the EIT the \textit{fluxes} are typically used as the extra state variables (in addition to the  conventional ones like mass, momentum and energy), which  usually leads to the result that the PDEs have no apparent structure. In the HTC formalism, only density fields may serve as state variables which, in fact, due to the fundamental \textit{conservation principle} allows to obtain equations in a rather complete form with an elegant structure, see Section~\ref{sec.HTC.Lagr}. 

For recent work on hyperbolic reformulations of the steady viscous and resistive MHD equations and time dependent convection-diffusion 
equations based on standard Maxwell-Cattaneo relaxation, see the papers of Nishikawa et al. \cite{Nishikawa1,Nishikawa2,NishikawaMHD} 
and Montecinos and Toro \cite{Montecinos2014a,Montecinos2014b,Montecinos2014c}. 

The plan of the paper is the following:  in the first part we concentrate on the mathematical principles of the  
HTC system (Sections \ref{sec.HTC.Lagr} and \ref{sec.EulerianFrame}), while in the second part we give an extensive numerical evidences of the applicability
of the model to a wide range of electromagnetic flows. 

In the rest of the paper we use the Einstein summation convention over repeated indices. 

\clearpage

\section{HTC formalism and the master system}
\label{sec.HTC.Lagr}

The hyperbolic dissipative theory discussed in this paper relies on the HTC formalism. The development of the formalism started in 1961 after it was observed by Godunov~\cite{God1961,God1962} that some systems of conservation laws admitting an \textit{extra conservation law} also admit an interesting parametrization 
\begin{equation}\label{eqn.formalism.parametr}
\dfrac{\pd M_{p_i}}{\pd t} + \dfrac{\pd N^j_{p_i}}{\pd y_j}=0
\end{equation}
which allows to rewrite the governing equations in a \textit{symmetric form}
\begin{equation}\label{eqn.formalism.symmetr}
M_{p_i p_k}\dfrac{\pd p_k}{\pd t} + N^j_{p_i p_k}\dfrac{\pd p_k}{\pd y_j}=0.
\end{equation}
Here, $ t $ is the time, $ y_j $ are the spatial coordinates, $ p_k $ is the vector of state variables, $ M(p_i) $ and $ N^j(p_i) $ are the scalar
potentials of the state variables. Here and in the rest of the paper, a potential with the state variables in the subscript should be understood as
the partial derivatives of the potential with respect to these state variables. 
Thus, for example, $ M_{p_k} $, $ N^j_{p_k} $, $ M_{p_i p_k} $ and 
$ N^j_{p_i p_k} $  
in~\eqref{eqn.formalism.parametr}--\eqref{eqn.formalism.symmetr} should be
understood as the first and second partial derivatives of the potentials $ M $ 
and $ N^j $ with respect to the state
variables $ p_i $, e.g. $ M_{p_k}=\pd M/\pd p_k $, $ M_{p_i p_k}=\pd^2M/\pd p_i \pd p_k $, etc.

In this parametrization, the extra conservation	law has
always the following form
\begin{equation}\label{eqn.formslism.energy}
\dfrac{\pd (p_iM_{p_i} - M)}{\pd t} + \dfrac{\pd (p_i N^j_{p_i} - N^j)}{\pd y_j}=0
\end{equation}
and, in fact, it is just a straightforward consequence of the governing equations~\eqref{eqn.formalism.parametr} and can be obtained as a \textit{linear combination} of these equations. Indeed,~\eqref{eqn.formslism.energy} can be obtained as a sum of the equations \eqref{eqn.formalism.parametr} multiplied by the corresponding factors $ p_i $.

If the potential $ M(p_i) $ is a strictly convex function
of the state variables then the symmetric matrix $ M_{p_i
  p_k} $ is positive definite
and~\eqref{eqn.formalism.symmetr} becomes a
\textit{symmetric hyperbolic} system of
equations~\cite{Friedrichs1958}. 

Usually, the \textit{generating potential} $ M $ has the meaning of the generalized pressure while its Legendre transformation $ p_i M_{p_i} - M$ has the meaning of the total energy and thus,~\eqref{eqn.formslism.energy} is the total energy conservation law\footnote{Note that the potentials $ N^j $ have no apparent physical meaning and play no role in the later developments of the HTC formalism.}. Hence, the observation of Godunov establishes the very important connection between the well-posedeness of the equations of mathematical physics and thermodynamics.
 
As it was understood later on the examples of the ideal MHD equations~\cite{God1972MHD}, 
that the original observation of Godunov~\cite{God1961} relates 
only to conservation laws written in the \textit{Lagrangian} 
frame which indeed admits a fully \textit{conservative} 
formulation\footnote{In this paper, under fully conservative  
  form of the equations we understand not only fully 
  divergence form of equations, i.e. generated by the 
  divergence differential operator, but rather that there 
  are no space derivatives multiplied by unknown 
  functions, while algebraic production source terms can
  be present.}, while the time evolution equations in the
Eulerian frame  have a more complicated structure, except
for the compressible Euler equations of ideal fluids.

The structure of the Eulerian equations and its relation to 
the fully conservative structure of the equations in Lagrangian form 
was revealed in a series of papers by Godunov and Romenski~\cite{GodRom1995,GodRom1996,godunov1996systems,GodRom1998,Rom1998,Rom2001,GodRom2003}. In particular, in~\cite{godunov1996systems}, based on the group representation theory~\cite{Gelfand1963}, a rather general form of first order PDEs with the following properties was proposed:
\begin{itemize}
\item PDEs are invariant under rotations
\item PDEs are compatible with an extra conservation law
\item PDEs are generated by only one potential like $ M $
\item PDEs are symmetric hyperbolic
\item PDEs are conservative and generated by invariant
  differential operators only, such as div, grad and curl.
\end{itemize}
One may naturally question how this class of PDEs, which
shall be referred 
to as as the \textit{master system}, relates to the
models that describe continuum mechanics and whether it is too 
restrictive to deal with dissipative processes such as
viscous momentum transfer, heat transfer, resistive MHD,
etc., typically described by second order parabolic equations.
First, it is important to emphasize that invariance under orthogonal transformations and the existence of an extra conservation law, which is typically the total energy conservation, is the compulsory requirement for continuum mechanics models. Second, as shown recently~\cite{PeshRom2014,DPRZ2016}, there is no physical reason imposing that the dissipative transport processes such as viscous momentum  transfer or heat conduction should be exclusively modeled by the second order parabolic diffusion theory, but they can also be very successfully modeled by a more general framework based 
on first order hyperbolic equations with relaxation source terms. 
Third, after analysis of a rather large number of particular examples of continuum models~\cite{GodRom1995,GodRom1996,GodRom1998,Rom1998,Rom2001}, it was shown that many models fall into the class of 
HTC systems. Among them are the compressible Euler equations of ideal fluids, the ideal MHD equations, the equations of nonlinear elasto-plasticity, the electrodynamics of moving media, a model describing  
superfluid helium, the equations governing compressible multi-phase flows, elastic superconductors, and finally also the unified first order hyperbolic formulation for fluid and solid mechanics   
introduced in \cite{PeshRom2014,DPRZ2016}. In this paper, we show that also the viscous and resistive MHD equations can be cast into the form of a first order HTC system. 

The starting point of the HTC formalism is a sub-system of the Lagrangian conservation laws given in eqs.~(1) of ~\cite{godunov1996systems}, 
which will be refereed to as the \textit{master system} from now on. 
The final governing PDEs written in the Eulerian frame will then be the result of the following system of Lagrangian master equations:  
\begin{subequations}\label{eqn.intro.mastersys}
	\begin{align}
	& \displaystyle\frac{{\rm d} M_{v_i}}{{\rm d} t} - \frac{\partial  P_{i j}}{\partial y_j} = 0, \label{eqn.intro.mastersys.a}\\[1mm]
	& \displaystyle\frac{{\rm d} M_{P_{ij}}}{{\rm d} t} - \frac{\partial  v_i}{\partial y_j} = 0, \label{eqn.intro.mastersys.b}\\[1mm]
	& \displaystyle\frac{{\rm d} M_{d_i}}{{\rm d} t}-\varepsilon_{ijk}\frac{\partial b_k }{\partial y_j}=0,\label{eqn.intro.mastersys.c}\\[1mm]
	& \displaystyle\frac{{\rm d} M_{b_i}}{{\rm d} t}+\varepsilon_{ijk}\frac{\partial d_k} {\partial y_j}=0.\label{eqn.intro.mastersys.d}
	\end{align}
\end{subequations}
Here, $ v_i $ is the velocity of the matter, $ P_{ij} $ is the stress tensor, while $ d_i $ and $ b_i $ are some vectors describing the electric and magnetic fields, respectively.

In contrast to the classical parabolic theory of dissipative processes, the 
governing equations in our approach are all \textit{first order} 
hyperbolic PDEs and the dissipative processes will \textit{not} 
be modeled by \textit{differential terms}, but \textit{exclusively} via 
\textit{algebraic relaxation source terms}, which will be specified later in the Eulerian case. This has the \textit{important consequence} that the structure of the differential terms and the type
of the PDE is the \textit{same} in both, the dissipative as well as in the non-dissipative case. We recall, that if the dissipation is excluded in the classical second order parabolic diffusion 
theory, this then changes not only the structure of the PDEs, but also their \textit{type}. 

Because of this fact, within the HTC formalism we can study the structure of the governing equations by restricting our considerations to the non-dissipative case only.  
We also note that if the dissipation source terms are switched off, then the model describes an elastic medium, see \cite{PeshRom2014,DPRZ2016}. 


%
%
%
%
%

\subsection{Variational nature of the field equations in a moving elastic medium}\label{sec.VariationNature}

It is well known that many equations of mathematical physics can be derived as 
the Euler-Lagrange equations obtained by the minimization of a Lagrangian.
As an example, one can consider the nonlinear elasticity equations in
Lagrangian coordinates~\cite{goldsteinclassical}. The classical Maxwell
equations of electrodynamics can also be derived by the minimization of a Lagrangian with the 
use of the gauge theory~\cite{gelfand1963calculus}. 
It turns out that the coupling of these two physical objects in a single
Lagrangian gives us a straightforward way to derive the equations for the 
electromagnetic field in a moving medium.
We start by introducing two vector potentials and a scalar potential:
\begin{equation}
\xx=[x_i(t,\yy)],\ \ \ \aaa=[a_i(t,\yy)],\ \ \ \varphi(t,\yy), 
\end{equation}
so that
\begin{align}
\hat{v}_i=\dfrac{\pd x_i}{\pd t},\ \ \  & \hat{F}_{ij}=\dfrac{\pd x_i}{\pd y_j},\label{eqn.PotentialDeriv1}\\
\hat{e}_i=-\dfrac{\pd a_i}{\pd t} - \dfrac{\pd \varphi}{\pd y_i},\ \ \ & \hat{h}_i=\varepsilon_{ijk}\dfrac{\pd a_k}{\pd y_j},\label{eqn.PotentialDeriv2}
\end{align}
Here, $ t $ is time, $ \yy=[y_i] $ and $ \xx=[x_i] $ are
the Lagrangian and Eulerian
spatial
coordinates respectively,
 while $\aaa$ and $\varphi$ are the
conventional electromagnetic potentials.

Then, we define the action integral
\begin{equation}\label{eqn.Lagrangian}
{\mathcal L}=\int \Lambda d\yy dt,
\end{equation}
where $ \Lambda=\Lambda(\hat{v}_i,\hat{F}_{ij},\hat{e}_i,\hat{h}_i,\hat{w}_i,\hat{c}) $ is the Lagrangian.

First variation of $\mathcal{L}$ gives us the Euler-Lagrange equations
\begin{align}
\dfrac{\partial \Lambda_{\hat{v}_i}}{\partial t} +\dfrac{\partial \Lambda_{\hat{F}_{ij}}}{\partial y_j}=&\ \ 0, \label{eqn.EulLagr.moment}\\[2mm]
\dfrac{\partial \Lambda_{\hat{e}_i}}{\partial t} +\varepsilon_{ijk}\dfrac{\partial \Lambda_{\hat{h}_k}}{\partial y_j}=&\ \ 0, \label{eqn.EulLagr.e}  \\[2mm]
 \dfrac{\pd \Lambda_{\hat{e}_j}}{\pd y_j}=&\ \ 0.\label{eqn.div.e}
\end{align}
To this system, the following compatibility constraints should be added (they are trivial consequences of the definitions \eqref{eqn.PotentialDeriv1} and \eqref{eqn.PotentialDeriv2})
\begin{eqnarray} 
	\frac{\pd \hat{F}_{ij}}{\pd t} - \frac{\pd \hat{v}_i}{\pd y_j} & =0\,, & \dfrac{\partial \hat{F}_{ij}}{\partial y_k}-\dfrac{\partial \hat{F}_{ik}}{\partial y_j}  = 0\,, \label{eqn.constr.F}\\
	\frac{\pd \hat{h}_i}{\pd t} +\varepsilon_{ijk}\dfrac{\pd \hat{e}_k}{\pd y_j}     &       =0\,, & \dfrac{\partial \hat{h}_j}{\partial y_j}                                               =0\,. 	\label{eqn.constr.h}
\end{eqnarray}



In order to rewrite equations \eqref{eqn.EulLagr.moment}--\eqref{eqn.constr.h} in the form of system \eqref{eqn.intro.mastersys}, let us introduce the potential $U$ as a partial Legendre transformation of the Lagrangian $ \Lambda $
\begin{eqnarray}
{\rm d}U&=&{\rm d}(\hat{v}_i \Lambda_{\hat{v}_i}+\hat{e}_i\Lambda_{\hat{e}_i}-\Lambda)=
\hat{v}_i{\rm d}\Lambda_{\hat{v}_i}+\hat{e}_i {\rm d}
\Lambda_{\hat{e}_i} -\Lambda_{\hat{F}_{ij}} {\rm d}
\hat{F}_{ij}-\Lambda_{\hat{h}_i} {\rm d} \hat{h}_i =
\nonumber \\
&&
\hat{v}_i{\rm d}\Lambda_{\hat{v}_i}+\hat{e}_i {\rm d} \Lambda_{\hat{e}_i} +  \Lambda_{\hat{F}_{ij}} {\rm d} (-\hat{F}_{ij})  + \Lambda_{\hat{h}_i} {\rm d} (-\hat{h}_i).
\end{eqnarray}
Hence, denoting $m_i=\Lambda_{\hat{v}_i} $, $e_i=\Lambda_{\hat{e}_i}$, $ F_{ij}=-\hat{F}_{ij} $, $ h_i=-\hat{h}_i $,
we get the thermodynamic identity 
\begin{equation*}
{\rm d}U=U_{m_i} {\rm d} m_i + U_{F_{ij}} {\rm d} F_{ij} + U_{e_i} {\rm d} e_i + U_{h_{i}}{\rm d} h_{i}.
\end{equation*}
Eventually, in terms of the variables
\begin{equation}\label{eqn.VarConsLagr}
\qq=(m_i,F_{ij},e_{i},h_i)
\end{equation}
and the potential $ U=U(\qq) $, equations \eqref{eqn.EulLagr.moment}, \eqref{eqn.EulLagr.e}, \eqref{eqn.constr.F}$ _1 $ and \eqref{eqn.constr.h}$ _1 $
become 
\begin{subequations}\label{eqn.MasterLagr}
	\begin{align}
	& \displaystyle\frac{{\rm d} m_i}{{\rm d} t} - \frac{\partial  U_{F_{i j}}}{\partial y_j} = 0, \label{eqn.MasterLagr.Momentum}\\[1mm]
	& \displaystyle\frac{{\rm d} F_{ij}}{{\rm d} t} - \frac{\partial  U_{m_{i}}}{\partial y_j} = 0, \label{eqn.MasterLagr.F}\\[1mm]
	& \displaystyle\frac{{\rm d} e_{i}}{{\rm d} t}-\varepsilon_{ijk}\frac{\partial {U}_{h_{k}} }{\partial y_j}=0,\label{eqn.MasterLagr.Electr}\\[1mm]
	& \displaystyle\frac{{\rm d} h_{i}}{{\rm d} t}+\varepsilon_{ijk}\frac{\partial {U}_{e_{k}} }{\partial y_j}=0\,,\label{eqn.MasterLagr.Magn}
	\end{align}
\end{subequations}
which should be supplemented by stationary constraints~\eqref{eqn.div.e}, \eqref{eqn.constr.F}$ _2 $ and \eqref{eqn.constr.h}$ _2 $ which now read as
\begin{equation}\label{eqn.constr.u}
\dfrac{\partial F_{ij}}{\partial y_k}-\dfrac{\partial F_{ik}}{\partial y_j} = 0\,,\qquad \dfrac{\pd e_i}{\pd y_i}=0\,, \qquad \dfrac{\pd h_i}{\pd y_i}=0\,.
\end{equation}

System~\eqref{eqn.MasterLagr} is, in fact, identical 
to~\eqref{eqn.intro.mastersys}. In order to see this, one needs to introduce 
fluxes as new (\textit{conjugate}) state variables
\begin{equation}\label{eqn.VarFluxLagr}
\pp=(U_{m_i},U_{F_{ij}},U_{e_i},U_{h_i}),
\end{equation}
which we denote as
\begin{equation}\label{eqn.intro.pvariables}
\begin{array}{cc}
v_i=U_{m_i}, & P_{ij}=U_{F_{ij}},\\[2mm]
d_{i}=U_{e_{i}}, & b_{i}=U_{h_{i}},
\end{array} 
\end{equation}
and a new potential $ M(\pp) $ as a Legendre transform of $ U(\qq) $, i.e.
\begin{equation}
M=m_i U_{m_i}+F_{ij} U_{F_{ij}}+e_i U_{e_i}+h_i U_{h_i} - U,
\end{equation}
or briefly
\begin{equation}
M(\pp) = \qq\cdot\pp - U(\qq).
\end{equation}
After that, system \eqref{eqn.MasterLagr} transforms exactly to \eqref{eqn.intro.mastersys},
while constraints \eqref{eqn.constr.u} read as
\begin{equation}\label{eqn.constr.m}
\dfrac{\partial M_{P_{ij}}}{\partial y_k}-\dfrac{\partial M_{P_{ik}}}{\partial y_j} = 0\,,\qquad \dfrac{\pd M_{d_i}}{\pd y_i}=0\,, \qquad \dfrac{\pd M_{b_i}}{\pd y_i}=0\,.
\end{equation}

One may clearly note, a similarity between the 
equations~\eqref{eqn.intro.mastersys.c}--\eqref{eqn.intro.mastersys.d} (or
\eqref{eqn.MasterLagr.Electr}--\eqref{eqn.MasterLagr.Magn}) and the Maxwell 
equations. However, because no assumptions about the Lagrangian $ \Lambda $, 
and thus, about the potentials  $ U(\qq) $ and $ M(\pp) $, has been done yet, 
these equations
should be considered as a \textit{nonlinear generalization} of the Maxwell
equations.

We note that equations 
\eqref{eqn.intro.mastersys.a}--\eqref{eqn.intro.mastersys.b} and 
\eqref{eqn.intro.mastersys.c}--\eqref{eqn.intro.mastersys.d} (or 
\eqref{eqn.MasterLagr.Momentum}--\eqref{eqn.MasterLagr.F} and 
\eqref{eqn.MasterLagr.Electr}--\eqref{eqn.MasterLagr.Magn}) are not independent 
as it may seem. They are coupled via the dependence of the potential $ M(\pp) $ 
(or $ U(\qq) $) on all the state variables~\eqref{eqn.VarConsLagr}. This 
coupling will emerge in a more transparent way when we shall consider these 
equations in the Eulerian frame in Section~\ref{sec.EulerianFrame}.

\subsection{Properties of the master system}

\subsubsection{Energy conservation}
A central role in the system formulation is played by the thermodynamic potential \begin{equation}\label{eqn.EnergyPotetnialLagr}
U=U(m_i,F_{ij},e_i,h_i).
\end{equation}
or its dual 
\begin{equation}\label{eqn.intro.MPotential}
M=M(v_i,P_{ij},d_i,b_i)
\end{equation}
as one of them generates the fluxes in \eqref{eqn.MasterLagr}, while the other generates the density fields in~\eqref{eqn.intro.mastersys}.
The potential $ U $ typically has the meaning of the total energy density of 
the system, while $ M $ has the meaning of a total pressure.

In addition, solutions of the system~\eqref{eqn.MasterLagr} satisfy an extra conservation law  
\begin{equation}\label{eqn.EnergyConsLagr}
\frac{{\rm d}{U}}{{\rm d}t} - \frac{\partial}{\partial 
y_j}\left(U_{m_i}U_{F_{ij}} + \varepsilon_{ijk}U_{e_i}{U}_{h_k} \right)=0
\end{equation}
which should be interpreted as the total energy conservation. In terms of the dual potential $ M $ and dual state variables~\eqref{eqn.intro.pvariables} it reads as
\begin{equation}\label{eqn.intro.energycons.M}
\frac{{\rm d}}{{\rm d}t}\left (v_i M_{v_i}+P_{ij} M_{P_{ij}}+d_i M_{d_i} + b_i M_{b_i} - M\right ) - \frac{\partial}{\partial y_j}\left(v_i P_{ij} + \varepsilon_{ijk}d_{i}{b}_{k} \right)=0.
\end{equation}
The energy conservation law~\eqref{eqn.EnergyConsLagr} is not independent but
a consequence of all the equations~\eqref{eqn.MasterLagr}. Indeed, if we multiply each equation in~\eqref{eqn.MasterLagr} by a corresponding factor and sum up the result, we obtain equation~\eqref{eqn.EnergyConsLagr} identically:
\begin{equation}\label{eqn.SumRule}
U_{m_i}\cdot(\ref{eqn.MasterLagr.Momentum})+U_{F_{ij}}\cdot(\ref{eqn.MasterLagr.F})+U_{e_i}\cdot(\ref{eqn.MasterLagr.Electr})+U_{h_i}\cdot(\ref{eqn.MasterLagr.Magn})
 \equiv(\ref{eqn.EnergyConsLagr}).
\end{equation}
The same is true for \eqref{eqn.intro.energycons.M} and \eqref{eqn.intro.mastersys}.

\subsubsection{Possible interpretation of the state variables}
Usually, the derivation of a model begins with the choice of state variables. 
In the context of classical hydrodynamics, the answer is universal. The state 
variables are the classical hydrodynamic fields such as mass, momentum, 
entropy, or total energy. In any case which is beyond the inviscid 
hydrodynamics settings, the choice of extra state variables is not universal. 
In the HTC formalism, we however follow a different strategy, which consists of 
two stages. In the first stage, the governing equations are formulated before 
any choice of extra state variables has been made. The structure of the 
governing PDEs is a consequence of the five fundamental requirements formulated 
earlier in this Section. The physical meaning of the state variables becomes 
clear at the second stage, when we try to compare a solution to the model with 
specific experimental observations. At this stage, we  simultaneously clarify 
the meaning of the state variables and look for an appropriate energy potential 
which can be seen also as the choice of the constitutive relations in the 
classical continuum mechanics. For the proposed model, this strategy is 
realized in Section~\ref{sec.Closure}.

Thus, in this Section, we give only approximate interpretations of the state variables while their precise meanings will be given in Section~\ref{sec.Closure}.
As stated above, the space variables  $ \yy=[y_i] $ can
be treated as the Lagrangian coordinates which are
connected to the Eulerian coordinates $ \xx(t)=[x_i(t)] $
measured relative to a laboratory frame by the equality $ y_i=x_i(0)
$. It is also implied that $ v_i =\frac{{\rm d}x_i}{{\rm
    d}t}$ in~\eqref{eqn.intro.mastersys} is the velocity
of the matter relative to the laboratory frame,
while $ m_i =M_{v_i}$ in~\eqref{eqn.MasterLagr} has a meaning of a 
generalized momentum density which may include contributions from
other physical processes and in general depends on the 
specification of the potential $ M $, or $ U $. As it will be shown
in Section~\ref{sec.Closure}, $ m_i $ couples the
material momentum and \textit{electromagnetic
  momentum}~(Poynting vector). The tensorial variable
$F_{ij}=\frac{\partial x_i}{\partial y_j} $ is the
deformation gradient, $ e_i $ and $ h_i $ are the
electric and magnetic fields, respectively. However, the
exact meaning of the fields $ e_i $ and $ h_i $ will be
clarified later in Section~\ref{sec.EulerianFrame} when
we shall distinguish among different reference frames.

\subsubsection{Symmetric hyperbolicity}

System \eqref{eqn.intro.mastersys} can be rewritten in a symmetric quasilinear from 
\begin{equation} 
\label{eqn.lag.symmetric}
\mathbb{M}\frac{\pd \pp}{\pd t} + \mathbb{N}_{j}\frac{\pd \pp}{\pd y_{j}}=0, 
\end{equation} 
with the symmetric matrix $ \mathbb{M}(\pp)=M_{\pp\pp}
=[\pd^2 M/\pd p_i \pd p_j ]$ and \textit{constant} symmetric matrices  
$ \mathbb{N}_{j} $ consisting only of $ 1 $, $ -1 $ and
zeros. Moreover, system (\ref{eqn.intro.mastersys}) is
symmetric hyperbolic if $ M(\pp) $ is convex. In other
words, the Cauchy problem for (\ref{eqn.intro.mastersys})
(as well as for  \eqref{eqn.MasterLagr}) is automatically
well posed locally in time for smooth initial data
\cite{Dafermos2005}. 
We recall that the convexity of $ M(\pp) $ is equivalent
to the convexity of $ U(\qq) $  
due to the properties of the Legendre transformation.

\subsubsection{$ \qq $ and $ \pp $-type state variables}\label{sec.qp}

We emphasize the very distinct nature of the variables $
\qq $ and $ \pp $. Namely, the variables $ \qq $ appear
in the time derivative and have the meaning of densities
(volume average quantities), we thus shall refer to
components of $\qq$ as \textit{density fields}. On the
other hand, the variables $ \pp $ appear as fluxes in the master system~\eqref{eqn.MasterLagr} (or~\eqref{eqn.intro.mastersys}), and thus will be referred to as \textit{flux fields} (surface defined quantities). See also discussion in~\cite{Pesh2015}.
In the HTC formalism, the potential $ M(\pp) $ (the generalized pressure) and
the flux fields $ \pp $ are conjugate quantities to the potential $ U(\qq) $
(total energy density) and the density fields $ \qq $, i.e. there are connected
by the following identities
\begin{equation}\label{eqn.formalism.qp}
p_i=U_{q_i},\qquad q_i=M_{p_i}
\end{equation}
and
\begin{equation}\label{eqn.formalism.UM}
M=q_i U_{q_i}-U,\qquad U=p_i M_{p_i}-M.
\end{equation}
Thus, it follows from~\eqref{eqn.formalism.qp} that, in order the nonlinear 
change of variables~\eqref{eqn.formalism.qp} be a one-to-one map, one should
require that the potentials $ U(\qq) $ and $ M(\pp) $ be convex functions 
because
\begin{equation}\label{eqn.formalism.convexity}
M_{\pp\pp}=\frac{\pd\qq}{\pd \pp}=\left [\frac{\pd \pp}{\pd \qq}\right 
]^{-1}=U_{\qq\qq}.
\end{equation}

\subsubsection{Stationary constraints}

Solutions to system \eqref{eqn.MasterLagr} satisfy some stationary conservation laws that are compatible with system \eqref{eqn.MasterLagr} and conditioned by the structure of the flux terms:
\begin{equation}\label{eqn.StationaryLaws}
\dfrac{\partial F_{ij}}{\partial y_k}-\dfrac{\partial F_{ik}}{\partial y_j}= 0,\ \ \ 	\dfrac{\partial e_k}{\partial y_k}= 0,\ \ \ \dfrac{\partial h_k}{\partial y_k}= 0.
\end{equation}
These stationary laws hold for every $ t > 0 $ if 
they are valid at $ t = 0 $, and thus should be considered as the constraints on the initial data. Indeed, applying the divergence operator, for instance, to equations \eqref{eqn.MasterLagr.Magn} we obtain 
\begin{equation}
\dfrac{\partial}{\partial t}\left (\dfrac{\partial h_k}{\partial y_k}\right )=0,
\end{equation}
which yields the third equation in~\eqref{eqn.StationaryLaws} if it was fulfilled at the initial time. The other laws can be obtained in a similar way. As we shall discuss later on the example of the Eulerian equations, the situation is rather different in the Eulerian setting, and the stationary constraints like~\eqref{eqn.StationaryLaws} are not separate but an intrinsic part of the structure 
of the governing equations written in the Eulerian frame.

\subsubsection{Complimentary structure}\label{sec.complimentary}
We also note a \textit{complimentary} structure of equations~\eqref{eqn.MasterLagr} and \eqref{eqn.intro.mastersys}, i.e. the PDEs are split into pairs. In each pair, a variable appearing in the time derivative, say $ u_i $ in~\eqref{eqn.MasterLagr.Momentum}, then appears in the flux of the complimentary equation as $ U_{u_i} $ in \eqref{eqn.MasterLagr.F}. Thus, $ u_i $ and $ F_{ij} $ are complimentary variables, as well as $ e_i $ and $ h_i $. This means, that a physical process should be always presented at least by two state variables and hence by two PDEs in the HTC formalism. One may note a close relation of such a complimentary structure of the HTC formalism and the odd and even parity of the state variables with respect to the time-reversal transformation in the context of the GENERIC (general equation of nonequilibrium reversible-irreversible coupling) formalism discussed in~\cite{Pavelka2014a}.

\section{Master system in the Eulerian frame}
\label{sec.EulerianFrame}

In this section, we formulate a system of governing equations describing motion 
of a heat conducting deformable medium (fluid or solid) in the electromagnetic 
field in the Eulerian frame. This system is obtained as a direct consequence of 
the master system~\eqref{eqn.MasterLagr} by means of the Lagrange-to-Euler 
change of variables: $ \yy\rightarrow\xx $. This transformation is a nontrivial 
task, and the details are given in~\ref{app.EulertoLagr} for the 
electromagnetic field equations and in~\ref{app.EulertoLagr.momentum} for the 
momentum conservation law while the details about the derivation of the other 
equations can be found in the Appendix in~\cite{Pesh2015} or 
in~\cite{godunov1996systems}. We give the Eulerian formulations using both 
density fields $ \qq $ and flux fields $ \pp $. As we shall see, the Eulerian 
equations do not have such a simple structure as the Lagrangian equations. 
Nevertheless, we stress that none of the differential terms was prescribed 
``\textit{by hand}'', but all of them are a direct consequence of the $ 
\yy\rightarrow\xx $ variable transformation solely.

\subsection{$ (\mcE,\qq)$-formulation }

The main system of governing equations studied in this paper is formulated in terms of $ \qq $-type state variables (density fields, see Section~\ref{sec.qp}) 
\begin{equation}\label{eqn.VarEulerCons}
\qq=(\rho,\m,\AAA,\ee,\hh,\w,\sigma),
\end{equation}
and the total energy density $ \mathcal{E}(\q)=w^{-1} U $,  where $ U $ is 
the Lagrangian total energy density introduced in 
Section~\ref{sec.HTC.Lagr}, $ 
\rho $ is the mass density, $ \sigma=\rho s $ is the
entropy density, $ s $ is the specific entropy, $ \m=[m_i]$ is a generalized
momentum density which couples the ordinary matter momentum density, $ \rho\vv 
$, with the electromagnetic momentum density, i.e. the Poynting vector.
The exact expression for $ \m $ will be given
later. Matrix $ \AAA=[A_{ik}] $ is the distortion field\footnote{Rigorously 
speaking, $ \AAA $ is not a tensor field of rank 2, since it transforms like a 
tensor of
rank 1 with respect to a change of coordinates. Thus, we shall avoid to call it 
the distortion tensor, but instead call it simply the distortion field.} (see 
Paper~I), $ \ee=[e_i] $ and $ \hh=[h_i] $ are the vector fields which relate to 
the electro-magnetic fields and will be specified later, $ \w=\rho\JJ $ is the 
thermal
impulse density (see Paper~I), which can be interpreted as an average momentum density 
of the heat carriers. The velocity of the media, $ \vv$, is not a primary state 
variable and should be computed from the generalized momentum $ \m$,
but also, according to the HTC formalism, the velocity and the generalized momentum relate to each other as $ v_i=\mcE_{m_i} $ (see~\eqref{eqn.intro.pvariables} and the discussion below). 
In the Eulerian coordinates $ x_k $, the system of governing equations reads as
\begin{subequations}\label{eqn.HPR}
	\begin{align}
	& \frac{\partial \rho}{\partial t}+\frac{\partial (\rho v_k)}{\partial x_k}=0,\label{eqn.HPR.conti}\\[2mm]
	&\displaystyle\frac{\partial m_i }{\partial t}+\frac{\partial }{\partial 
	x_k}\left(m_i v_k + \delta_{ik} \left( \rho \mcE_\rho + m_l\mcE_{m_l} + e_l 
	\mcE_{e_l} + h_{l} \mcE_{h_l} - \mcE \right)  + A_{li}\mcE_{A_{lk}} - e_k 
	\mcE_{e_i} -  h_{k} \mcE_{h_i} \right)=0, \label{eqn.HPR.momentum}\\[2mm]
	&\displaystyle\frac{\partial A_{i k}}{\partial t}+\frac{\partial (A_{il} v_l)}{\partial x_k}+v_j\left(\frac{\partial A_{ik}}{\partial x_j}-\frac{\partial A_{ij}}{\partial x_k}\right)
	=-\dfrac{ \mcE_{A_{ik}} }{\rho \,\theta_1(\tau_1)},\label{eqn.HPR.deformation}\\[2mm]
	&\displaystyle\frac{\partial e_i}{\partial t} +  \frac{\partial \left( e_i v_k - v_i e_k - \varepsilon_{ikl} \mcE_{h_l} \right)}{\partial x_k} + v_i\dfrac{\pd e_k}{\pd x_k} = - \frac{ \mcE_{e_i}}{\eta}, \label{eqn.HPR.Efield}\\[2mm]
	&\displaystyle\frac{\partial h_i}{\partial t} + \frac{\partial \left( h_i v_k - v_i h_k + \varepsilon_{ikl} \mcE_{e_l} \right)}{\partial x_k} + v_i\dfrac{\pd h_k}{\pd x_k} = 0, \label{eqn.HPR.Hfield}\\[2mm]
	&\displaystyle\frac{\partial w_i}{\partial t}+\frac{\partial \left(w_i v_k + \mcE_{\sigma} \delta_{ik}\right)}{\partial x_k}=-\dfrac{ \rho\, \mcE_{w_i}}{\theta_2(\tau_2)}, \label{eqn.HPR.heatflux}\\[2mm]
	&\displaystyle\frac{\partial \sigma}{\partial t}+\frac{\partial \left(\sigma v_k + \mcE_{w_k} \right)}{\partial x_k}=\frac{1}{\mcE_{\sigma}}\left (\dfrac{1}{\rho \theta_1} \mcE_{A_{ik}} \mcE_{A_{ik}} + \dfrac{\rho}{\theta_2} \mcE_{w_i} \mcE_{w_i} + \frac{1}{\eta} \mcE_{e_i} \mcE_{e_i}\right )  \geq 0. \label{eqn.HPR.entropy}
	\end{align}
\end{subequations}
The energy conservation law
\begin{multline}
\frac{\partial \mcE}{\partial t}+\frac{\partial}{\partial x_k} \left( v_k \mcE 
+ v_i 
\left[ \left ( \rho \mcE_\rho + m_l\mcE_{m_l} + e_l \mcE_{e_l} + h_{l} 
\mcE_{h_l} - \mcE\right ) \delta_{ik} +  
A_{li} \mcE_{A_{lk}}  - e_k \mcE_{e_i}- h_{k} \mcE_{h_i} \right] 
\phantom{1^1}\right.\\ 
\left.   + \varepsilon_{ijk} \mcE_{e_i} \mcE_{h_j} + \mcE_{\sigma} \mcE_{w_k} \right)=0 \label{eqn.Euler.energy}
\end{multline}
is a consequence of equations~\eqref{eqn.HPR}, i.e. it can be obtained by means of the summation rule~\eqref{eqn.SumRule}.
We emphasize that in the numerical computations shown later in Section~\ref{sec:model}, we solve the energy equation (\ref{eqn.Euler.energy}) instead of the entropy equation \eqref{eqn.HPR.entropy}, but from the point of view of the model formulation, the entropy should be considered among the vector of unknowns because it is the complementary variable to the thermal impulse $ \w=\rho\JJ $, see the remark in Section \ref{sec.complimentary} and Paper~I.

All equations in system~\eqref{eqn.HPR} except the continuity equation\footnote{The continuity equation is, in fact, a consequence of the distortion equation \eqref{eqn.HPR.deformation}, e.g. see~\cite{GodRom2003,Pesh2015}, but it is convenient to consider density as an independent state variable with the compatibility constraint $ \rho=\rho_0\det(\AAA) $.}~\eqref{eqn.HPR.conti} and the heat conduction equation\footnote{A different form of the hyperbolic heat conduction is possible, see system~(38) in \cite{Rom1998}, which is fully compatible with the HTC formalism in the sense that its Lagrangian equations belong to the master system~\cite{godunov1996systems}. However, both forms are consistent in the Fourier approximation and because we do not consider non-Fourier heat conduction we follow the hyperbolic heat conduction formulation from Paper~I in this study. The detailed comparison of the heat conduction \eqref{eqn.HPR.heatflux}--\eqref{eqn.HPR.entropy} and \cite{Rom1998} is the subject of an ongoing research and will be presented somewhere else.}~\eqref{eqn.HPR.heatflux}--\eqref{eqn.HPR.entropy} originate from the Lagrangian equations with the structure~\eqref{eqn.MasterLagr}. The momentum  equation and the distortion equation are derived from the pair~\eqref{eqn.MasterLagr.Momentum}--\eqref{eqn.MasterLagr.F}, the electromagnetic field equations~\eqref{eqn.HPR.Efield}--\eqref{eqn.HPR.Hfield} are derived from the pair \eqref{eqn.MasterLagr.Electr}--\eqref{eqn.MasterLagr.Magn}. 

The energy conservation law \eqref{eqn.Euler.energy} is the consequence of equations~\eqref{eqn.HPR}, since it can be obtained as a linear combination of all equations~\eqref{eqn.HPR} with 
coefficients introduced in the following section. As in the Lagrangian frame, these coefficients (multipliers) are the thermodynamically conjugate state variables and have the meaning of fluxes. 

As discussed in~\cite{PeshRom2014,DPRZ2016}, the distortion field $ \AAA $ describes deformability and orientation of the continuum particles which assume to have a finite (non-zero) length scale. Macroscopic flow is naturally considered as the process of continuum particles rearrangements in the HTC model. Because of the rearrangements of particles, the field $ \AAA $ is not integrable in the sense that it does not relate Eulerian and Lagrangian coordinates of the continuum. As a result, the field $ \AAA $ is local and it relates to the deformation gradient $ \FF $ introduced in Section~\ref{sec.VariationNature} only via
\begin{equation}
\det(\FF)=1/\det(\AAA).
\end{equation}
However, if we consider a particular case of
system~\eqref{eqn.HPR} when the dissipation term in the
right hand side of~\eqref{eqn.HPR.deformation} is absent,
which corresponds to an elastic solid (e.g. see the last
numerical example in~Paper~I), then we have that $ \AAA=\FF^{-1} $.

For simplicity, we use the same notations $ m_i $, $ e_i $ and $ h_i $ for the 
generalized momentum, electric and magnetic fields in both the Lagrangian and 
the Eulerian framework. However, these fields are different, 
see~\ref{app.EulertoLagr}. 
For example, if we denote by $ \m_L $, $ \ee_L$ and $ \hh_L$ the Lagrangian 
fields, i.e. exactly those fields which are used in 
equations~\eqref{eqn.MasterLagr.Momentum}, \eqref{eqn.MasterLagr.Electr} and 
\eqref{eqn.MasterLagr.Magn}, 
then they are related to $ \m $, $ \ee $ and $ \hh $ appearing in the Eulerian 
equations~\eqref{eqn.HPR.momentum}, \eqref{eqn.HPR.Efield} and 
\eqref{eqn.HPR.Hfield} as 
\begin{equation}\label{eqn.eh_prime}
\m_L=w\,\m,\ \ \ \ee_L = w \FF^{-1}\ee,\ \ \ \hh_L = w \FF^{-1}\hh,
\end{equation}
where $ w=\det(\FF)=1/\det(\AAA) $, and $\FF $ is the deformation gradient 
introduced in Section~\ref{sec.VariationNature}. Subsequently, the Lagrangian 
total energy density $ U $ relates to the Eulerian total energy density $ 
\mcE $  as
\begin{equation}\label{eqn.UE}
w^{-1}U(\m_L,\FF,\ee_L,\hh_L)=w^{-1}U(w\,\m,\FF,w \FF^{-1}\ee,w 
\FF^{-1}\hh)=\mcE(\rho,\m,\FF,\ee,\hh).
\end{equation}
Here, for brevity, we omit other state variables. For example, if we denote by $ \boldsymbol{\Sigma} = [\Sigma_{ij}] $ all the terms in the momentum flux~\eqref{eqn.HPR.momentum} except the advective term $ m_i v_k\, $, it can be shown (e.g. see~\cite{godunov1996systems} or Appendix in~\cite{Pesh2015}) that, after the Lagrange-to-Euler transformation, the Lagrangian momentum flux $ U_{F_{ij}} $, see \eqref{eqn.MasterLagr.Momentum},
transforms to the Eulerian momentum flux $ \Sigma_{ik}=\rho F_{kj}U_{F_{ij}} $ 
which in turn,  because of the change of the state variables like~\eqref{eqn.eh_prime}, expands as
\begin{equation}\label{eqn.EulerStress}
\Sigma_{ik} = -\delta_{ik} P  - A_{li} \mcE_{A_{lk}} + e_k \mcE_{e_i}+ h_{k} 
\mcE_{h_i},
\end{equation}
where the scalar
\begin{equation}\label{eqn.GeneralPressure}
P=\rho \mcE_\rho + m_{l} \mcE_{m_l} + e_l \mcE_{e_l} + h_{l} \mcE_{h_l} - \mcE
\end{equation}
is the \textit{generalized pressure} which includes the hydrodynamic pressure and electromagnetic pressure.
Nevertheless, $ P $ is not a \textit{total pressure}\footnote{We emphasize, however, that the definition of the pressure plays no role in the model formulation and introduced merely for convenience.} in the media given by the trace $ \Sigma_{ii} $ of the total stress tensor $ \boldsymbol{\Sigma}=[\Sigma_{ik}] $ and the rest of the terms in~\eqref{eqn.EulerStress} may also contribute to the total pressure $ \Sigma_{ii} $.

In~\eqref{eqn.EulerStress}, we also use that $ \partial w/\partial F_{ij} = w F_{ji}^{-1}$, $ \partial F_{jk}^{-1}/\partial F_{il}=-F_{ji}^{-1}F_{lk}^{-1} $ and that $ \rho=\rho_0 w^{-1} $, where $ \rho_0 $ is the reference mass density and $ F_{ij}^{-1} $ are the entries of the inverse deformation gradient $ \FF^{-1} $ (should not be confused with $ 1/F_{ij} $). 
It is important to emphasize that the \textit{structure} of \eqref{eqn.EulerStress} does not depend on the specification of the total energy, 
but it is only conditioned by the structure of the governing equations. 

Note that strictly speaking it is not correct to call $ \boldsymbol{\Sigma} $ the stress tensor because, classically, by the stress tensor the non-advective flux of  the \textit{matter  momentum} conservation law is understood, while equation~\eqref{eqn.HPR.momentum} is the \textit{conservation law} for the \textit{matter-field} momentum $ \m $. Nevertheless, for brevity, $ \boldsymbol{\Sigma} $ shall be referred to as the stress tensor in the rest of the paper.  Recall that the matter momentum $ \rho\v $ is not a conservative quantity in the case of a medium moving in an electromgnetic field.
In general, the question of definition of the force acting on a matter moving in an electromagnetic field seems to have no universally accepted solution and several forms of the stress tensor are known~\cite{eringen2012electrodynamics,MakarovRukhadze,DOROVSKY199491}.
It is thus necessary to emphasize that the form of the
momentum flux \eqref{eqn.HPR.momentum}, quite abstract yet, is fully determined by the structure of the fluxes of the entire system~\eqref{eqn.HPR} and, at this moment, it does not depend on the physical settings. The only remaining degree of freedom to fulfill experimental observations is to specify the total energy potential $ \mcE $ and define a proper meaning of the state variables which then completely determine the force acting on the matter moving in an electromagnetic field. An example of such a potential and state variables will be given in Section~\ref{sec.Closure}.


We now make a very important remark about the divergence terms, $ \frac{\pd e_k}{\pd x_k} $ and $ \frac{\pd h_k}{\pd x_k} $, in equations~\eqref{eqn.HPR.Efield} and \eqref{eqn.HPR.Hfield} as they have a clear physical meaning of electric and magnetic (hypothetical though) charge, respectively. It is necessary to emphasize that these terms are not added by hand, but they emerge during the Lagrange-to-Euler change of variables, see~\ref{app.EulertoLagr}. Moreover, as can be checked by 
taking the divergence of Eqs.~\eqref{eqn.HPR.Efield} and \eqref{eqn.HPR.Hfield}, the following formal relations hold
\begin{equation}\label{eqn.euler.charges}
\dfrac{\pd R}{\pd t} + \dfrac{\pd (v_i R + \mathcal{J}_i)}{\pd x_i} = 0,\qquad \dfrac{\pd Q}{\pd t} + \dfrac{\pd (v_i Q + K_i)}{\pd x_i} = 0\,,
\end{equation}
where $ R = \frac{\pd e_i}{\pd x_i}$ and $ \mathcal{J}_i = \frac{1}{\eta}E_{e_i} $ are the volume electric charge and the electric current, while
$ Q = \frac{\pd h_i}{\pd x_i}$ and $ K_i \equiv 0 $ are, at least formally, the analogous magnetic charge and magnetic current.
Because it is assumed that $ K_i\equiv0 $, it follows from~\eqref{eqn.euler.charges}$_2$ that $ Q = \pd h_i/\pd x_i\equiv 0 $ if it was so at the initial moment of time. Nevertheless, we stress that the term $ \pd h_i/\pd x_i $ should not be dropped out from the equation~\eqref{eqn.HPR.Hfield} because, first, this would destroy the Galilean invariance of the system and, secondly, such a system will be not compatible with the energy conservation. 

\subsection{$(L,\pp)$-formulation and symmetric hyperbolicity}

In the previous section, the governing equations were formulated in terms of the state variables $ \qq $ and the total energy potential $ \mcE(\qq) $. In this section, we also provide another, dual, formulation in terms of $ \pp $-type state variables (flux fields)  and a potential $ L $ whose physical meaning is, in fact, identical to the generalized pressure~\eqref{eqn.GeneralPressure}. This formulation is not used for the numerical solution, but allows to emphasize an exceptional role of the generating potentials $ L $  and $ \mcE $ in our formalism. 

If we introduce new state variables as the following partial derivatives of the total energy density $ \mcE=\rho E $ with respect to the conservative state variables
\begin{equation}
\pp=(r,v_i,\alpha_{ik},d_i,b_i,\eta_k,T)
\end{equation} 
where
\begin{equation*}
\begin{array}{c}
r=\mcE_\rho,\qquad v_i=\mcE_{m_i},\qquad  \alpha_{ik}=\mcE_{A_{ik}},\qquad  d_i=\mcE_{e_i},\qquad b_i=\mcE_{h_i}, \qquad
\eta_k=\mcE_{w_k},\qquad T=\mcE_{\sigma},
\end{array} 
\end{equation*}
and a new potential $ L(\pp)=r\rho + v_im_i + \alpha_{ij}A_{ik} + d_i e_i + b_i h_i + \eta_i w_i-\mcE $ as the Legendre transformation of $ \mcE $ then system~\eqref{eqn.HPR} can be rewritten as (see details in~\cite{GodRom2003,Rom1998,Rom2001,Pesh2015})
\begin{subequations}\label{eqn.MasterEulerL}
	\begin{align}
	& \displaystyle\frac{\pd L_r}{\pd t} + \frac{\partial  [(v_k L)_r]}{\partial x_k} = 0, \label{eqn.MasterEulerL.Mass}\\[2mm]
	& \displaystyle\frac{\pd L_{v_i}}{\pd t} + \frac{\partial}{\partial x_k}  \left[(v_k L)_{v_i}  + \alpha_{mk}L_{\alpha_{mi}}-\delta_{ik}\alpha_{mn}L_{\alpha_{mn}} 
	- d_i L_{d_k}-b_i L_{b_k} \right] = 0, \label{eqn.MasterEulerL.Momentum}\\[2mm]
	& \displaystyle\frac{\pd L_{\alpha_{ik}}}{\pd t} + \frac{\partial  [(v_m L)_{\alpha_{im}}]}{\partial x_k} + \varepsilon_{klj}\varepsilon_{lmn}v_j\dfrac{\pd L_{\alpha_{in}}}{\pd x_m}= 
	 -\frac{1}{\rho \theta_1} \alpha_{ik}, \label{eqn.MasterEulerL.F}\\[2mm]
	& \displaystyle\frac{\pd L_{d_{i}}}{\pd t}+\frac{\partial[ {(v_kL)_{d_i}} - v_iL_{d_k}-\varepsilon_{ikl}b_l]}{\partial x_k}+v_i \dfrac{\pd L_{d_k}}{\pd x_k} = 
	-\frac{1}{\eta} d_i, \label{eqn.MasterEulerL.Electr}\\[2mm]
	& \displaystyle\frac{\pd L_{b_{i}}}{\pd t}+\frac{\partial[ {(v_kL)_{b_i}} - v_iL_{b_k}+\varepsilon_{ikl}d_l]}{\partial x_k}+v_i \dfrac{\pd L_{b_k}}{\pd x_k}=0,\label{eqn.MasterEulerL.Magn}\\[2mm]
	& \displaystyle\frac{\pd L_{\eta_i}}{\pd t}+\frac{\partial[ (v_kL)_{\eta_i}+\delta_{ik}T]}{\partial x_k} = -\frac{\rho}{\theta_2} \eta_i,  \label{eqn.MasterEulerL.Heat}\\[2mm]
	& \displaystyle\frac{\pd L_{T}}{\pd t}+\frac{\partial[ (v_kL)_T+\eta_k]}{\partial x_k}=0.  \label{eqn.MasterEulerL.Entropy}
	\end{align}
\end{subequations}
In terms of the potential $L$, the symmetric Cauchy stress tensor $\Sigma_{ik}$ reads 
\begin{equation}\label{eqn.EulerStress.L}
 \Sigma_{ik} = -\delta_{ik} \left( L - \alpha_{mn} L_{\alpha_{mn}} \right)  - \alpha_{mk} L_{\alpha_{mi}} + d_i L_{d_k} + b_i L_{b_k}.
\end{equation}

As in the Lagrangian framework, the parametrization of the governing equations in terms of the flux fields $ \pp $ and generating potential $ L(\pp) $ allows to rewrite the system in a symmetric quasilinear form. Moreover, if $ L $ is a convex potential and if we neglect the algebraic source terms on the right hand side, then the system \eqref{eqn.MasterEulerL} is \textit{symmetric hyperbolic}, since it can be written as  
\begin{equation}\label{eqn.symm}
    \mathbf{\mathcal{M}}(\pp)\dfrac{\partial\pp}{\partial t} + \mathbf{\mathcal{H}}_k(\pp)\dfrac{\partial \pp}{\partial x_k}=0, 
\end{equation} 
with $ \mathbf{\mathcal{M}}^\mathsf{T}=\mathbf{\mathcal{M}} = L_{\pp \pp} > 0$ and $ \mathbf{\mathcal{H}}_k^\mathsf{T} =\mathbf{\mathcal{H}}_k$. 
This was discussed many times in~\cite{GodRom1995,GodRom1996,Rom1998,GodRom1998,Rom1998} and we omit these details.


\subsection{Closure relation and the choice of state variables\label{sec.Closure}}

As we have seen in the previous sections and Paper I, the total energy potential $ \mcE(\qq) $ or its dual potential $ L(\pp) $ has the meaning of a \textit{generating potential}, and in order to close system \eqref{eqn.HPR} or \eqref{eqn.MasterEulerL}, one needs to specify one of these potentials (the other one then can be obtained as the Legendre transformation). In this section, we provide a particular example for the energy $ \mcE $ which completely defines equations~\eqref{eqn.HPR} and which then will be used in the numerical part of the paper. Of course, other specifications of the energy $ \mcE $ are possible.

One should however note that there is still no universally accepted set of equations describing the motion of a deformable dielectric medium in electromagnetic fields, and we do not have a 
reference system of PDEs to compare with. At the same time, we note that the resulting equations are not in contradiction with the existing theories, e.g.~\cite{Landau1984electrodynamics,GinzbUgarov,MakarovRukhadze,eringen2012electrodynamics} but rather generalize them.

The HTC formalism described above and in the
papers~\cite{godunov1996systems,GodRom1995,GodRom1996,Rom1998,Rom2001,GodRom2003}
gives us the information about the general structure of
the macroscopic time evolution equations, while the
question of the choice of the state variables remains
unaddressed. However, we are interested only in those
state variables whose time evolution equations can be
cast into the forms~\eqref{eqn.HPR} or
\eqref{eqn.MasterEulerL}. In this section we introduce
such state variables for the electromagnetic field.

After the choice of the state variables has been done, another nontrivial task is to specify the generating potentials $ \mcE $ or $ L $. According to the HTC formalism, such potentials should be convex functions of the  chosen state variables and depend on these variables only through their invariants. In general, these potentials should be derived from microscopic theories such as nonequilibrium statistical physics, kinetic theory, etc. To the best of our knowledge, there were no successful attempts to derive such potentials from microscopic theories for such a general case considered here. In this section, we complete the model formulation by specifying state variables and the potential~$ \mcE=\rho E $.


We assume the following additive decomposition of the total energy
\begin{equation}\label{eqn.EnergyAdditive}
\mcE(\rho,\m,\AAA,\ee,\hh,\JJ,s)=\\
\mcE_{\rm micro}(\rho,s) + \mcE_{\rm meso}(\AAA,\JJ) + \mcE_{\rm macro}( \m,\ee,\hh).
\end{equation}
These three terms are refereed to as the part of the
total energy distributed on the microscale ($ \mcE_{\rm
  micro} $ is the kinetic energy of the molecular
motion), on the mesoscale which is the scale of the
continuum particles, $ \mcE_{\rm meso} $, and on the
observable macroscale represented by $ \mcE_{\rm macro}
$. The first two terms were specified in Paper~I, while
the third term, macroscopic energy, was represented by
the macroscopic kinetic energy. 
In the presence of the electromagnetic field, the
macroscopic energy carries also the contribution of the
electromagnetic field.

In this paper, we use the same expressions for the
energies $ \mcE_{\rm micro} $ and $ \mcE_{\rm meso} $ as
specified in~\cite{PeshRom2014} and Paper~I. 
Moreover, for the rest of this section, we ignore the dissipative effects and thus we assume that $ \mcE_{\rm meso}=0 $ as it has no influence on the specification of the energy $ \mcE_{\rm macro} $. We are now in position to specify the macroscopic part $ \mcE_{\rm macro}( \m,\ee,\hh) $ of the total energy and to give a certain meaning to the electromagnetic fields $ \ee $ and $ \hh $ and to the generalized momentum $ \m $. The following strategy will be used. Because no exact physical meaning is assigned yet to the fields $ \m $, $ \ee $ and $ \hh $, we cannot construct the potential $ \mcE_{\rm macro}(\m,\ee,\hh) $ directly. However, as usual, the flux fields $ \vv $, $ \dd $ and $ \bb $ have more intuitive meaning\footnote{This, however, should not be a reason to use flux fields as the state variables because usually time evolution equations for the flux fields are highly complex and, which is more important, have no apparent structure.} and thus we first construct the dual potential $ L(r,\vv,\dd,\bb) $ and then, according to the HTC formalism, we define density fields and the potential $ \mcE $ as
\begin{equation}\label{eqn.qpEL}
\rho = L_r,\qquad m_i=L_{v_i},\qquad e_i = L_{d_i}, \qquad h_i=L_{b_i}, \qquad \mcE = r L_r + v_i L_{v_i}+d_i L_{d_i} + b_i L_{b_i} - L.
\end{equation}

Following the paper~\cite{Rom1998} and the discussion made in \ref{app.max.gpr}, we define the potential $ L $ (for brevity, we also assume that the medium is isentropic and omit entropy in this section) as the following function
\begin{equation}\label{eqn.Lpotetnial}
L(r,\vv,\dd,\bb)=\rho^{2} e_{\rho} + \frac{1}{2}\left( \epsilon' d_i d_i + \mu' b_i b_i \right) + \epsilon' \mu' \, \varepsilon_{ijk}v_i d_j b_k,
\end{equation}
or
\begin{equation*}
L(r,\vv,\dd,\bb)=\rho^2 e_{\rho} + \frac{1}{2}\left( \epsilon' \dd^{2} + \mu' \bb^{2} \right) + \epsilon' \mu' \, \mathbf{v} \cdot ( \dd \times \bb ) 
\label{eqn.Lpotetnial.vector}
\end{equation*}
of the state variables
\begin{equation}\label{eqn.qvdb}
\pp=(r,\vv,\dd,\bb),
\end{equation}
where $ e=\rho^{-1}\mcE_{\rm micro} $ is the specific internal energy, $ r=e+\rho e_{\rho} - v_i v_i/2 $ is a scalar state variable dual to the density (see~\eqref{eqn.qpEL}$ _1 $),  $ \vv $ is the velocity of the medium, $\mu' = \mu_0 \mu_r$ denotes the magnetic permeability (we use $\mu'$ throughout this paper, to avoid confusion with the fluid viscosity $\mu$, which is also used later) and $\epsilon' = \epsilon_0 \epsilon_r$ is the electric permittivity of the continuum. As usual, $\mu_0$ and $\epsilon_0$ are respectively the permeability and the 
permittivity of vacuum, while $\epsilon_r$ and $\mu_r$ are dimensionless parameters that depend on the material.  
We furthermore use the standard relation between the speed of light in the medium $c$, the magnetic permeability and the electric permittivity of the medium: 
\begin{equation}
  c^2 = \frac{1}{\epsilon' \mu'}. 
\end{equation}
Eventually, the fields $\dd=[d_i]$ and $\bb=[b_i]$ are the electric and magnetic fields  in the \textit{comoving} frame\footnote{

We note that it is necessary to distinguish a comoving frame from the Lagrangian frame of reference used in Section~\ref{sec.VariationNature}. The both frames move with the mater, however the distance between two points in the comoving frame changes in time (because it is measured with respect to the Laboratory frame) while it is constant in the Lagrangian frame (because the distance is measured with respect to the reference frame itself). That is why the transformation of fields between the comoving and the Laboratory frame involves only the velocity, like in~\eqref{eqn.dbDefinition.d} and \eqref{eqn.dbDefinition.b}, but it involves the deformation gradient $ F_{ij}=\frac{\pd x_i}{\pd y_j} $ in the other case, see\eqref{eqn.eh_prime}.} with velocity $\vv$, which are related to the electric field $\EE=[E_i]$ and the magnetic field $\BB=[B_i]$ in the \textit{laboratory frame} by 
\begin{eqnarray} 
	\dd = [E_i + \phantom{\frac{1}{c^2}}\varepsilon_{ijk} v_j B_k] &=& \EE + \vv \times \BB, \label{eqn.dbDefinition.d}\\[2mm]
	\bb = \frac{1}{\mu'} [B_i - \frac{1}{c^2} \varepsilon_{ijk}v_j E_k] &=& \frac{1}{\mu'} \left( \BB - \frac{1}{c^2} \vv \times \EE \right). \label{eqn.dbDefinition.b}
\end{eqnarray}

The motivation for this choice of the potential $ L $ is 
in the field  equations for a slowly moving medium (see
 \textsection 76 of \cite{Landau1984electrodynamics}
 and~\cite{MakarovRukhadze})  obtained under the
 assumption of small $ \vv^{2}/c^{2} $ ratios.  To show that the equations from~\cite{Landau1984electrodynamics,MakarovRukhadze} are indeed recovered in our formalism, one needs to put the partial derivatives  $ L_{d_i}=\pd L/\pd d_i $ and $ L_{b_i} =\pd L/\pd b_i $ 
\begin{eqnarray}
\ee = [L_{d_i}] = [\epsilon' d_i - \epsilon' \mu' \, \varepsilon_{ijk} v_j b_k] = \epsilon' \dd - \frac{1}{c^2}\v\times\bb, \label{eqn.e}\\[3mm]
\hh = [L_{b_i}] = [\mu' b_i      + \epsilon' \mu' \, \varepsilon_{ijk} v_j d_k] = \mu'      \bb + \frac{1}{c^2}\v\times\dd,\label{eqn.h}
\end{eqnarray}
into the equations \eqref{eqn.MasterEulerL.Electr}
and
\eqref{eqn.MasterEulerL.Magn}. 

According to the momentum
equation~\eqref{eqn.MasterEulerL.Momentum}, the
field-medium momentum density  $ \rho\uu=\m=[m_i] $ is defined as the partial derivative of $ L $ with respect to $ \vv $ 
\begin{equation}\label{eqn.rhou}
\m = [L_{v_i}] = [\rho v_i + \epsilon' \mu' \, \varepsilon_{ijk}d_j b_k] = \rho \v + \frac{1}{c^2} \dd\times\bb \,,
\end{equation}
which includes the Poynting vector.
Here, one needs to take into account that the potential $ L $ is not a fully explicit function of unknowns~\eqref{eqn.qvdb}. Namely, the hydrodynamic pressure $ \rho^2 e_\rho $ is the implicit part, and to compute the derivative $ L_{v_i} $ one needs also to express $ \rho^2 e_\rho $ in terms of $ r=e + \rho e_\rho -v_i v_i/2 $, $ v_i $, $ d_i $ and $ b_i $, e.g. see~\cite{GodRom2003}.


Note that fields $ \ee $ and $ \hh $ can be obtained from the electric and magnetic fields $ \EE $ and $ \BB $ in the Eulerian frame as
\begin{eqnarray}
\ee = & \epsilon' \left( \EE + \dfrac{1}{c^2} \vv \times ( \vv \times \EE) \right) \,, \\
\hh = &  \phantom{\mu'}\BB + \dfrac{1}{c^2} \vv \times ( \vv \times \BB).
\end{eqnarray}
An apparent difference between the fields $ (\dd,\bb) $ and $ (\ee,\hh) $ is that each of the fields $ (\dd, \bb) $ depends on both $ \EE $ and $ \BB $ while the fields $ (\ee,\hh) $ depend on either $ \EE $ or $ \BB $.

The unknown functions \eqref{eqn.qvdb} belong to the $ \pp $-type state variables (flux fields) in our classification, while the energy potential $ \mcE $ depends on $ \qq $-type state variables (density fields) which are $ \rho $, $ \m $, $ \ee $ and $ \hh $. Therefore, to complete the formulation of the model, we need to find the expression for $ \mcE(\rho,\m,\ee,\hh) $. 

According to the HTC formalism, total energy density $ \mcE(\rho,\m,\ee,\hh) $ 
is the Legendre transformation of $ L(r,\vv,\dd,\bb) $:
\begin{equation}
\mcE=rL_r + v_iL_{v_i} + d_i L_{	d_i} + b_i L_{b_i}  - L=
(e+\rho e_{\rho}-v_i v_i/2)\rho+v_i m_i + d_i e_i  + b_i h_i  - L.
\end{equation}

It appears that to express $\mcE=\rho E $ explicitly in terms of
$ \rho $, $ \m $, $ \ee $ and $ \hh $ only is a nontrivial task. However, using formulas~\eqref{eqn.rhou}, \eqref{eqn.e} and \eqref{eqn.h}, it can be easily expressed in terms of the dual variables $ \vv $, $ \dd $ and $ \bb $ as
\begin{equation}
\mcE(\rho,\vv,\dd,\bb) = \rho e + \frac{1}{2}\left( \rho \vv^2 +  \epsilon' \dd^{2} + \mu' \bb^{2} \right) + 
2 \epsilon' \mu' \left| 
\begin{array}{ccc}
v_1 & d_1 & b_1 \\ 
v_2 & d_2 & b_2 \\ 
v_3 & d_3 & b_3
\end{array}  \right |,
\label{eqn.Edual} 
\end{equation}
Moreover, because we restrict ourselves to flows for which $ \vv^2/c^2 \ll 1$ is small and terms of the order $ c^{-4} $ can be ignored, 
then an approximate expression for $ \mcE $ can be obtained in terms of $ \rho $, $ \m $, $ \ee $ and $ \hh $. Under such assumptions, $ \mcE $ can be approximated as 
\begin{equation}\label{eqn.Eapproax}
\mcE(\rho,\m,\ee,\hh) = \rho e + \halb \left ( \frac{1}{\rho} \m^2 +  \frac{1}{\epsilon'} \ee^{2} + \frac{1}{\mu'} \hh^{2} \right) - \\
  \frac{1}{2\rho}\left| 
\begin{array}{ccc}
m_1 & e_1 & h_1 \\ 
m_2 & e_2 & h_2 \\ 
m_3 & e_3 & h_3
\end{array}  \right |.
\end{equation}
When compared with \eqref{eqn.EnergyAdditive} then $\rho e $ is $ \mcE_{\rm micro} $ and the rest of the terms in~\eqref{eqn.Eapproax} are $ \mcE_{\rm macro} $, while we recall that $ \mcE_{\rm meso} $ was omitted in this section. Therefore, system~\eqref{eqn.HPR} together with the energy potential~\eqref{eqn.Eapproax} supplemented by $ \mcE_{meso} $ from Paper~I form the closed system of PDEs. 

Note that \eqref{eqn.Edual} can be also rewritten in terms of $ \vv $, $ \EE $ and $ \BB $ as 
\begin{equation}\label{eqn.energyEB}
\mcE = \rho e + \halb \left ( \rho \vv^2 + \epsilon' \EE^2 + \frac{1}{\mu'}\BB^2\right ),
\end{equation}
where we again ignore quadratic terms in $ \vv/c $ and terms of the order of $ c^{-4} $. Remark that the determinant term is not present in $ \mcE $ if the fields $ \EE $ and $ \BB $ are used.

We accomplish the model formulation by giving an explicit formula for the total stress tensor $ \boldsymbol{\Sigma} $ in the case 
of an inviscid medium moving in the electromagnetic field. Note that this expression for the stress tensor is, of course, the result 
of a particular definition of the energy potential~\eqref{eqn.Eapproax}. Thus, using formulas~\eqref{eqn.Eapproax} and \eqref{eqn.EulerStress}, one can write~\eqref{eqn.EulerStress} as
\begin{gather}\label{eqn.stressExplicit.meh}
\boldsymbol{\Sigma}(\rho,\m,\ee,\hh)=-P\II - \frac{1}{\rho}\m\otimes\m + \frac{1}{\epsilon'} \ee \otimes \ee + \frac{1}{\mu'} \hh\otimes\hh + \frac{1}{\rho}\m\otimes(\ee\times\hh),\\
P =  p + \frac{1}{2}\left (\frac{1}{\epsilon'} \ee^2 + \frac{1}{\mu'} \hh^2 \right).\label{eqn.Pressure.meh}
\end{gather}
where $ \otimes $ means the dyadic product, $ \II $ is
the identity tensor, $ p=\rho^2 e_\rho $ is the matter pressure.

If required, the stress tensor can be expressed in terms of $ \vv $, $ \dd $ and $ \bb $ using formulas~\eqref{eqn.EulerStress.L} and \eqref{eqn.Lpotetnial} as
\begin{gather}\label{eqn.stressExplicit.vdb}
\boldsymbol{\Sigma}(\rho,\vv,\dd,\bb)=-P\II + \epsilon' \dd \otimes\dd + \mu' \bb\otimes\bb + \epsilon'\mu' [\bb\otimes(\v\times\dd) - \dd\otimes(\v\times\bb)],\\
P =L=p + \frac{1}{2}\left( \epsilon' \dd^{2} + \mu' \bb^{2} \right) + \epsilon' \mu' \, \mathbf{v} \cdot ( \dd \times \bb ).\label{eqn.Pressure.vdb}
\end{gather}
Eventually, for the slowly moving media, the above formulas are equivalent to the following one written in terms of $ \v $, $ \EE $ and $ \BB $
\begin{gather}
\boldsymbol{\Sigma}(\rho,\vv,\EE,\BB)=-P\II + \epsilon' \EE \otimes\EE + \frac{1}{\mu'} \BB\otimes\BB ,\label{eqn.stressExplicit.vEB}\\
P =p + \frac{1}{2}\left( \epsilon' \EE^{2} + \frac{1}{\mu'} \BB^{2} \right) .\label{eqn.Pressure.vEB}
\end{gather}
We recall that the scalar $ P $ is introduced merely for convenience and should not be understood as the total pressure $ {\rm tr}\boldsymbol{\Sigma} $, trace of the stress tensor, but just as a part of it.

\section{The mathematical model for the numerical solution} 
\label{sec:model} 

The form of the equations adopted for the numerical simulations is obtained from system~\eqref{eqn.HPR} closed by \eqref{eqn.Eapproax} by dropping the terms of the order 
$ \v^2/c^2 $ and by expressing the contributions to the momentum and total energy equation in conventional form, i.e. by using the fluid pressure, the electro-magnetic pressure, the viscous stress tensor and the Maxwell stress tensor. From $ \v^2/c^2 \ll 1 $ 
it also follows that $ e_i = E_i / ( \mu' c^2) = \epsilon' E_i = D_i$ and $ h_i=B_i$, see also \ref{app.max.gpr}.  
In this notations, the relations between $ \v $, $ \dd $ and $ \bb $ and $ 
\m=\rho\uu $, $ \DD $ and $ \BB $ read as 
(see~\eqref{eqn.dbDefinition.b}, \eqref{eqn.dbDefinition.d} and \eqref{eqn.rhou})
\begin{eqnarray} 
	\dd = \frac{1}{\epsilon'}\DD + \vv \times \BB, \label{eqn.db2DB.d}\\[2mm]
	\bb = \frac{1}{\mu'}     \BB - \vv \times \DD, \label{eqn.db2DB.b}
\end{eqnarray}
and
\begin{equation}\label{key}
\rho\uu = \rho \vv + \epsilon' \mu' \, \dd \times \bb = \rho\vv + \DD \times \BB - \mu'\DD\times(\vv\times\DD) - \epsilon'\BB\times(\vv\times\BB).
\end{equation}

Furthermore, the compatibility condition $\nabla \cdot \BB = 0$ in general holds exactly only at the continuous level. Within a numerical method, discretization errors can
lead to a violation of this constraint, which is a well-known problem in computational electro-magnetics. Therefore, in this paper the divergence constraint on the magnetic 
field is imposed by making use of the hyperbolic generalized Lagrangian multiplier (GLM) approach of Dedner et al. \cite{Dedneretal}, which introduces an evolution equation for an 
\textit{additional auxiliary field} variable $\varphi$ with associated propagation speed $c_h$, that is supposed to propagate errors in the $\nabla \cdot \BB = 0$ constraint 
out of the computational domain. 
However, it has to be pointed out that there are also very important and widely-used numerical methods that guarantee the divergence condition \textit{exactly} even at the discrete
level by using a proper discretization of the equations on \textit{staggered} grids, see for example the Yee scheme \cite{YeeEM} for the time domain Maxwell equations and its 
extension to the MHD equations proposed by Balsara and Spicer in \cite{BalsaraSpicer1999}. For very recent developments concerning exactly divergence-free high order schemes, 
see \cite{balsarahllc2d,ADERdivB,BalsaraMUSIC1}. We also note that within this paper, we do not take any measures to enforce the compatibility conditions on the distortion $\AAA$ or
on the electric field $\DD$.  
After that, system~\eqref{eqn.HPR} reads as 
\begin{subequations}\label{eqn.GPR}
	\begin{align}
	& \frac{\partial \rho}{\partial t}+\frac{\partial \rho v_k}{\partial x_k}=0,\label{eqn.conti}\\[2mm]
	&\displaystyle\frac{\partial \left( \rho u_i \right)}{\partial t}+\frac{\partial \left(\rho u_i v_k + p \delta_{ik} - \sigma_{ik} + \beta_{ik} \right)}{\partial x_k}=0, \label{eqn.momentum}\\[2mm]
	&\displaystyle\frac{\partial A_{i k}}{\partial t}+\frac{\partial A_{im} v_m}{\partial x_k}+v_j\left(\frac{\partial A_{ik}}{\partial x_j}-\frac{\partial A_{ij}}{\partial x_k}\right)
	=-\dfrac{ \psi_{ik} }{\theta_1(\tau_1)},\label{eqn.deformation}\\[2mm]
	&\displaystyle\frac{\partial (\rho J_i)}{\partial t}+\frac{\partial \left(\rho J_i v_k+ T \delta_{ik}\right)}{\partial x_k}=-\dfrac{\rho H_i}{\theta_2(\tau_2)}, \label{eqn.heatflux}\\[2mm]
	&\displaystyle\frac{\partial D_i}{\partial t} + \frac{\partial \left( v_k D_i - v_i D_k - \varepsilon_{ikl} b_l \right)}{\partial x_k} + v_i \frac{\partial D_k}{\partial x_k} 
	= - \frac{1}{\eta} d_i, \label{eqn.Efield}\\[2mm]
	&\displaystyle\frac{\partial B_i}{\partial t} +
        \frac{\partial \left(v_k B_i - v_i B_k + \varepsilon_{ikl} d_l + \varphi \delta_{ik} \right)}{\partial x_k} 
        + v_i \frac{\partial B_k}{\partial x_k}=
        0, \label{eqn.Bfield}\\[2mm]
&\frac{\partial( \rho E)}{\partial t}+\frac{\partial \left(v_k \rho E + v_i [ p \delta_{ik} - \sigma_{ik} + \beta_{ik}] + \varepsilon_{ijk} d_i b_j  + q_k \right)}{\partial x_k}=0, \label{eqn.energy} 
	 \\[2mm] 
& \frac{\partial \varphi}{\partial t} + \frac{\partial \left( c_h^2 B_k \right)}{\partial x_k} = 0, \label{eqn.dedner.B} 	
\end{align}
\end{subequations}
where $\varepsilon_{ijk}$ is the three dimensional Levi--Civita tensor. 
Neglecting the presence of the artificial scalar $\varphi$, the entropy production equation is given according to 
\eqref{eqn.HPR.entropy} by  
\begin{equation}
\displaystyle\frac{\partial (\rho s)}{\partial t}+\frac{\partial \left(\rho s v_k + H_k \right)}{\partial x_k}=\dfrac{\rho}{\theta_1(\tau_1) T} \psi_{ik} \psi_{ik} + \dfrac{\rho}{\theta_2(\tau_2) T} H_i H_i + \frac{1}{\eta T} d_i d_i  \geq 0, \label{eqn.entropy}
\end{equation}
In the following, we will refer to the above model given by \eqref{eqn.conti}-\eqref{eqn.entropy} also as the Godunov-Peshkov-Romenski (GPR) model. 
These equations are the mass conservation (\ref{eqn.conti}), the momentum conservation~(\ref{eqn.momentum}), the time evolution for the distortion~(\ref{eqn.deformation}), 
the time evolution equations for the electric and magnetic field \eqref{eqn.Efield} and \eqref{eqn.Bfield}, the evolution equations for the thermal impulse~(\ref{eqn.heatflux}) and 
the entropy ~(\ref{eqn.entropy}) as well as the total energy conservation law given by~(\ref{eqn.energy}). 
The PDE governing the time evolution of the thermal impulse~\eqref{eqn.heatflux} looks formally very similar to the momentum equation \eqref{eqn.momentum}, where the temperature $T$ takes the role of the pressure $p$.  Due to this similarity, it will also be called the \textit{thermal momentum equation} in the following.

%

According to the definitions made in the previous sections of this paper,  
$ [A_{ik}]=\AAA $ is the distortion, $ [J_i]=\JJ $ is the 
thermal impulse vector, $ s $ is the entropy, $   \mathcal{E} = \rho E = \rho E(\rho,s,\vv,\AAA,\JJ,\DD,\BB)$ is the total energy density, 
$ p = \rho^2 e_{\rho} $ is the fluid pressure, $ e(\rho,s)=E_{\rm int}(\rho,s) $ is the internal energy defined below, $ \delta_{ik} $ is the Kronecker delta and 
$ [\sigma_{ik}]=\boldsymbol{\sigma} = [- A_{mi} (\rho E)_{A_{mk}}] $ is the symmetric stress tensor. Because we are interested in flows for which the ratio $ \v^2/c^2 \ll 1 $ is 
small (Newtonian limit), the Maxwell stress reduces to (see~\eqref{eqn.stressExplicit.vEB}--\eqref{eqn.Pressure.vEB})
%
 
\begin{eqnarray} 
\boldsymbol{\beta } 
&=& \frac{1}{2}\left(\frac{1}{\epsilon'} \DD^2 + \frac{1}{\mu'} \BB^2\right)  \II -  \frac{1}{\epsilon'} \DD \otimes \DD - \frac{1}{\mu'} \BB\otimes\BB.
\end{eqnarray}
 
Moreover, $ T = E_s = (\rho E)_{\rho s} $ is the temperature, 
$[q_k] = \mathbf{q} = [E_s E_{J_k}]$  is the heat flux vector; $ \theta_1 = \theta_1(\tau_1) > 0$ and $ \theta_2 = \theta_2(\tau_2) > 0$ are positive scalar functions, 
which will be specified below, and which depend on the strain dissipation time $ \tau_1 > 0$ and on the thermal impulse relaxation time $ \tau_2 > 0$. 
The parameter $\eta > 0$ is the electric resistivity of the medium. 
The viscous stress tensor and the heat flux vector are directly related to the dissipative terms on the right hand side via 
$ \boldsymbol{\sigma} = - \rho \AAA^T \boldsymbol{\psi} $ and $ \mathbf{q} = T \, \mathbf{H}$.

The first two non-conventional dissipative terms $\psi_{ik}$ and $H_i$ on the right hand side of the evolution equations for $\AAA$, $\JJ$, $E_i$ and $s$ are 
given by $[\psi_{ik}] = \boldsymbol{\psi} = [E_{A_{ik}}]$ and $[H_i] = \mathbf{H} = [E_{J_i}]$, respectively. 
The algebraic source term on the right-hand side of equation~(\ref{eqn.deformation}) describes the shear strain dissipation due to material element rearrangements, 
see \cite{PeshRom2014} for a detailed discussion. The source term in (\ref{eqn.heatflux}) describes the relaxation of the thermal impulse due to heat exchange 
between material elements, while the one in the governing equations for the electric field \eqref{eqn.Efield} is the well-known law of Ohm \cite{OhmLaw}. 

We stress that in the GPR model \eqref{eqn.conti}-\eqref{eqn.entropy} derived within the HTC framework, \textit{all} dissipative processes have the \textit{same}  
structure and take the form of algebraic relaxation source terms. The structure of these terms is a \textit{result} of the HTC formalism, in order to 
guarantee energy conservation and consistency with the second principle of thermodynamics, see also \cite{DPRZ2016}. 

As a result of this observation, it is very interesting to note that the dissipative term $-\frac{1}{\eta} d_i$ in the PDE for the \textit{electric field} 
is given by the well-known \textit{Ohm law} \cite{OhmLaw} discovered in 1826, which already \textit{has} a suitable structure that directly fits into the 
HTC framework. It is a question of mere philosophical nature, but from the viewpoint of hyperbolic thermodynamically compatible systems, it seems that a 
mathematically more consistent and perhaps even more profound insight into the physics of dissipative processes has first been discovered in the equations 
of electro-dynamics rather than in the classical standard laws of dissipative transport processes, such as the Newtonian law for viscous fluids and the 
Fourier law for heat transfer. The latter lead to \textit{parabolic} differential terms in the governing equations, while Ohm's law in the Maxwell equations 
does \textit{not} generate parabolic terms. 

As detailed in the previous sections, $E_\rho $, $E_s $, $E_{A_{ik}} $ and $E_{J_i} $ should be understood as the partial derivatives 
$ \partial E/\partial \rho$, $ \partial E/\partial s $,  $ \partial E/\partial A_{ik}$ and $ \partial E/\partial J_i$; 
they are the so-called \textit{energy gradients in the state space} or the \textit{thermodynamic forces}. 

One can clearly see that in order to close the system, it is necessary to specify the total energy potential $ \rho E(\rho,s,\vv,\AAA,\JJ,\dd,\bb) $. This potential then generates all the constitutive fluxes (\textit{i.e.} non advective fluxes)  and source terms by means of its partial derivatives with respect to the state variables. Hence, the energy specification is one of the key steps in the model formulation.

The total energy density (i.e. energy per unit volume) $ \rho E$ is the sum of four terms, i.e.
\begin{equation}\label{eq:total_energy}
 \rho E(\rho,s,\vv,\AAA,\JJ,\dd,\bb)=  \rho E_{\rm int}(\rho,s) +  \rho E_{\rm mes}(\AAA,\JJ) +  \rho E_{\rm kin}(\vv) + \mathcal{E}_{\rm em}(\dd,\bb)\,,
\end{equation}
where 
\begin{itemize}
\item
$E_{\rm int}$ is the specific (i.e. per unit mass) internal energy, which depends on the equation of state chosen, and which in the rest of the  paper we assume to be that of an ideal gas
\begin{equation}\label{eq:ideal_gas_eos}
E_{\rm int}(\rho,s)=\frac{c_0^2}{\gamma(\gamma-1)},\ \ c_0^2=\gamma\rho^{\gamma-1}e^{s/c_V}\,,
\end{equation}
or the stiffened gas equation of state
\begin{equation}\label{eq:stiff_gas_eos}
E_{\rm int}(\rho,s)=\dfrac{c^2_0}{\gamma(\gamma-1)}\left ( \dfrac{\rho}{\rho_0}\right )^{\gamma-1}e^{s/c_V}+\dfrac{\rho_0 c_0^2-\gamma p_0}{\gamma\rho},\ \ c_0^2=const.
\end{equation}
In both cases, $c_0$ has the meaning of the adiabatic sound speed, $ c_V $ is the specific heat capacity at constant volume, $ \gamma $ is the ratio of the specific heats, \textit{i.e.} $ \gamma=c_P/c_V $, if $ c_P $ is the specific heat capacity at constant pressure. In~\eqref{eq:stiff_gas_eos}, $ \rho_0 $ is the reference mass density, $ p_0 $ is the reference (atmospheric) pressure. 
\item
$ E_{\rm mes}$ is the specific energy density at the mesoscale level
\begin{equation}\label{eq:e_2}
E_{\rm mes}(\AAA,\JJ)=\dfrac{c_s^2}{4}G^{\rm TF}_{ij}G^{\rm TF}_{ij}+\frac{\alpha^2}{2}J_i J_i,
\end{equation}
with 
\begin{equation}
 [G_{ij}^{\rm TF}] = \dev(\GG) = \GG-\frac{1}{3} {\rm tr}(\GG) \II,  \qquad \textnormal{ and } \qquad \GG=\AAA^\mathsf{T}\AAA. 
\end{equation} 
Here, $[G_{ij}^{\rm TF}] = \dev(\GG)$ is the deviator, or the \textit{trace-free} part, of the tensor $\GG=\AAA^\mathsf{T}\AAA$ and ${\rm tr}(\GG)=G_{ii}$ is its trace, 
$ \II $ is the unit tensor and $ c_s $ is the characteristic velocity of propagation of transverse perturbations. 
In the following we shall refer to it as the \textit{shear sound velocity}. 
\item
$E_{\rm kin} = \dfrac{1}{2} v_i v_i$ is the specific kinetic energy and, finally
\item
$\mathcal{E}_{\rm em}$ is the energy density of the electromagnetic field, which is given  by (see \eqref{eqn.Edual})
\begin{equation} 
\label{eqn.electricenergy}
\mathcal{E}_{\rm em}(\dd,\bb) = \frac{1}{2}\left( \epsilon' \dd^{2} + \mu' \bb^{2} \right) + 2 \epsilon' \mu' \vv \cdot \left( \dd \times \bb \right), 
\end{equation}
or by the following approximate relation in terms of quantities in the laboratory frame (see \eqref{eqn.energyEB})  
\begin{equation} 
\label{eqn.electricenergy.lab}
\mathcal{E}_{\rm em}(\DD,\BB) = \frac{1}{2} \left ( \frac{1}{\epsilon'} \DD^2 + \frac{1}{\mu'} \BB^2\right )\,. 
\end{equation}
In our implementation, we have used \eqref{eqn.electricenergy}, since it uses less assumptions.

\end{itemize}

After the total energy potential has been specified, one can write all fluxes and source terms in an explicit form. Thus, for the energy $ E_{\rm mes}(\AAA,\JJ) $ given by~(\ref{eq:e_2}), 
we have $\boldsymbol{\psi}= E_{\AAA}= c_s^2 \AAA \dev(\GG)$, hence the shear stresses are 
\begin{equation}
\label{eqn.stress} 
\boldsymbol{\sigma}= -\rho\AAA^\mathsf{T} \boldsymbol{\psi} = -\rho\AAA^\mathsf{T} E_{\AAA} = -\rho c_s^2 \GG \dev(\GG),\ \ \qquad {\rm tr}(\boldsymbol{\sigma})=0,
\end{equation}
and the strain dissipation source term is 
\begin{equation}
\label{eqn.psi} 
-\dfrac{\boldsymbol{\psi}}{\theta_1(\tau_1)} = -\dfrac{E_{\AAA}}{\theta_1(\tau_1)}=-\dfrac{3}{\tau_1 } \left| \AAA \right|^{\frac{5}{3}} \AAA \dev(\GG),
\end{equation}
where we have chosen $ \theta_1(\tau_1) = \tau_1 c_s^2 / 3 \, |\AAA|^{-\frac{5}{3}} $, with $|\AAA|=\det(\AAA) > 0$ the determinant of $\AAA$ and $\tau_1$ being 
the strain relaxation time, or, in other words, the time scale that characterizes how long a continuum particle is connected with its neighbor elements before 
rearrangement.\footnote{Following Frenkel~\cite{Frenkel1955},  this relaxation time was called particle-settled-life (PSL) time in \cite{PeshRom2014}.
}
Note, that the determinant of $\AAA$ must satisfy the \textit{constraint}    
\begin{equation} 
\label{eqn.compatibility} 
 |\AAA| = \frac{\rho}{\rho_0},    
\end{equation} 
where $\rho_0$ is the density at a reference configuration, see \cite{PeshRom2014}. 
Furthermore, from the energy potential $E_{\rm mes}(\AAA,\JJ)$ the heat flux vector follows with $E_{\JJ} = \alpha^2 \JJ$ directly as 
\begin{equation}
\label{eqn.hyp.heatflux}
 \mathbf{q} = T \, \mathbf{H} = E_s E_{\JJ} = \alpha^2 T \JJ.    
\end{equation} 
For the thermal impulse relaxation source term, we choose $ \theta_2=\tau_2\alpha^2 \frac{\rho}{\rho_0} \frac{T_0}{T}$, and hence
\begin{equation}
-\dfrac{\rho \mathbf{H}}{\theta_2(\tau_2)} = -\dfrac{\rho E_{\JJ}}{\theta_2(\tau_2)}= - \frac{T}{T_0} \frac{\rho_0}{\rho} \dfrac{\rho\JJ}{\tau_2}.
\end{equation}
It contains another characteristic relaxation time $\tau_2$ that is associated to heat conduction.  

\section{Formal asymptotic analysis} 
\label{sec.asymptotics} 

In \cite{DPRZ2016} we have studied in detail the behaviour of the GPR model in the stiff relaxation limit $\tau_1 \to 0$ and $\tau_2 \to 0$ without  
the presence of electro-magnetic forces. Here, we briefly present the main results of this analysis and extend it also to the case when 
$\eta \to 0$ and $c \to \infty$. In all cases, the employed technique is a formal asymptotic analysis based on the Chapman-Enskog expansion. 

\subsection{Asymptotic limit of the viscous stress tensor} 

Here we briefly recall the main results found in \cite{DPRZ2016} when expanding the tensor $\GG = \AAA^\mathsf{T} \AAA$ in a series of the relaxation parameter $\tau_1$,   
\begin{equation}
  \GG = \GG_0 + \tau_1 \GG_1 + \tau_1^2 \GG_2 + ... 
	\label{eqn.Gseries} 
\end{equation} 
To analyze the stress tensor in the stiff relaxation limit we start from the derivation of an evolution equation for $\GG$. Using the definition of $\GG = \AAA^\mathsf{T} \AAA$ 
and the product rule, we obtain $\dot{\G} = \A^\mathsf{T} \dot{\A} + \dot{\A}^\mathsf{T} \A$, where the dot denotes the \textit{Lagrangian} or \textit{material} derivative 
$\dot{\GG} = d \GG / d t = \partial \GG / \partial t + \mathbf{v} \cdot \nabla \GG$.  
Summing up equation \eqref{eqn.deformation} multiplied by $ \A^\mathsf{T} $ from the left and transposing equation \eqref{eqn.deformation} multiplied by $ \AAA $ from the right, 
and using that $ \BS=-\rho\A^\mathsf{T}E_\A=-\rho(E_\A)^\mathsf{T} \AAA =\BS^\mathsf{T}$ we obtain the sought evolution equation under the following form:  
\begin{equation}\label{eqn.GODE}
	\dot{\G} = - \left( \G \nabla \v + \nabla \v^\mathsf{T} \G \right) +\dfrac{2}{\rho\,\theta_1} \, \BS,  
\end{equation} 
where $\nabla \v$ is the velocity gradient. 
As in \cite{DPRZ2016} we define $ \theta_1 = \tau_1 |\A|^{-\frac{5}{3}} c_s^2/3 = \tau_1 |\G|^{-\frac{5}{6}} c_s^2/3$. With $E_{\AAA} = c_s^2 \AAA \dev(\G)$ and after 
inserting \eqref{eqn.Gseries} into \eqref{eqn.GODE} and collecting terms of the same power in $\tau_1$ one has  
\begin{equation}
 \tau_1^{-1} \, \underbrace{\left(6|\G|^{\frac{5}{6}}\G\dev(\G_0)\right )}_{0} + 
  \tau_1^0  \, \underbrace{\left( \frac{ {\rm d} \G_0}{{\rm d} t}  + ...\right )}_{0} + ... = 0.  
 \label{eqn.aux3}
\end{equation}

Relation \eqref{eqn.aux3} holds for any $\tau_1$, hence the coefficients which multiply powers of $\tau_1$ must be equal to zero. Since $\rho =\rho_0|\A|=\rho_0|\G|^{\frac{1}{2}}> 0$ we have $|\G| > 0$, which
means that $\G$ is invertible. Thus, the leading order term ($\tau_1^{-1}$) in \eqref{eqn.aux3}, yields 
\begin{equation}
  \dev \left( \G_0 \right) = 0,  \quad \Rightarrow  \quad \G_0 - \frac{1}{3} \tr \left( \GG_0 \right) \Id = 0, \quad \Rightarrow 
	\G_0 = \frac{1}{3} \tr \left( \GG_0 \right) \Id. 
	\label{eqn.devg0} 
\end{equation} 
Introducing the definition $g := \frac{1}{3} \tr \left( \GG_0 \right)$ and neglecting higher order terms, we obtain  
\begin{equation}
  \G = g \Id + \tau_1 \G_1 = \AAA^{\mathsf{T}} \AAA, 
	\label{eqn.gtemp} 
\end{equation} 
i.e. in the stiff limit $\tau_1 \ll 1$, the distortion matrix $\AAA$ tends to an \textit{orthogonal matrix}. 
The coefficient $g$ can be easily computed from the determinant of $\G$ and the compatibility condition $\rho = \rho_0 | \AAA |$ as  
$ g = | \G |^{\frac{1}{3}} = | \AAA |^{\frac{2}{3}} = \left( {\rho} / {\rho_0} \right)^{\frac{2}{3}}$. 
Retaining only the leading term $\G_0$ in the expansion \eqref{eqn.Gseries} we get $ \G=\G_0=g\II $ and thus $ \sigma=-\rho c_s^2\G_0 \dev(\G_0)=0 $, 
hence as zeroth order approximation one retrieves the \textit{inviscid} case in the limit $\tau_1 \to 0$. 

To get a first order approximation of the viscous stress tensor $\BS$ in terms of $\tau_1$ one needs to expand the stress tensor \eqref{eqn.stress} in a series of $ \tau_1 $. 
With $ \G=g\II+\tau_1\G_1 $ one has that $ \rho=\rho_0|\A|=\rho_0|g\II+\tau_1\G_1|^\frac{1}{2}=\rho_0 (g^{3/2}+\frac{\tau_1}{2} g^{1/2} \tr(\G_1)+\mathcal{O}(\tau_1^2))$ and 
$ \dev(\G)=\dev(g\II+\tau_1\G_1)=\tau_1\dev(\G_1)$. Then, the viscous stress tensor can be written as  
\begin{equation}
\BS = - \rho c_s^2 \G \dev(\G)=-\rho_0 c_s^2\left  (g^{3/2}+\frac{\tau_1}{2} g^{1/2} \tr(\G_1)\right )\left (g\II+\tau_1\G_1\right )\tau_1\dev(\G_1).
\label{eqn.sdot} 
\end{equation} 
Retaining only the leading terms  $ \tau_1 $ yields the simple expression    
\begin{equation}\label{eqn.sigma.expan}
\BS = - \tau_1  \rho_0 c_s^2 g^{5/2}\dev(\G_1).
\end{equation}

After applying the ``$ \dev $'' operator to \eqref{eqn.GODE} one gets the following evolution equation for $ \dev(\G) $: 
\begin{equation}\label{eqn.devG}
\dfrac{{\rm d}}{{\rm d}t}\dev(\G)+\G\nabla \v+\nabla \v^\mathsf{T}\G-\dfrac{1}{3}\tr(\G\nabla \v+\nabla \v^\mathsf{T}\G)\II=-\dfrac{6}{\tau_1}|\G|^{5/6}\dev(\G\dev(\G)).
\end{equation}
Inserting \eqref{eqn.Gseries} into \eqref{eqn.devG} and recalling from \eqref{eqn.devg0} that $\dev{\G_0}=0$, one gets the following relation for the leading order terms ($ \tau_1^0 $): 
\begin{equation*}
\G_0\nabla \v+\nabla \v^\mathsf{T}\G_0-\dfrac{2}{3}\tr(\G_0\nabla \v)\II=-6|\G_0|^{7/6}\dev(\G_1).
\end{equation*}
Since $ \G_0=g\II $, the last relation can be rewritten as 
\begin{equation}\label{eqn.devG1}
g\left  (\nabla \v+\nabla \v^\mathsf{T}-\dfrac{2}{3}\tr(\nabla \v)\II\right  )=-6\, g^{7/2}\dev(\G_1).
\end{equation}
After inserting \eqref{eqn.devG1} into \eqref{eqn.sigma.expan} one obtains the following final expression for the first order approximation of the viscous stress tensor in terms of $\tau_1$: 
\begin{equation}
\BS = \frac{1}{6} \tau_1 \rho_0 c_s^2 \left( \nabla \v + \nabla \v^T  - \frac{2}{3} \tr ( \nabla \v ) \Id  \right) := \mu \left( \nabla \v + \nabla \v^T  - \frac{2}{3} ( \nabla \cdot \v ) \Id  \right),  
\label{eqn.stresslimit} 
\end{equation} 
This is nothing else than the classical viscous stress tensor that is known from the compressible Navier-Stokes equations (using Stokes' hypothesis), where the dynamic viscosity coefficient is given in 
terms of the relaxation time $\tau_1$ and the shear sould speed $c_s$ as 
\begin{equation}
	\mu = \frac{1}{6} \tau_1 \rho_0 c_s^2,   
	\label{eqn.newton.mu} 
\end{equation} 
see \cite{PeshRom2014,DPRZ2016}. For a comment on the possible experimental measurement of $\tau_1$ and $c_s$ see \cite{DPRZ2016}. We stress at this point that the usual form of the viscous stress tensor 
of the compressible Navier-Stokes equations is a result of the model, which is obtained by the mere choice of a quadratic energy potential in terms of $\GG^{TF}$. The special choice of $\theta_1$ was 
only made to produce a \textit{constant} viscosity coefficient $\mu$. More general relations of $\mu$ (say, e.g., the well-known law of Sutherland) can be obtained by a suitable choice of $\theta_1$. 

\subsection{Asymptotic limit of the heat flux} 

In \cite{DPRZ2016} a formal asymptotic analysis was also carried out for the heat flux $\mathbf{q} = E_S E_{\JJ} = \alpha^2 T \JJ$. The Chapman-Enskog expansion of the thermal impulse vector $\JJ$ 
in terms of the small relaxation parameter $\tau_2 \ll 1$ reads 
\begin{equation}
  \JJ = \JJ_0 + \tau_2 \JJ_1 + \tau_2^2 \JJ_2 + ...   
 \label{eqn.jseries} 
\end{equation} 
With $ \theta_2=\tau_2\alpha^2 \frac{\rho}{\rho_0} \frac{T_0}{T}$ the PDE \eqref{eqn.heatflux} for $\JJ$ becomes:  
\begin{equation}
  \frac{\partial \rho \JJ}{\partial t} + \nabla \cdot \left( \rho \JJ \otimes \u \right) + \nabla T = -\frac{1}{\tau_2} \, \frac{T}{T_0} \frac{\rho_0}{\rho} \, \rho \JJ. 
 \label{eqn.j} 	
\end{equation} 
Inserting \eqref{eqn.jseries} into \eqref{eqn.j} and proceeding with the collection of the $\tau_2$ terms as in the previous section yields 
\begin{equation}
 \tau_2^{-1} \underbrace{\left( \frac{T}{T_0} \frac{\rho_0}{\rho}  \, \rho \JJ_0 \right)}_{0} 
+ \tau_2^0 \underbrace{\left( \frac{\partial \rho \JJ_0}{\partial t} + \nabla \cdot \left( \rho \JJ_0 \otimes \u \right) + \nabla T +  \frac{T}{T_0} \frac{\rho_0}{\rho} \, \rho \JJ_1 \right)}_{0} + ... = 0.  
\end{equation} 
As a consequence one obtains the following relations for the first two terms in the expansion of $\JJ$:   
\begin{equation} 
  \JJ_0 = 0, \qquad \textnormal{ and } \qquad \JJ_1 = - \frac{T_0}{T \rho_0} \nabla T.  
  \label{eqn.asyj0} 
\end{equation} 
As a result of  \eqref{eqn.jseries} and \eqref{eqn.asyj0} the heat flux vector $\mathbf{q} = \alpha^2 T \JJ$ becomes for small relaxation times $\tau_2 \ll 1$ 
\begin{equation}
 \mathbf{q} = \alpha^2 T \JJ = - \alpha^2 \tau_2 \frac{T_0}{\rho_0} \nabla T := -\kappa \nabla T,
\label{eqn.fourier.kappa} 
\end{equation} 
which is the familiar form of the Fourier heat flux with heat conduction coefficient $\kappa = \alpha^2 \tau_2 \frac{T_0}{\rho_0}$.  

\subsection{Asymptotic limit of the electro-magnetic stresses} 

In this paper the extended GPR model \eqref{eqn.conti} - \eqref{eqn.energy} also accounts for the presence of electro-magnetic forces and effects. We therefore analyze the model in the stiff relaxation limit for
$\eta \to 0$ and $c \to \infty$. From \eqref{eqn.dbDefinition.b} and \eqref{eqn.rhou} we immediately obtain $\mathbf{b} \to \frac{1}{\mu'} \mathbf{B}$ and $\mathbf{u} \to \mathbf{v}$ for $c \to \infty$. Furthermore, 
the governing PDE system for the electric field reduces for $c \to \infty$ to the simple relation  
\begin{equation} 
	- \frac{1}{\mu'}  \nabla \times \mathbf{B} = - \frac{1}{\eta} \mathbf{d}. 
	\label{eqn.Ecinf} 
\end{equation} 
A Chapman-Enskog expansion of $\mathbf{d}$ in terms of the small parameter $\eta$ reads 
\begin{equation}
\mathbf{d} = \mathbf{d}_0 + \eta \mathbf{d}_1 + \eta^2 \mathbf{d}_2 + ... 
\label{eqn.dseries} 
\end{equation} 
and thus eqn. \eqref{eqn.Ecinf} becomes
\begin{equation}
   \eta^{-1} \left( \mathbf{d}_0 \right) +  \eta^{0} \left( \mathbf{d}_1 - \frac{1}{\mu'} \nabla \times \mathbf{B} \right) + ... = 0.  
\end{equation}
Since the above equation must be valid for any $\eta$, we set all coefficients multiplying terms with $\eta$ to zero and get as a result 
\begin{equation} 
   \mathbf{d}_0 = 0, \qquad \textnormal{ and } \qquad \mathbf{d}_1 = \frac{1}{\mu'} \nabla \times \mathbf{B}. 
   \label{eqn.d01} 
\end{equation} 
From $\mathbf{d}_0 = 0$ and \eqref{eqn.dbDefinition.d} follows immediately that at leading zeroth order the electric field behaves as 
$\mathbf{E} = - \v \times \mathbf{B}$, which is a well known relation for the ideal MHD equations. It also means that in the comoving 
frame, the electric field vanishes. Inserting \eqref{eqn.d01} into the PDE for the magnetic field \eqref{eqn.Bfield} we obtain 
\begin{equation}
\displaystyle\frac{\partial B_i}{\partial t} + \frac{\partial}{\partial x_k} \left( v_k B_i - v_i B_k  \right) 
                                             + v_i \frac{\partial B_k}{\partial x_k}= 
                                             - \frac{\eta}{\mu'}  \frac{\partial}{\partial x_k} \varepsilon_{ikl} \frac{\partial}{\partial x_m} \varepsilon_{lmp} B_p, \label{eqn.Bfield2}	 
\end{equation} 
i.e. we obtain the classical dissipative term of the type 
$- \frac{\eta}{\mu'} \nabla \times \left( \nabla \times \mathbf{B}\right) = \frac{\eta}{\mu'} \nabla \cdot \left( \nabla \mathbf{B} - \nabla \mathbf{B}^T \right)$ that is present in the viscous and 
resistive MHD equations 
\cite{WarburtonVRMHD,ADERVRMHD}. For $\eta \ll 1$ and $c \to \infty$ the flux term $\varepsilon_{ijk} d_i b_j$ in the energy equation \eqref{eqn.energy} becomes up to first order  
terms in $\eta$ 
\begin{equation}
	  \mathbf{d} \times \mathbf{b} = \frac{\eta}{(\mu')^2} \nabla \times \left( \mathbf{B} \times \mathbf{B}\right) = -\frac{\eta}{(\mu')^2} \mathbf{B}^T 
		\left( \nabla \mathbf{B} - \nabla \mathbf{B}^T \right), 
\end{equation} 
see \cite{WarburtonVRMHD,ADERVRMHD}. 
Finally, the Maxwell stress tensor $\boldsymbol{\beta}$ reduces to 
\begin{equation}
    \boldsymbol{\beta} = \frac{1}{\mu'} \left( \halb \mathbf{B}^2 \, \mathbf{I} - \mathbf{B} \otimes \mathbf{B} \right), 
\end{equation}    
which is the usual relation for the ideal MHD equations. 

\subsection{Asymptotically reduced limit system} 

Combining the results of the previous sections, we get the following asymptotically reduced system for the quantities $\rho$, $\rho \mathbf{u} \to \rho \mathbf{v}$, $\rho E$ and 
$\mathbf{B}$, in the stiff relaxation limit when $\eta \to 0$, $\tau_1 \to 0$ and $\tau_2 \to 0$: 
\begin{eqnarray}
	  \frac{\partial}{\partial t} \left( \begin{array}{c} \rho \\ \rho \mathbf{v} \\ \rho E \\ \mathbf{B} \\ \psi \end{array} \right) 
		+ \nabla \cdot \left( \begin{array}{c} \rho \mathbf{v} \\ \rho \mathbf{v} \otimes \mathbf{v} + p \mathbf{I} - \boldsymbol{\sigma} + \boldsymbol{\beta} \\ 
		 \mathbf{v}^T \left( (\rho E + p) \mathbf{I} - \boldsymbol{\sigma} + \boldsymbol{\beta} \right) + \mathbf{q}  
		- \frac{\eta}{(\mu')^2} \mathbf{B}^T \left( \nabla \mathbf{B} - \nabla \mathbf{B}^T \right)  \\ 
		\mathbf{B} \otimes \mathbf{v} - \mathbf{v} \otimes \mathbf{B} - \frac{\eta}{\mu'} \left( \nabla \mathbf{B} - \nabla \mathbf{B}^T \right) + \psi \mathbf{I}  \\ 
		c_h^2 \mathbf{B} \end{array} \right) = 0, 
		\label{eqn.vrmhd} 
\end{eqnarray} 
with the viscous shear stress tensor of the fluid  
\begin{equation}
   \boldsymbol{\sigma} = \mu \left( \nabla \mathbf{v} + \nabla \mathbf{v}^T - \frac{2}{3} \nabla \cdot \mathbf{v} \right),  \qquad \textnormal{ with } \qquad 
	\mu = \frac{1}{6} \tau_1 c_s^2 \rho_0 , 
\end{equation} 
the heat flux
\begin{equation}
   \mathbf{q} = -\kappa \nabla T,  \qquad \textnormal{ with } \qquad \kappa = \tau_2 \alpha^2  \frac{T_0}{\rho_0}, 
\end{equation} 
and the Maxwell stress tensor of the electro-magnetic forces 
\begin{equation}
	   \boldsymbol{\beta} = \frac{1}{\mu'} \left(  \halb \mathbf{B}^2 \, \mathbf{I} - \mathbf{B} \otimes \mathbf{B} \right). 
\end{equation} 
The above system \eqref{eqn.vrmhd} is the classical viscous and resistive MHD system based on conventional parabolic terms for the description of dissipative momentum and 
heat transfer. In this system, also the electric resistivity of the medium is modeled by parabolic terms, which is in contrast to the original Maxwell equations.

\section{The numerical scheme} 
\label{sec.scheme} 
As in Paper~I, the governing equations of the HTC formulation of the GPR model 
can be written as  a nonlinear system of hyperbolic PDEs with non-conservative products and stiff source terms: 
\begin{equation}	
\label{eqn.pde.nc}
	\frac{\partial \Q}{\partial t} + \nabla \cdot \bf F(\Q) + \mathbf{\mathcal{B}}(\Q) \cdot \nabla \Q = \S(\Q), 
\end{equation}
where $\Q=\Q(\x,t)$ is the state vector,
 ${\bf F}(\Q) = (\f, \g, \h)$ is the nonlinear flux tensor expressing
the conservative part of the PDE system, while $\S(\Q)$ contains the potentially stiff algebraic relaxation source terms and 
$\mathbf{\mathcal{B}}(\Q) \cdot \nabla \Q$ is a purely non-conservative term. 
The system (\ref{eqn.pde.nc}) can also be written in quasilinear form as
\begin{equation}	\label{eq:Csyst}
\frac{\partial \Q}{\partial t}+ \mathbf{\mathcal{A}} (\Q) \cdot\nabla \Q = \S(\Q)\,,
\end{equation}
where $\mathbf{\mathcal{A}}(\Q)=\partial {\bf F}(\Q)/\partial \Q + \mathbf{\mathcal{B}}(\Q)$ includes both 
the Jacobian of the conservative flux, as well as the non-conservative product. 

We choose to solve the 
PDE system \eqref{eqn.pde.nc} by using high order one-step ADER-DG methods, which
evolve in time the degrees of freedom with respect to a given basis, rather than the point values or the cell averages  of the solution, 
like in finite difference or in finite volume methods.
Our description of the numerical scheme is limited to the most relevant aspects, 
while all the details can be found in 
\cite{Dumbser2008,DumbserZanotti,HidalgoDumbser,Balsara2013,Dumbser2014,Zanotti2015a,Zanotti2015b}. 
At the generic time $t^n$, the numerical solution of the PDE
is represented within each cell $T_i$ 
by polynomials of maximum degree $N \geq 0$, namely
\begin{equation}
\label{eqn.ansatz.uh}
  \u_h(\x,t^n) = \sum_{l=0}^{\mathcal{N}}\Phi_l(\x) \hat{\u}^n_l= \Phi_l(\x) \hat{\u}^n_l, \qquad \x \in T_i\,,
\end{equation}
where the coefficients
$\hat{\u}^n_l$ are sometimes called the \emph{degrees of freedom}.
The functions $\Phi_l(\x)$ form 
a nodal basis, which is given by the Lagrange interpolation polynomials 
passing through the Gauss-Legendre quadrature nodes associated with element $T_i$, 
see \cite{stroud}. The symbol $\mathcal{N}$ denotes the number of degrees of freedom 
per element and is given by $\mathcal{N}=(N+1)^d$ for tensor-product elements in $d$ 
space dimensions. 

\subsection{The Discontinuous Galerkin scheme}
\label{sec.ADERNC}

A fully discrete one-step ADER-DG scheme is 
obtained after multiplying the governing PDE \eqref{eqn.pde.nc} by  test functions  
$\Phi_k$ identical to the spatial basis functions of Eq.~\eqref{eqn.ansatz.uh}.
After that, we integrate over the space-time control volume $T_i \times [t^n;t^{n+1}]$. 
Following the idea of path-conservative schemes, see \cite{Castro2006,Pares2006,ADERNC}, one obtains: 
\begin{equation}
\label{eqn.pde.nc.gw2}
\begin{split}
\left( \int \limits_{T_i} \Phi_k \Phi_l d\x \right) \left( \hat{\u}_l^{n+1} -  \hat{\u}_l^{n} \right) +
\int \limits_{t^n}^{t^{n+1}} \int \limits_{\partial T_i} \Phi_k \, \mathcal{D}_h^-\left(\q_h^-, \q_h^+ \right)\cdot\mathbf{n} \, dS dt 
\\ 
+\int\limits_{t^n}^{t^{n+1}} \int \limits_{T_i \backslash \partial T_i} \Phi_k \left( \nabla \cdot \F\left(\q_h \right) + \mathbf{\mathcal{B}}(\q_h) \cdot \nabla \q_h \right) d\x dt  
= \int \limits_{t^n}^{t^{n+1}} \int \limits_{T_i} \Phi_k \S(\q_h) d\x dt\,,
\end{split} 
\end{equation}
where $\mathbf{n}$ is the outward pointing unit normal vector on the surface $\partial T_i$ of element $T_i$.
There are a couple of aspects worth mentioning in the above expression. First of all, we used the symbol $\q_h$
to denote a \textit{predictor} state available 
at any intermediate time between $t^n$ and $t^{n+1}$ and with the same spatial accuracy of the initial DG polynomial.
The calculation of $\q_h$ is briefly described in the next Section.
Secondly, 
the element mass matrix appears in the first integral of \eqref{eqn.pde.nc.gw2}, the second term is a Riemann solver 
(written in terms of fluctuations) that accounts for the jump in the discrete 
solution at element boundaries and the third term takes into account the smooth part of the non-conservative product. 
Third, due to the presence of non-conservative products, the jumps of $\q_h$ across element boundaries are taken into account in 
the framework of path-conservative schemes put forward by Castro and Par\'es in the finite volume context 
\cite{Castro2006,Pares2006}. Finally, as for the choice of the Riemann problem, in this paper we have 
used the simple Rusanov method \cite{Rusanov:1961a} (also called the local Lax Friedrichs
method), although any other kind of Riemann solver could be adopted in principle.

The ADER-DG method described above refers to the \textit{unlimited} scheme. In the presence of discontinuities, a proper nonlinear 
limiting strategy is needed. Here, we use the \textit{a posteriori} finite volume subcell limiter proposed in \cite{Dumbser2014,Zanotti2015a,Zanotti2015b},
which is based on the MOOD framework developed in \cite{CDL1,CDL2,CDL3}.  

\subsection{Local space-time predictor}
\label{sec.predictor}

The computation of the \textit{predictor} state $\q_h$
is obtained after resorting to an element-local weak formulation of the governing PDE in space-time, 
see \cite{DumbserEnauxToro,Dumbser2008,HidalgoDumbser,DumbserZanotti,Balsara2013,Dumbser2014,Zanotti2015a,Zanotti2015b}. 
Since this procedure is performed locally for each computational element, irrespective of neighbouring elements, no Riemann problem is implied in that.
To simplify notation, we define  
\begin{equation}
 \label{eqn.operators1}
  \left<f,g\right> =
      \int \limits_{t^n}^{t^{n+1}} \int \limits_{T_i}  f(\x, t)  g(\x, t)  \, d \x \, d t,
\qquad 
  \left[f,g\right]^{t} =
      \int \limits_{T_i}f(\x, t) g(\x, t) \,  d \x,
\end{equation}
which denote the scalar products of two functions $f$ and $g$ over the space-time element $T_i \times \left[t^n;t^{n+1}\right]$ and  
over the spatial element $T_i$ at time $t$, respectively. 
The discrete representation of $\q_h$ in element $T_i \times [t^n,t^{n+1}]$ is assumed to have the following form 
\begin{equation}
\label{eqn.st.state}
 \q_h = \q_h(\x,t) =
 \sum \limits_l \theta_l(\x,t) \hat{\q}^n_{l,i} := \theta_l \hat{\q}^n_{l,i},
\end{equation}
where it is importatn to stress that $\theta_l(\x,t)$ is now a space-time basis function, of degree $N$. 
At this point we
multiply \eqref{eqn.pde.nc} with a space-time test function $\theta_k=\theta_k(\x,t)$ and subsequently integrate over 
the space-time control volume $T_i \times \left[t^n;t^{n+1}\right]$. Replacing for  $\q_h$, the following weak formulation of 
the PDE is obtained: 
\begin{equation}
\label{eqn.pde.nc.weak1}
 \left< \theta_k, \frac{\partial \q_h}{\partial t}  \right>
    + \left< \theta_k, \nabla \cdot \F \left(\q_h\right) + \mathbf{\mathcal{B}}(\q_h) \cdot \nabla \q_h \right> = \left< \theta_k, \S \left( \q_h \right)  \right>.
\end{equation}
After integration by parts in time of the first term, eqn. \eqref{eqn.pde.nc.weak1} reads 
\begin{equation}
\label{eqn.pde.nc.dg1}
 \left[ \theta_k, \q_h \right]^{t^{n+1}} - \left[ \theta_k, \u_h(\x,t^n) \right]^{t^n} - \left< \frac{\partial}{\partial t} \theta_k, \q_h \right> 
    + \left< \theta_k, \nabla \cdot \F \left(\q_h\right) + \mathbf{\mathcal{B}}(\q_h) \cdot \nabla \q_h \right> = \left< \theta_k, \S \left( \q_h \right)  \right>. 
\end{equation}
Eq.~\eqref{eqn.pde.nc.dg1} represents a nonlinear system to be solved in the unknown expansion 
coefficients $\hat{\q}^n_{l,i}$.
We recall that, unlike the original ADER approach based on the Cauchy-Kovalewski procedure, 
 the discontinuous Galerkin predictor just described remains valid  even in the presence of stiff source terms, as it has been done
for various physical systems in 
\cite{DumbserZanotti,HidalgoDumbser,Zanotti2011,Lagrange1D,DPRZ2016} and as it is the case for the equations considered in this paper.

\section{Numerical results} 
\label{sec.tests} 

In this section on numerical results, we will assume that the magnetic permeability of the medium is $\mu'=1$ for all test problems, and only the speed of
light $c$ is explicitly specified. If not stated otherwise, the ideal gas equation of state (EOS) is used. 
Within this section, we will specify the standard material parameters that are conventionally used in continuum mechanics, i.e. the fluid viscosity $\mu$ 
and the heat conduction coefficient $\kappa$. Together with the associated wave speeds $c_s$ (shear sound speed) and $\alpha$ (heat propagation wave speed), 
one can calculate the corresponding characteristic times $\tau_1$ and $\tau_2$ used in the GRP model according to the results \eqref{eqn.newton.mu} and 
\eqref{eqn.fourier.kappa} given by the formal asymptotic analysis presented in Section \ref{sec.asymptotics}. Note that the model parameter $\eta$ is already 
a well-known quantity, namely the electric resistivity of the medium used in the Ohm law. 

\subsection{Numerical convergence results in the stiff relaxation limit} 

As shown in the formal asymptotic analysis carried out in \cite{DPRZ2016} and Section \ref{sec.asymptotics} of this paper, 
the governing PDE system \eqref{eqn.conti}-\eqref{eqn.energy} relaxes to the classical ideal MHD equations in the case where $c \to \infty$ and when 
the relaxation times and the resistivity tend to zero, i.e. for $\tau_1 \to 0$, $\tau_2 \to 0$ and $\eta \to 0$. 
We can use this knowledge in order to design a test case that allows us to verify numerically the order of accuracy of our high order one-step ADER-DG schemes in 
the stiff relaxation limit of the first order hyperbolic GPR model by comparing against known exact solutions of the ideal MHD system. For that purpose, we use the initial 
condition proposed by Balsara in \cite{Balsara2004} that consists of a smooth magnetized vortex. The computational domain used for this test is $\Omega = [-10,+10]^2$ and 
the initial condition is given by 
\begin{equation*}
  \rho = 1, \qquad u = 1 + \delta u, \qquad v = 1 + \delta v, \qquad w = 0, \qquad p = 1 + \delta p, 
\end{equation*}  
\begin{equation*}
  \mathbf{E} = - \mathbf{v} \times \mathbf{B}, \qquad \AAA = \sqrt[3]{\rho} \, \mathbf{I}, \qquad \mathbf{J}=0, 
\end{equation*}
with $r = \sqrt{x^2+y^2}$, $\epsilon = \frac{1}{2\pi}$ and the perturbations 
\begin{equation*}
   \delta u = -y \epsilon \exp \left( \halb(1-r^2) \right), \quad  
   \delta v = +x \epsilon \exp \left( \halb(1-r^2) \right), \quad  
   \delta p = - \halb \epsilon^2  r^2 \exp(1-r^2),
\end{equation*} 
\begin{equation*}     
   B_x = -y \epsilon \exp \left( \halb (1-r^2) \right), \quad    
   B_y = +x \epsilon \exp \left( \halb (1-r^2) \right), \quad 
   B_z = 0.     
\end{equation*} 
For the governing PDE system, we use the following parameters: $\gamma = 1.4$, $\rho_0=1$, $c_s=0.8$, $\alpha^2=0.8$, $c=100$, $\mu=\kappa=\eta=10^{-6}$. 
The speed for the hyperbolic divergence cleaning in the GLM approach \cite{Dedneretal} is set to $c_h=2$. 
The simulations are carried out with a fourth and fifth order accurate ADER-DG scheme until a final simulation time of $t=0.1$ using a sequence of successively 
refined meshes. The exact solution of the underlying ideal MHD problem consists in a mere transport of the initial condition translated with velocity 
$\mathbf{v}=(1,1,0)$. 
This test is very difficult for the GPR model, since the system is run in a very stiff regime and with $c \gg 1$, so that the resulting time step is very 
small due to the CFL condition. For that reason, only a small final simulation time has been chosen.  

The obtained numerical convergence rates are reported in Table \ref{tab.conv}, where we can observe that the schemes reach their designed 
order of accuracy even in the stiff relaxation limit, which is a very important property of the numerical method used here. 

\begin{table}  
\caption{Numerical convergence results for the magnetized vortex obtained with ADER-DG $P_3$ and $P_4$ schemes applied to the GPR model 
($c_s=0.8$, $\alpha^2=0.8$, $c=100$) in the stiff relaxation limit ($\mu \ll 1, \kappa \ll 1, \eta \ll 1$).  
Results are shown for the magnetic field component $B_x$ at a final time of $t=0.1$. The reference solution is given by the exact solution of the ideal MHD equations.} 
\begin{center} 
\begin{small}
\renewcommand{\arraystretch}{1.0}
\begin{tabular}{ccccccc} 
\hline
  $N_x$ & $\epsilon({L_1})$ & $\epsilon({L_2})$ & $\epsilon({L_\infty})$ & $\mathcal{O}(L_1)$ & $\mathcal{O}(L_2)$ & $\mathcal{O}(L_\infty)$ \\ 
\hline
  \multicolumn{7}{c}{ADER-DG $P_3$ ($\mu = \kappa = \eta = 10^{-6}$)}  \\ 
\hline
20	& 2.9646E-03	& 7.9141E-04	& 7.0490E-04	&     	&     	&        \\ 
30	& 3.7070E-04	& 1.0288E-04	& 9.3677E-05	& 5.13	& 5.03	& 4.98   \\ 
40	& 8.9193E-05	& 2.4785E-05	& 2.1279E-05	& 4.95	& 4.95	& 5.15   \\ 
50	& 2.9814E-05	& 8.2723E-06	& 7.8091E-06	& 4.91	& 4.92	& 4.49   \\ 
\hline
  \multicolumn{7}{c}{ADER-DG $P_4$ ($\mu = \kappa = \eta = 10^{-6}$)}   \\
\hline
8	  & 2.4129E-02	& 5.2803E-03	& 4.2926E-03	& 		  &       &      \\		
10	& 6.2946E-03	& 1.4070E-03	& 1.0618E-03	& 6.02	& 5.93	& 6.26 \\ 
12	& 1.4985E-03	& 3.2871E-04	& 2.9132E-04	& 7.87	& 7.97	& 7.09 \\
16	& 3.1902E-04	& 7.1927E-05	& 5.8008E-05	& 5.38	& 5.28	& 5.61 \\
\hline 
\end{tabular}
\end{small}
\end{center}
\label{tab.conv}
\end{table} 

\subsection{Current sheet} 

\begin{figure}[!htbp]
  \begin{center}
      \includegraphics[draft=false,width=0.75\textwidth]{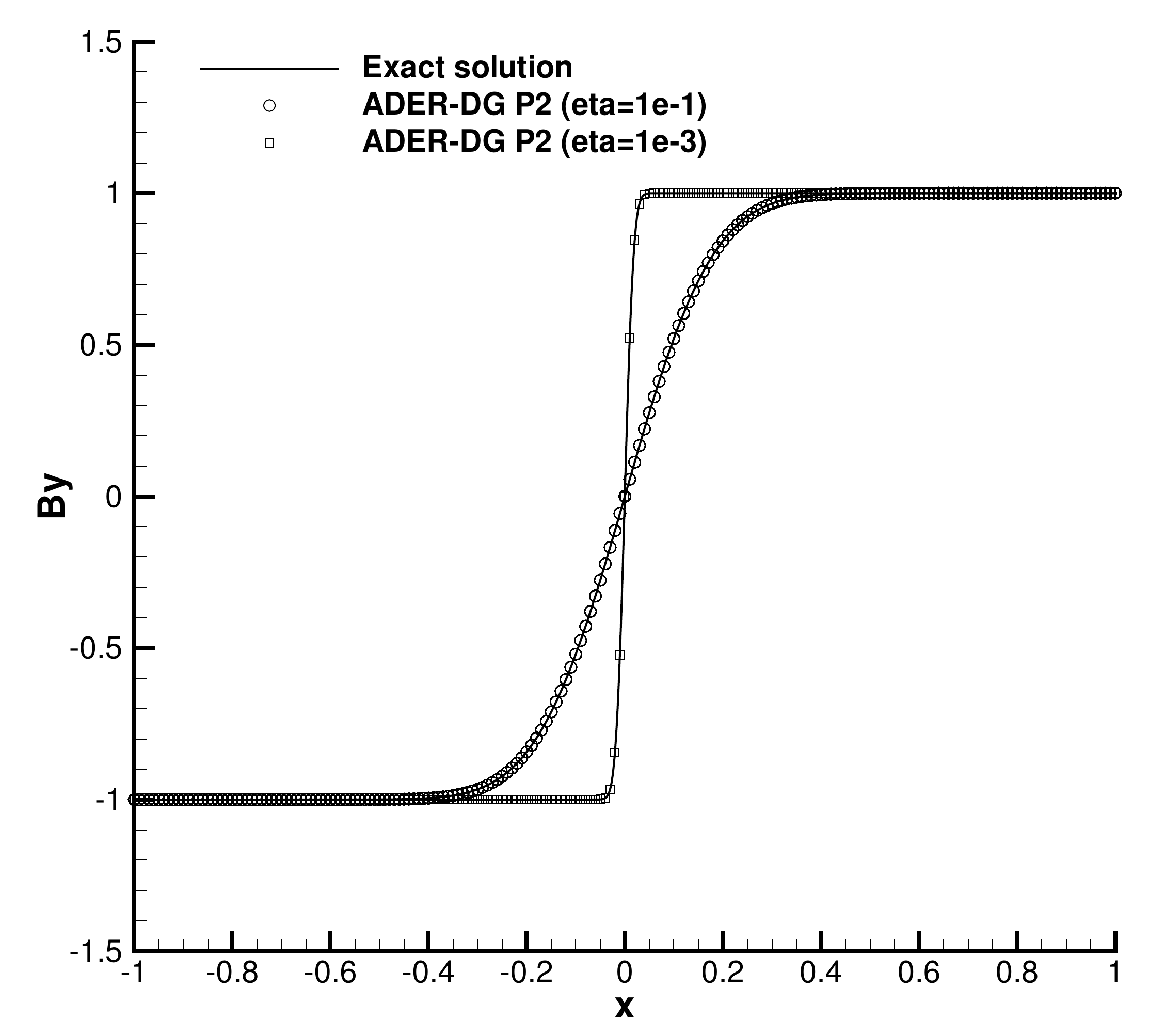}
       
    \caption{Current sheet at time $t=0.1$ simulated for different resistivities ($\eta = 10^{-1}$ and $\eta = 10^{-3}$) with the GPR model using an ADER-DG $P_2$ scheme.} 
    \label{fig.currentsheet}
	\end{center}
\end{figure}

Here, we simulate a simple current sheet, see \cite{Komissarov2007,DumbserZanotti}, in order to verify the correct description of resistive effects by the model. 
The computational domain is defined as $\Omega = [-1,+1] \times [-0.1, +0.1]$ and the initial condition is given by $\rho=1$, $\mathbf{v}=0$, $\AAA=\mathbf{I}$, $\mathbf{J}=0$, 
$p=1$, $B_x=B_z=0$ and $B_y = \textnormal{sign}(x)$. The parameters for the simulation are $\rho_0=1$, $\gamma=1.4$,  
$c_s=0.8$, $\alpha^2=0.8$, $c=10$, $\mu=\kappa=\eta$. In this test, two different values for the resistivity have been used, namely $\eta=10^{-1}$ and $\eta=10^{-3}$, 
respectively. The exact solution for the time evolution of the magnetic field component $B_y$ in the current sheet is given by (see \cite{Komissarov2007,DumbserZanotti}): 
\begin{equation*}
   B_y(x,t) = \textnormal{erf}\left( \frac{x}{2 \sqrt{\eta t}} \right).  
\end{equation*}   
A comparison between the exact solution and the numerical solution obtained with an ADER-DG $P_2$ 
method on a uniform grid composed of $ 100 \times 5$ grid points for both values of the resistivity is depicted 
at time $t=0.1$ in Fig. \ref{fig.currentsheet}, where an excellent agreement can be observed in both cases. 

\subsection{Riemann problems} 
\begin{figure}[!htbp]
  \begin{center}
	\begin{tabular}{cc} 
      \includegraphics[draft=false,width=0.45\textwidth]{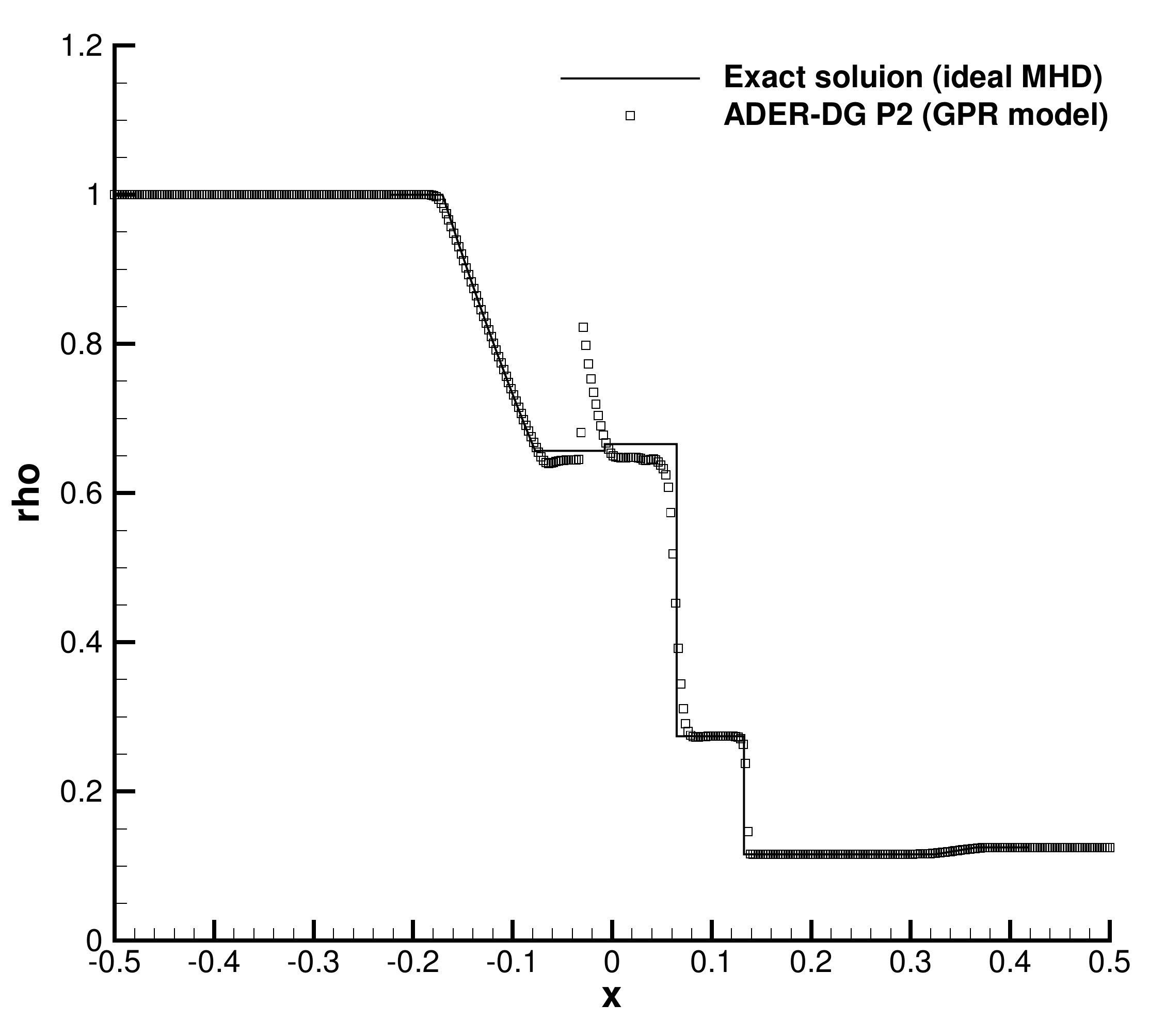} & 
      \includegraphics[draft=false,width=0.45\textwidth]{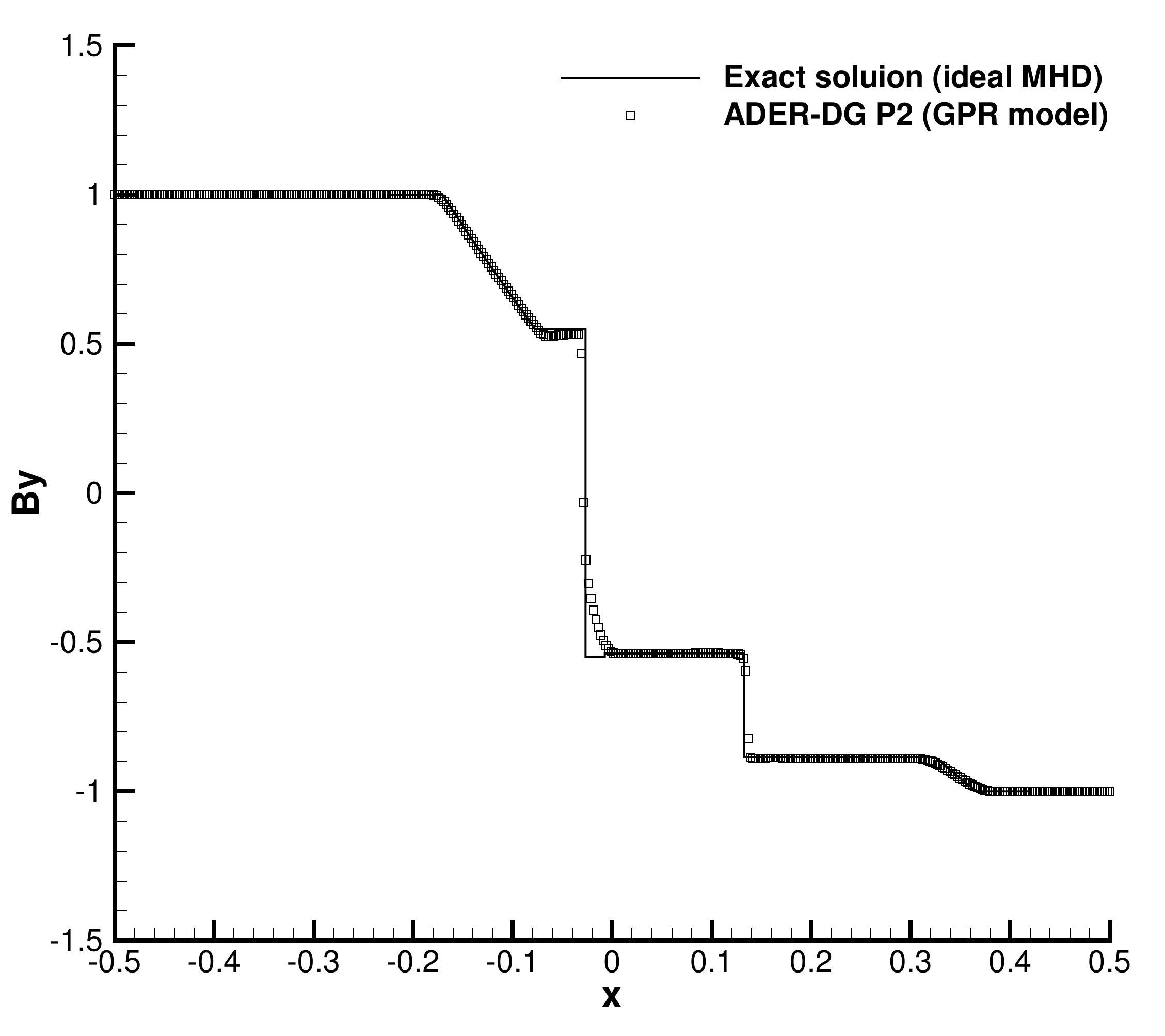}  \\  
      \includegraphics[draft=false,width=0.45\textwidth]{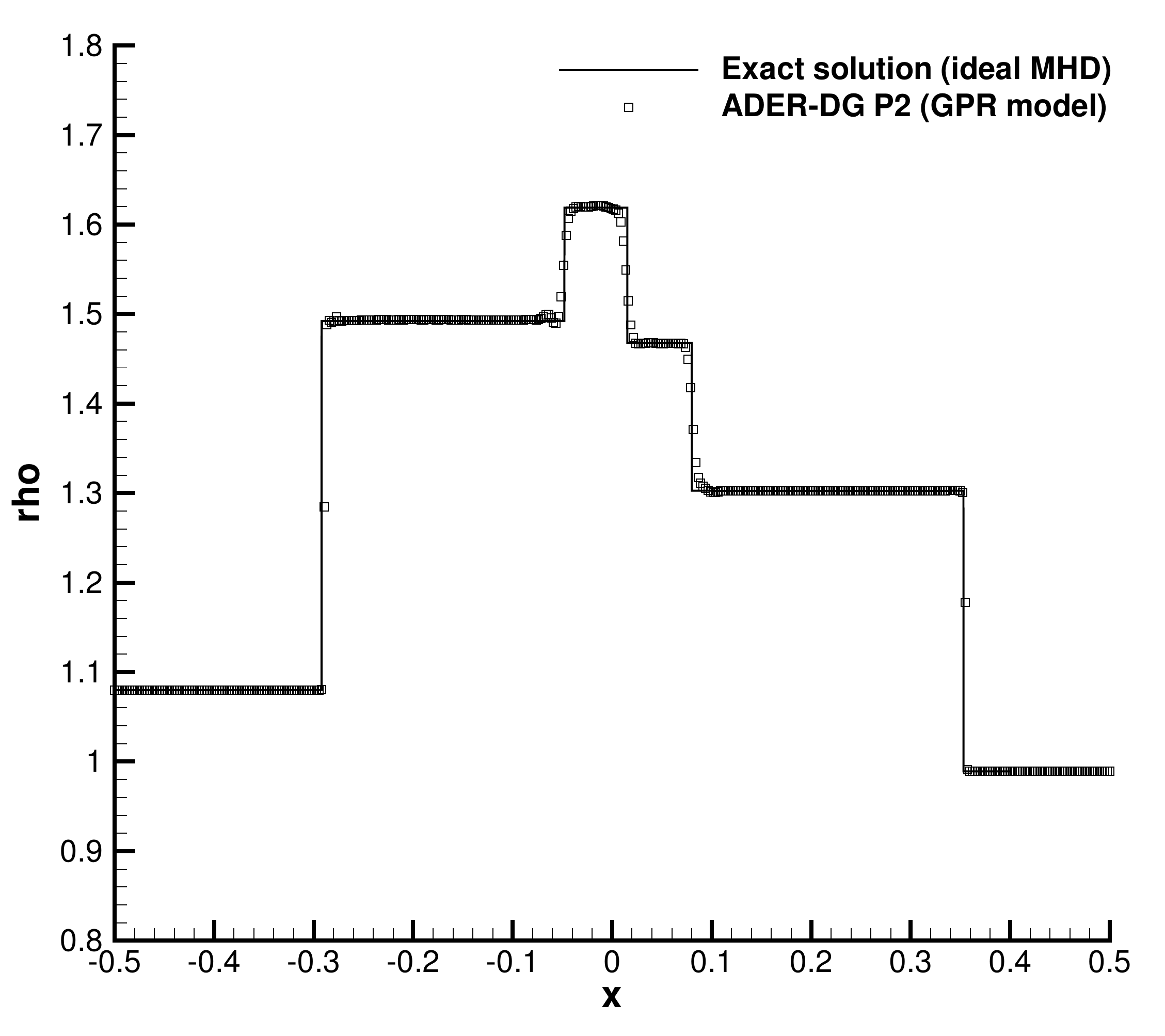} & 
      \includegraphics[draft=false,width=0.45\textwidth]{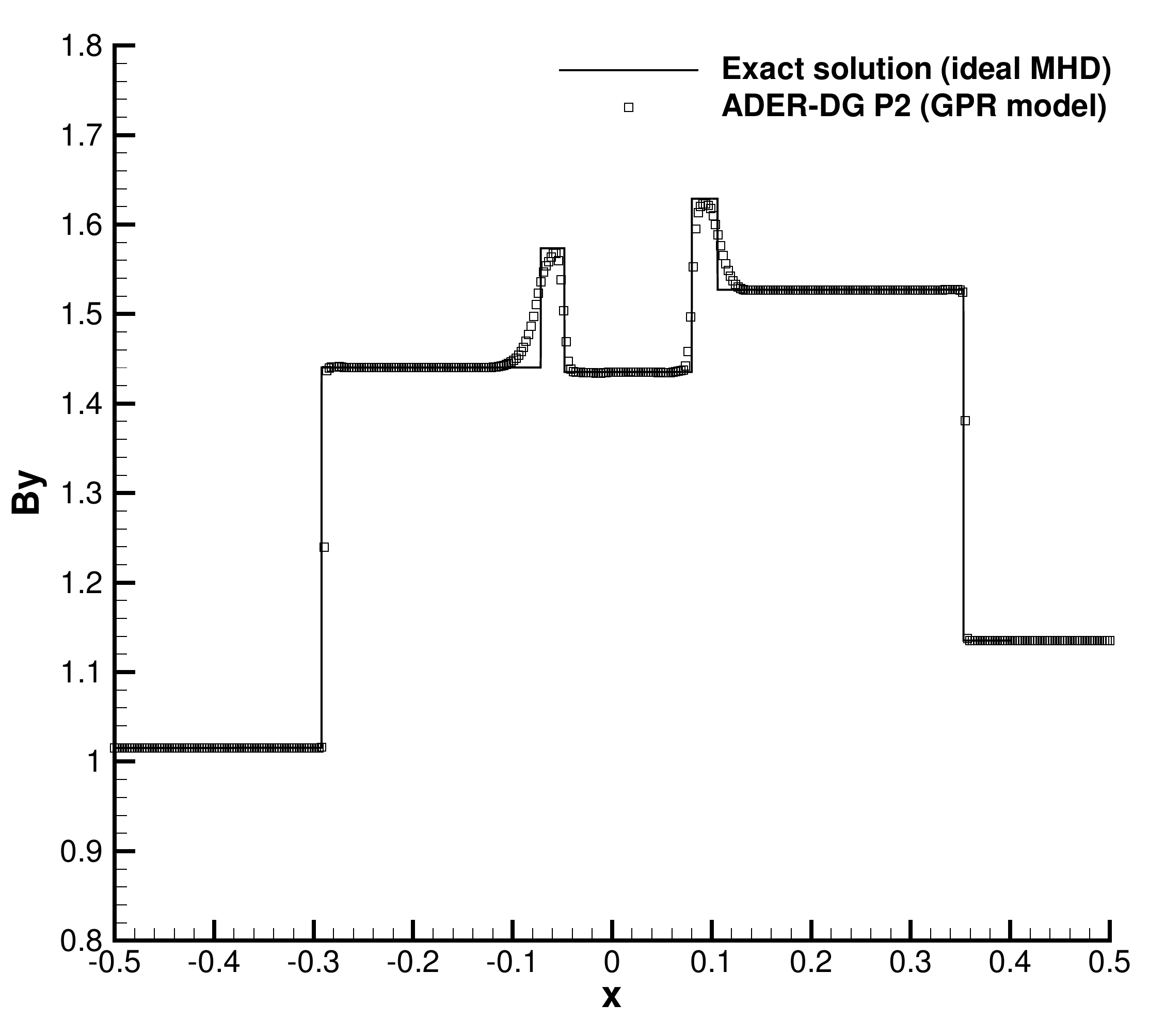}  \\    
      \includegraphics[draft=false,width=0.45\textwidth]{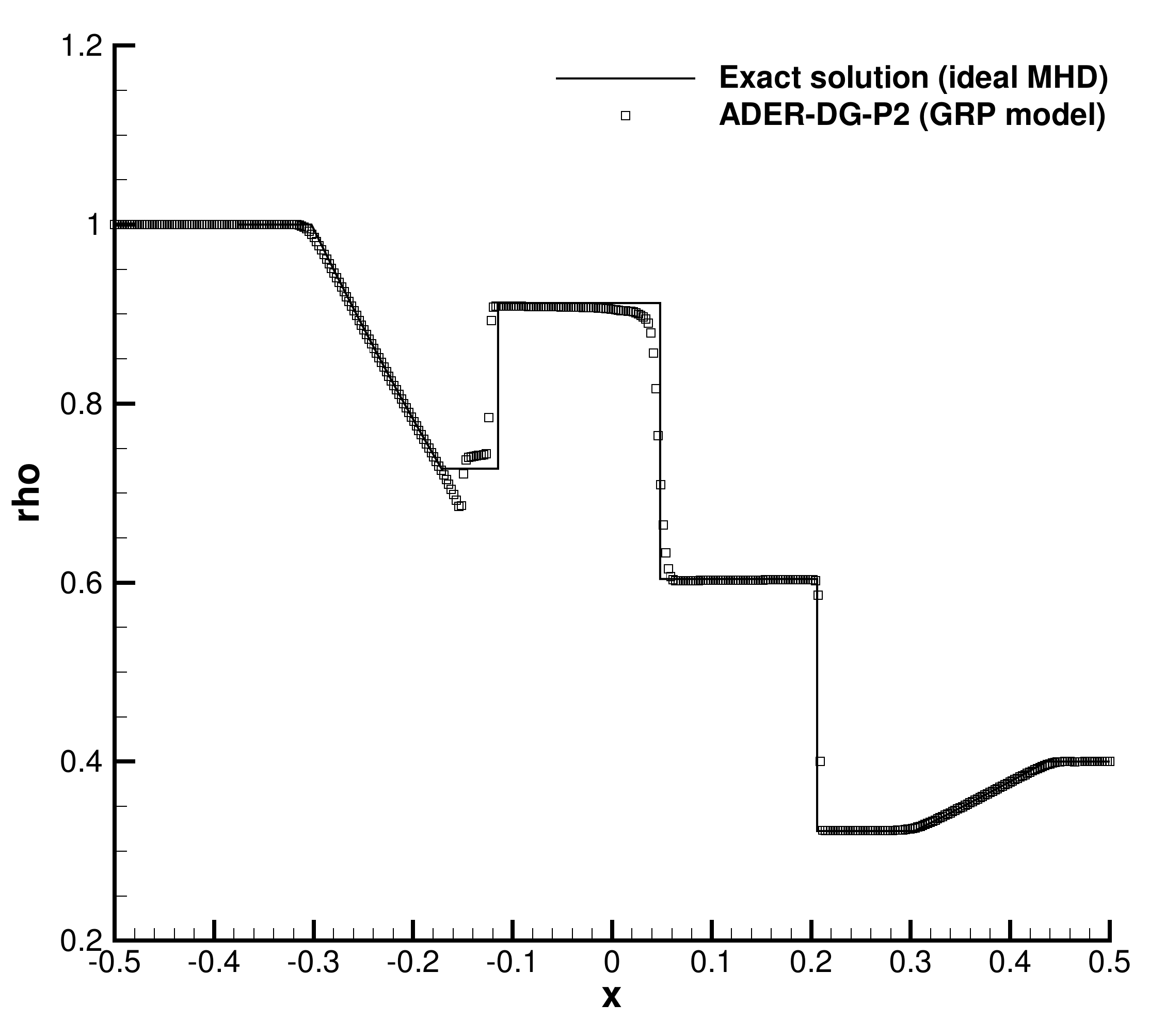} & 
      \includegraphics[draft=false,width=0.45\textwidth]{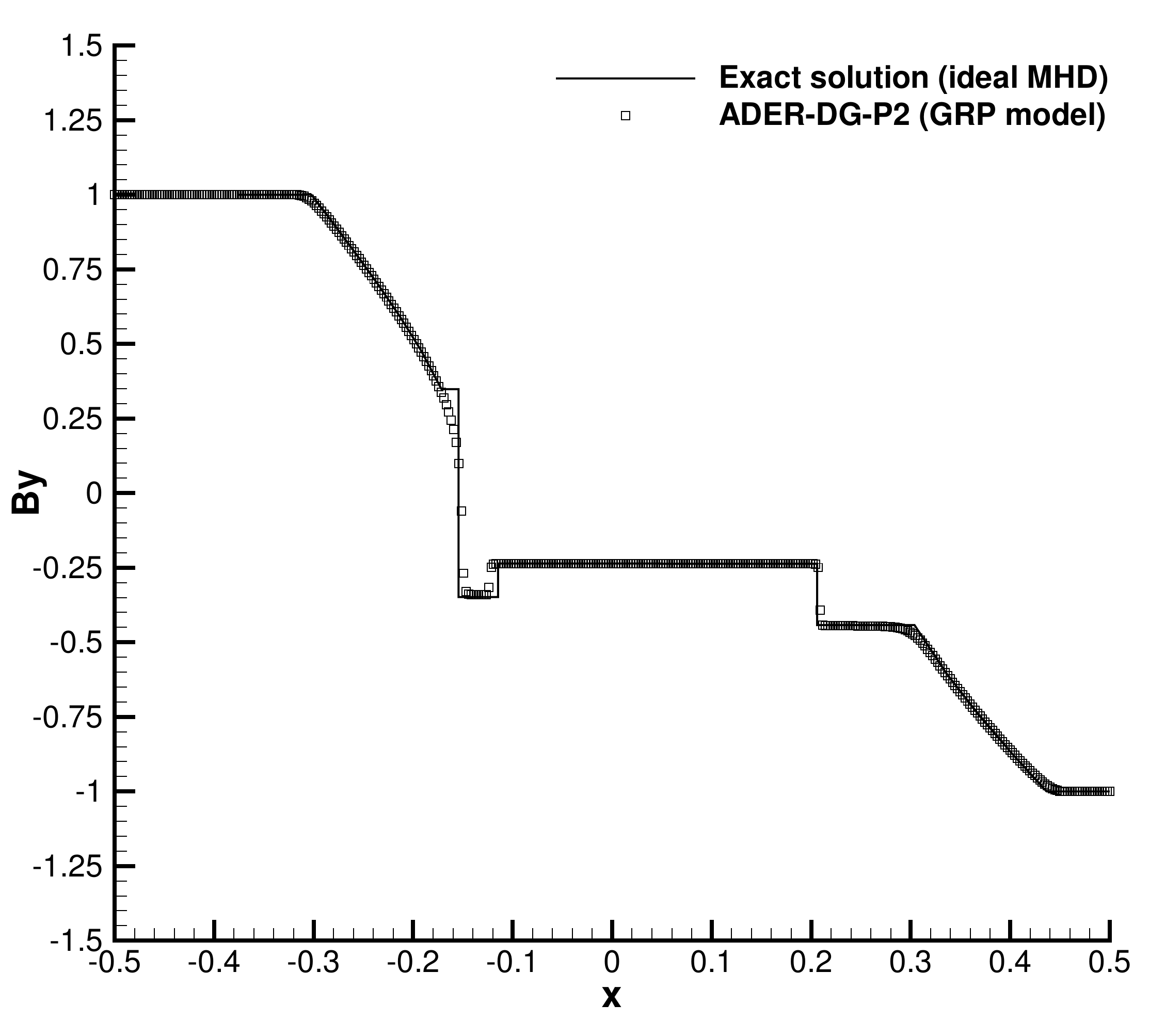}    
	\end{tabular} 
    \caption{MHD Riemann problems RP1 (top), RP2 (center) and RP3 (bottom) simulated with the GPR model ($\mu=\kappa=\eta=10^{-4}$) using an ADER-DG $P_2$ scheme with AMR
		 and comparison with the exact solution of the ideal MHD equations. The fluid density $\rho$ (left) and the magnetic field component $B_y$ (right) are depicted.} 
    \label{fig.rp}
	\end{center}
\end{figure}

While the previous test cases involved only smooth solutions, the Riemann problems solved in this section contain all different kinds of elementary flow discontinuities. 
The initial conditions for density, velocity, pressure and magnetic field together with the final simulation time $t_{\textnormal{e}}$ as well as the position of the initial
discontinuity $x_d$ are summarized in Table \ref{tab.ic.mhd}, while the initial data for the remaining variables are  
given by $\AAA=\sqrt[3]{\rho} \, \mathbf{I}$, $\mathbf{J}=0$ and $\mathbf{E} = -\mathbf{v} \times \mathbf{B}$. 
The computational domain is $\Omega = [-0.5,+0.5] \times [-0.1,0.1]$ and is discretized with an ADER-DG $P_2$ method using an adaptive Cartesian mesh (AMR) with initial 
resolution on the level zero grid of $100 \times 4$ cells. 
Two levels of refinement are admitted ($\ell_{\max} = 2$) with a refinement factor of $r=3$ between two adjacent levels. Refinement and recoarsening are based 
on the density as indicator variable. 
For more details on the AMR implementation, in particular for high order ADER schemes in combination with time-accurate local time stepping (LTS), see \cite{AMR3DCL,Zanotti2015a}. 
The model parameters used for the simulation are $\gamma = \frac{5}{3}$, $\rho_0 = 1$, $c_s=0.8$, $\alpha^2 = 0.8$, $c=10$ and $\mu = \kappa = \eta = 10^{-4}$, i.e. we are in a rather
stiff regime of the model where a comparison with the ideal MHD equations is appropriate. The results obtained with the GPR model for density $\rho$ and the magnetic field 
component $B_y$ are depicted in Fig. \ref{fig.rp} for all three problems, together with the exact solution of the Riemann problem of the ideal MHD equations. Overall, a good agreement 
can be noted, apart from the compound wave which can be observed in the density for RP1. This phenomenon is visible also in standard finite volume schemes applied to the ideal 
MHD equations, see e.g. \cite{OsherUniversal}. 

\begin{table}[!t]
 \caption{Initial states left and right for the density $\rho$, velocity vector $\mathbf{v} = (u,v,w)$, 
 the pressure $p$ and the magnetic field vector $\mathbf{B} = (B_x,B_y,B_z)$. 
 The final output times,   ($t_{\textnormal{end}}$) and the initial position of the discontinuity ($x_d$) are also given. } 
\begin{center} 
 \begin{tabular}{rccccccccc}
 \hline
 Case & $\rho$ & $u$ & $v$ & $w$ & $p$ & $B_x$ & $B_y$ & $B_z$ & $t_{\textnormal{e}}$, $x_d$      \\ 
 \hline
 RP1 L: &  1.0    &  0.0     & 0.0    & 0.0      &  1.0     & $\frac{3}{4}$ &  $ 1$  & 0.0   & 0.1    \\  
     R: &  0.125  &  0.0     & 0.0    & 0.0      &  0.1     & $\frac{3}{4}$ &  $-1$  & 0.0   & 0.0    \\
 RP2 L: &  1.08   &  1.2     & 0.01   & 0.5      &  0.95    & 0.564189 &  1.015541 & 0.564189 & 0.2         \\
     R: &  0.9891 &  -0.0131 & 0.0269 & 0.010037 &  0.97159 & 0.564189 &  1.135262 & 0.564923 & -0.1        \\
 RP3 L: &  1.0    &  0.0     & 0.0    & 0.0      &  1.0     & $1.3$ &  $ 1$   & 0.0   & 0.16          \\
     R: &  0.4    &  0.0     & 0.0    & 0.0      &  0.4     & $1.3$ &  $-1$   & 0.0   & 0.0           \\
 \hline
 \end{tabular}
\end{center} 
 \label{tab.ic.mhd}
\end{table}
 
\subsection{MHD rotor problem} 
\label{sec.mhdrotor} 

\begin{figure}[!htbp]
  \begin{center}
	\begin{tabular}{cc} 
      \includegraphics[draft=false,width=0.45\textwidth]{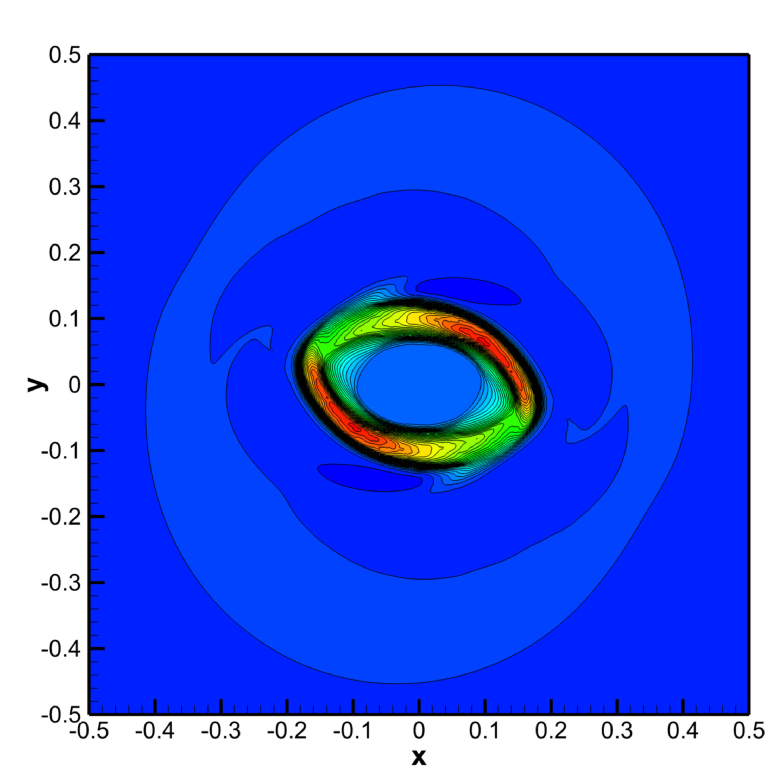} & 
      \includegraphics[draft=false,width=0.45\textwidth]{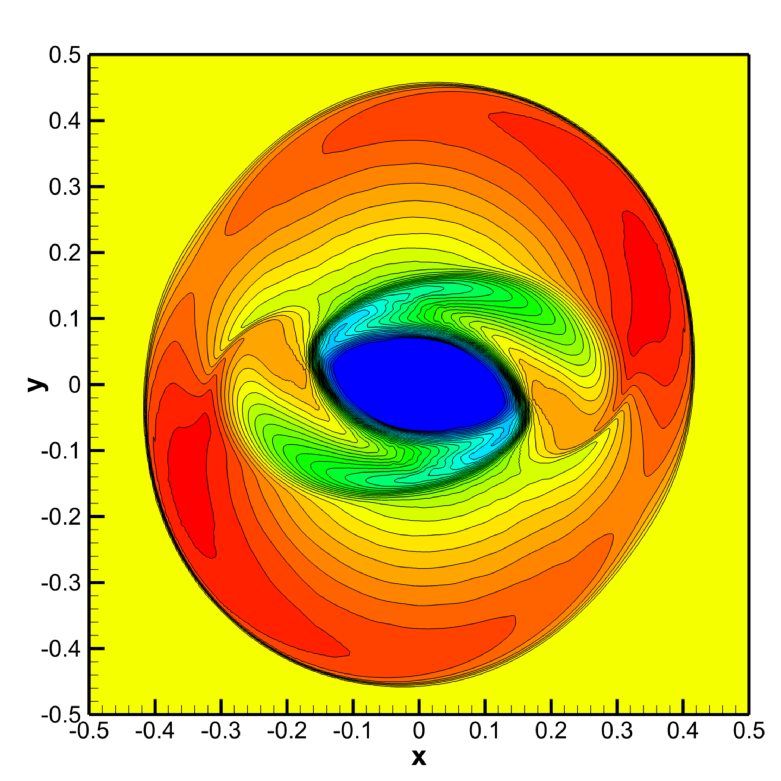}   \\  
      \includegraphics[draft=false,width=0.45\textwidth]{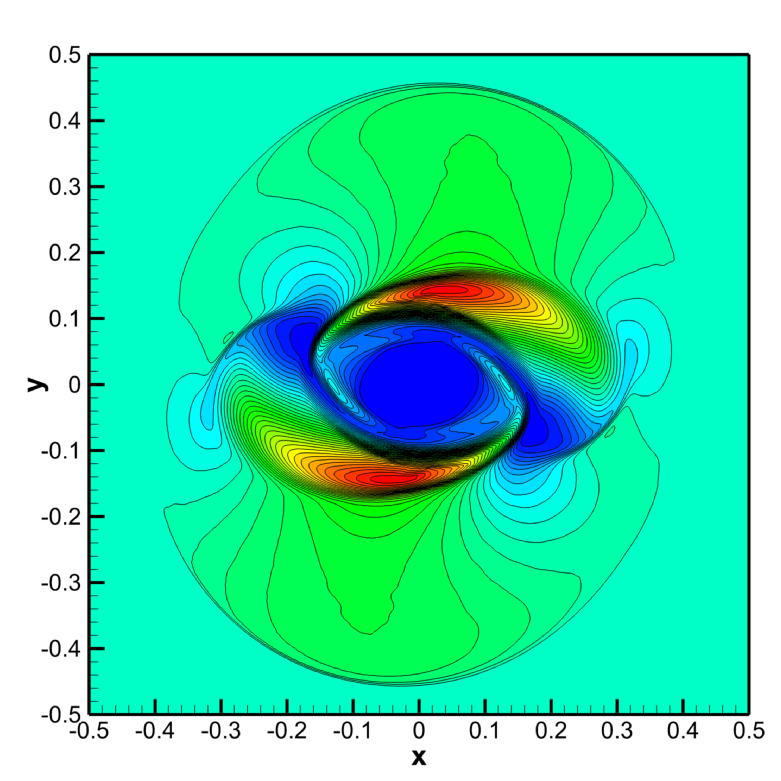} & 
      \includegraphics[draft=false,width=0.45\textwidth]{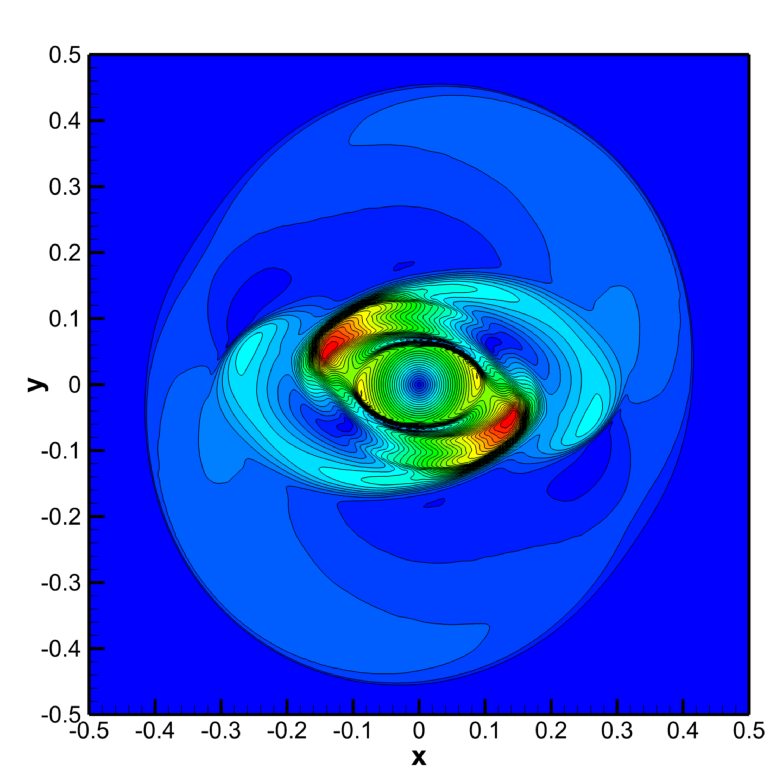}     
	\end{tabular} 
    \caption{MHD rotor problem at time $t=0.25$ simulated with the GPR model ($\mu=\kappa=\eta=10^{-4}$) using an ADER-DG $P_2$ scheme. The contours of fluid density $\rho$ (top left), fluid pressure $p$ (top right), magnetic pressure (bottom left) and Mach number (bottom right) are shown.}
    \label{fig.rotor}
	\end{center}
\end{figure}

Here we solve the well-known MHD rotor problem originally proposed by Balsara and Spicer in \cite{BalsaraSpicer1999} and later also used in many other papers on 
numerical methods for the ideal MHD equations. In this test problem, a rapidly rotating high density fluid (the rotor) is embedded in a low density atmosphere at rest. 
The fluid pressure and the magnetic field are initially constant everywhere. 
The rotor produces torsional Alfv\'en waves which travel into the outer fluid. The computational domain is defined as $\Omega = [-0.5,+0.5]^2$ and we use
an ADER-DG $P_2$ scheme on a uniform Cartesian grid composed of $200 \times 200$ elements. 
The initial density is $\rho=10$ inside the rotor ($0 \leq r \leq 0.1$) and $\rho=1$ for the outer fluid. The velocity in the outer fluid is initially
set to zero, while it is given by $\mathbf{v} = \boldsymbol{\omega} \times \mathbf{x}$ inside the rotor, with $\boldsymbol{\omega}=(0,0,10)$. 
The initial pressure is $p=1$ and the magnetic field vector is set to $\mathbf{B} = ( B_0, 0, 0)^T$ in the entire computational 
domain $\Omega$, with $B_0 = \frac{2.5}{\sqrt{4\pi}}$. As proposed by Balsara and Spicer, a linear taper is applied to the velocity and density field between $0.1 \leq r \leq 0.105$. 
The other variables are initially set to $\AAA = \sqrt[3]{\rho} \, \mathbf{I}$, $\mathbf{J}=0$ and $\mathbf{E} = -\mathbf{v} \times \mathbf{B}$. 
The speed for the hyperbolic divergence cleaning is set to $c_h=2$ and the model parameters are $\gamma=1.4$, $c=10$, $\alpha^2=0.8$, $\rho_0=1$, $c_s=0.8$ and $\mu=\kappa=\eta=10^{-4}$, 
which make the system sufficiently stiff so that a comparison with the ideal MHD equations is possible. The results are depicted at time $t=0.25$ in Fig. \ref{fig.rotor} for the usual 
quantities density, pressure, Mach number and magnetic pressure. The results agree qualitatively well with those reported by Balsara and Spicer in \cite{BalsaraSpicer1999}, 
as well as those reported in other papers on high order numerical methods for the ideal MHD equations, see e.g. \cite{Dumbser2008,AMR3DCL}. 

\subsection{MHD blast wave problem} 

The MHD blast wave problem is a very challenging test problem even for numerical schemes applied to the ideal MHD equations. Here, we solve the GPR model \eqref{eqn.conti}-\eqref{eqn.energy} 
in a rather stiff regime so that comparisons with numerical results obtained with the ideal MHD equations are possible. The computational domain is given by 
$\Omega=[-0.5,+0.5]^2$ and is discretized with an ADER-DG $P_2$ scheme on a uniform Cartesian grid using $200 \times 200$ elements. 
The initial data are $\rho=1$, $\mathbf{v}=0$ and $\mathbf{B}=(B_0,0,0)$ with $B_0 = \frac{100}{\sqrt{4 \pi}}$. The pressure is set to $p=1000$ in a small internal 
circular region ($r<0.1$) and is $p=0.1$ elsewhere, hence the pressure jumps over four orders of magnitude in this test problem. Furthermore, the fluid is highly 
magnetized due to the presence of a very strong magnetic field in the entire domain.  

\begin{figure}[!htbp]
  \begin{center}
	\begin{tabular}{cc} 
      \includegraphics[draft=false,width=0.45\textwidth]{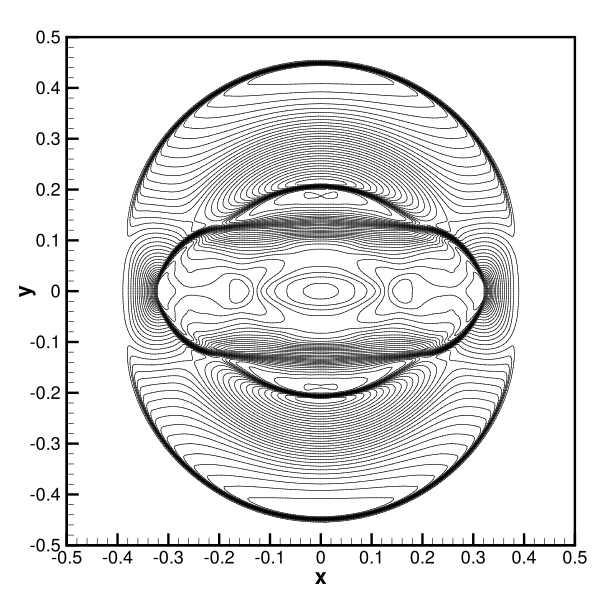} & 
      \includegraphics[draft=false,width=0.45\textwidth]{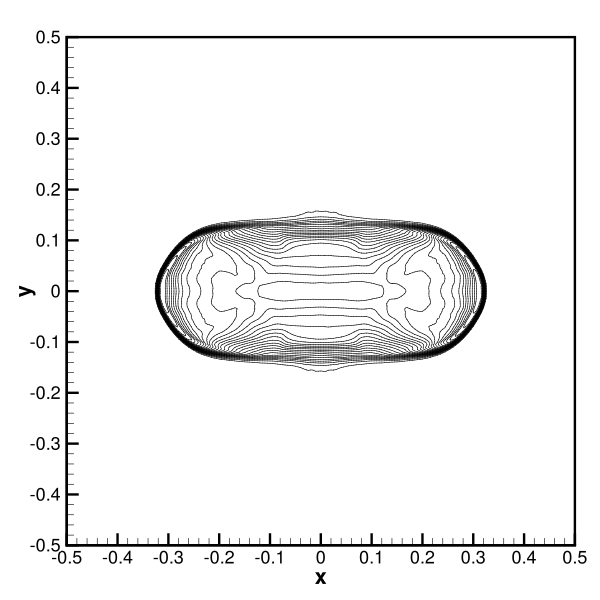}   \\  
      \includegraphics[draft=false,width=0.45\textwidth]{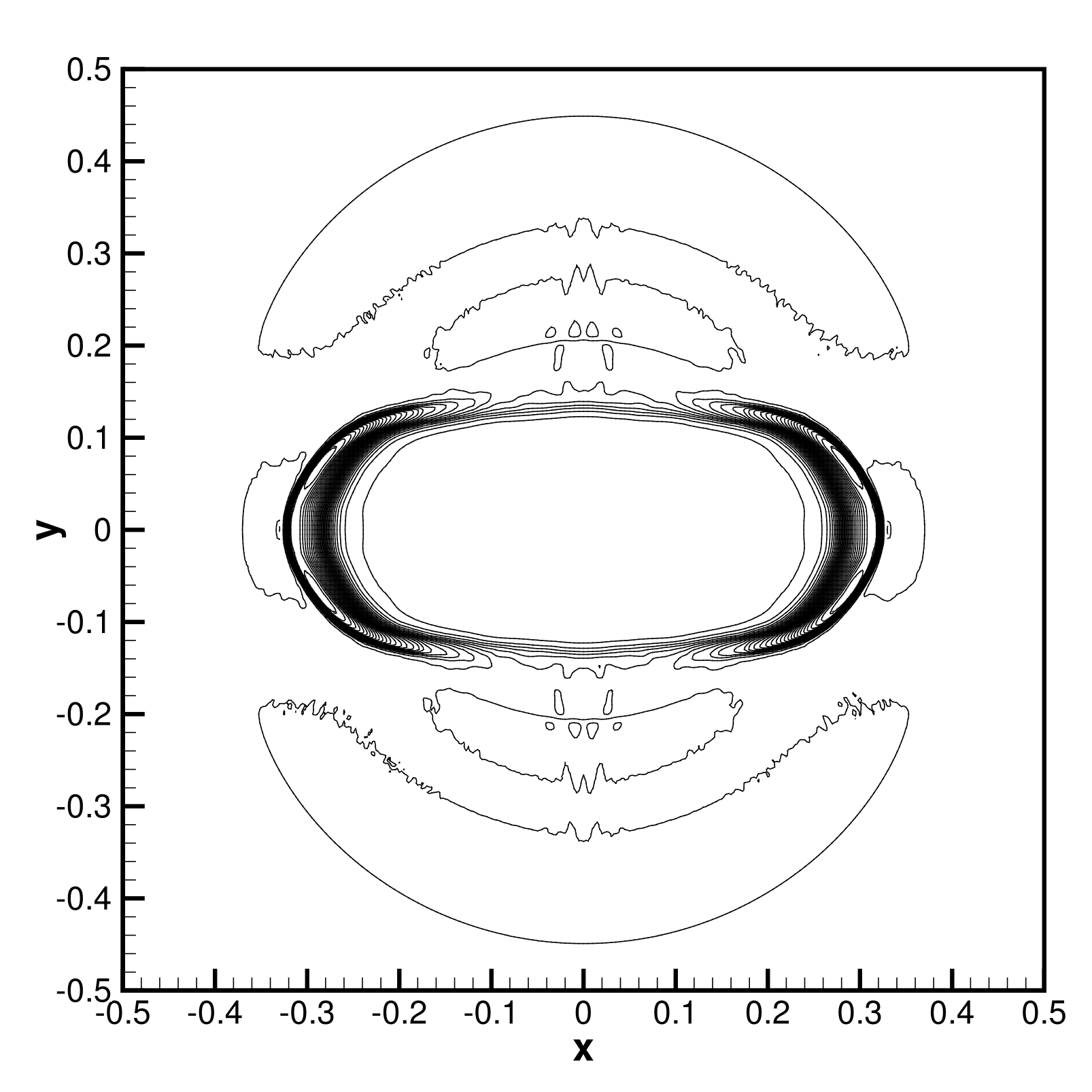} & 
      \includegraphics[draft=false,width=0.45\textwidth]{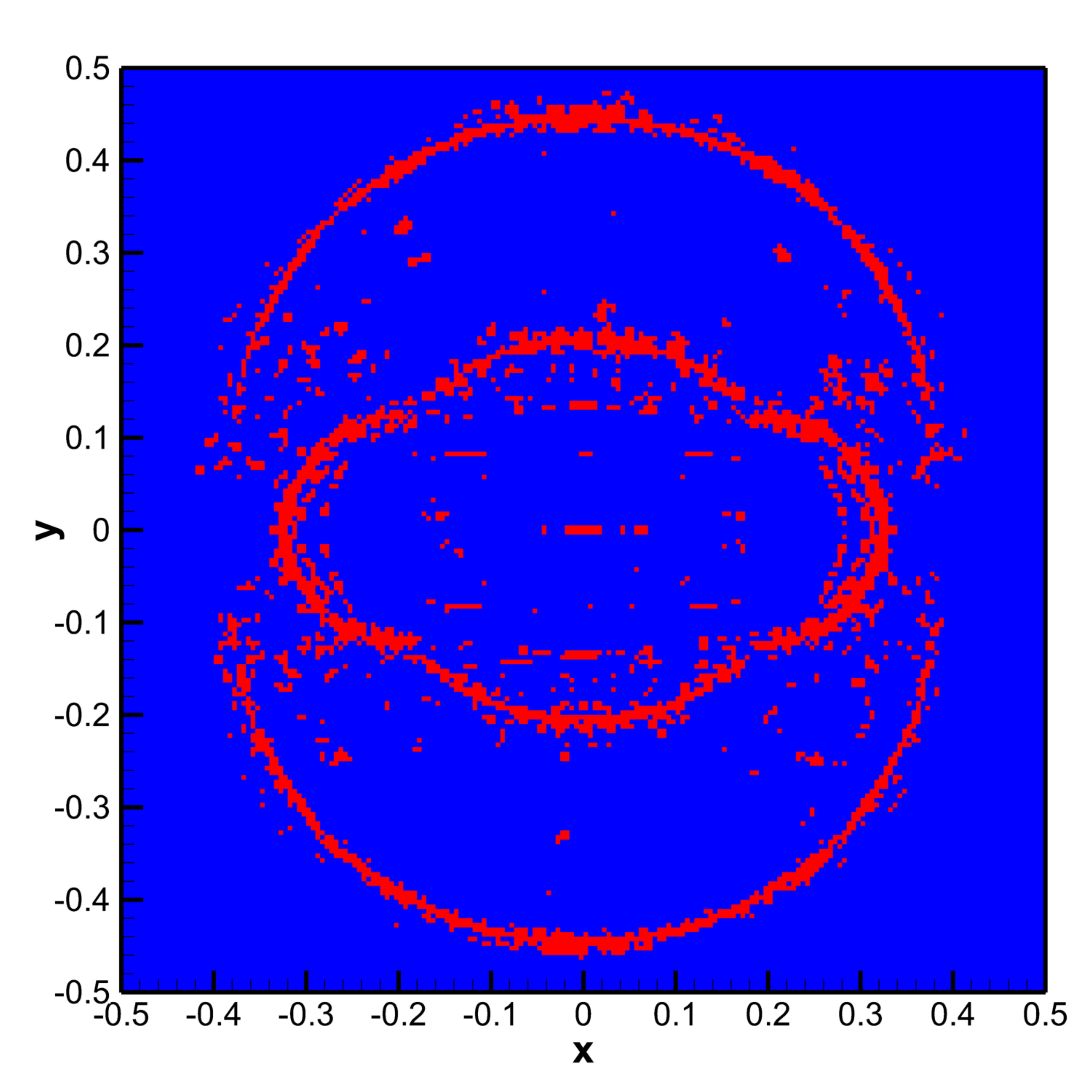} 
          
	\end{tabular} 
    \caption{MHD blast wave problem at time $t=0.01$ using the GPR model ($\mu=\kappa=\eta=10^{-3}$) and an ADER-DG $P_2$ scheme. The contours of the magnetic field 
		component $b_x$ (top left), the fluid pressure $p$ (top right) and the fluid density $\rho$ (bottom left) are shown, together with a map of troubled zones in red  
		that use the subcell finite volume limiter of the ADER-DG $P_2$ scheme, while unlimited cells are colored in blue (bottom right).}
    \label{fig.blast}
	\end{center}
\end{figure}

The other variables of the model are initially set to $\AAA = \sqrt[3]{\rho} \, \mathbf{I}$, $\mathbf{J}=0$ and $\mathbf{E} = -\mathbf{v} \times \mathbf{B}$. 
The speed for the hyperbolic divergence cleaning is chosen as $c_h=2$ and the model parameters are given by $\gamma=1.4$, $c=10$, $\alpha^2=0.8$, $\rho_0=1$, 
$c_s=0.8$ and $\mu=\kappa=\eta=10^{-3}$. 
The computational results are depicted at time $t=0.01$ in Fig. \ref{fig.blast} for the magnetic field component $B_x$, the fluid pressure $p$, the density $\rho$ and the color map 
of the limited cells and unlimited cells in red and blue, respectively. For details on the finite volume subcell limiter, see \cite{Dumbser2014,Zanotti2015a,Zanotti2015b}. 
The computational results agree qualitatively with those reported by Balsara and Spicer in \cite{BalsaraSpicer1999}. 

\subsection{Inviscid Orszag-Tang vortex system} 

\begin{figure}[!htbp]
  \begin{center}
	\begin{tabular}{cc} 
      \includegraphics[draft=false,width=0.4\textwidth]{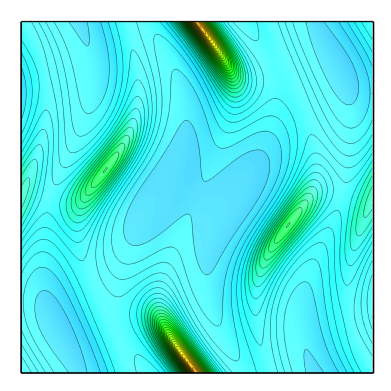} 
      & 
      \includegraphics[draft=false,width=0.4\textwidth]{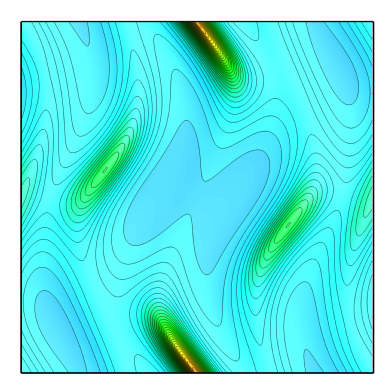}
       \\ 
      \includegraphics[draft=false,width=0.4\textwidth]{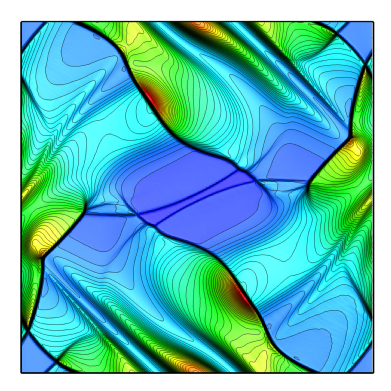} 
      &   
      \includegraphics[draft=false,width=0.4\textwidth]{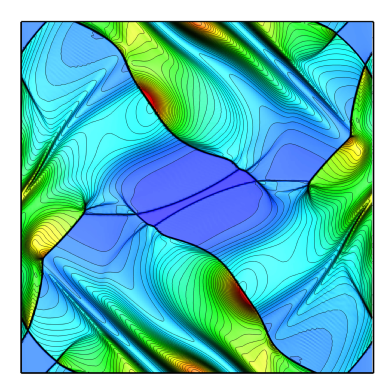}
       \\   
      \includegraphics[draft=false,width=0.4\textwidth]{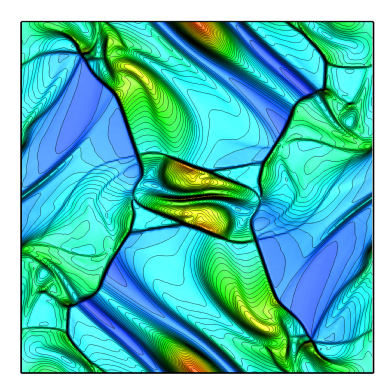} 
      &    
      \includegraphics[draft=false,width=0.4\textwidth]{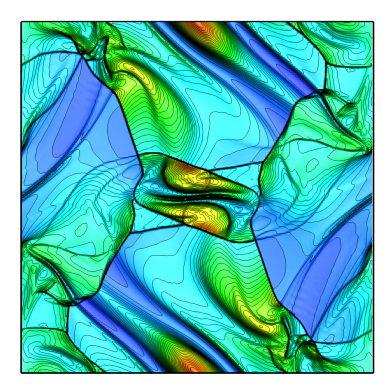}
          
	\end{tabular} 
    \caption{Orszag-Tang vortex problem at output times $t=0.5$, $t=2.0$ and $t=3.0$ from top to bottom. 
		         Left column: GPR model ($\mu=\kappa=\eta=10^{-4}$). 
		         Right column: ideal MHD equations for a direct comparison.}
    \label{fig.ot}
	\end{center}
\end{figure}

In this section we study the well-known Orszag-Tang vortex system for the MHD equations \cite{OrszagTang,DahlburgPicone,PiconeDahlburg}, comparing the numerical results 
of the GPR model with those obtained with the ideal MHD equations. The setup is the one used in \cite{JiangWu} and \cite{Dumbser2008}. 
In both computations, the numerical method and the 
computational grid used are identical, as well as the initial conditions for the density, velocity, pressure and the magnetic field. The other variables of 
the GPR model are set to $\AAA=\sqrt[3]{\rho} \, \mathbf{I}$, $\mathbf{J}=0$ and $\mathbf{E}=-\mathbf{v} \times \mathbf{B}$. The computational 
domain under consideration is $\Omega = [0,2\pi]^2$ with four periodic boundary conditions and the initial conditions are given by
$\rho = \gamma^2$, $\mathbf{v} = (-\sin(y),\sin(x),0)$, $p=\gamma$ and $\mathbf{B}=(-\sin(y),\sin(2x),0)$, with 
$\gamma = \frac{5}{3}$. 
The remaining parameters of the GPR model are $c=10$, $\alpha^2=0.8$, $c_s=0.8$, $c_h=2$, $\rho_0=1$ and $\mu=\kappa=\eta=10^{-4}$, so that the system is 
sufficiently stiff in order to allow a comparison with the ideal MHD equations. 
Simulations are carried out on a uniform Cartesian grid composed of $200 \times 200$ elements using an ADER-DG $P_2$ scheme until a final 
time of $t=3$. The comparison between the computational results obtained with the GPR model and the ideal MHD equations is provided in Fig. \ref{fig.ot}. 
A very good agreement between the two solutions can be noted, even for later times when the solution has already developed many small scale structures. We stress that
in both cases two completely different PDE systems have been solved. 

\subsection{Viscous and resistive Orszag-Tang vortex} 

We now solve the Orszag-Tang vortex system in a highly viscous and resistive regime, where the computational results of the GPR model are 
compared with those of the \textit{viscous} and \textit{resistive} MHD equations (VRMHD). 
The computational setup of this test case is taken from \cite{WarburtonVRMHD} and \cite{ADERVRMHD}, where also the governing PDE of the classical 
viscous and resistive MHD equations have been detailed. 
The computational domain is again $\Omega=[0,2\pi]^2$ with four periodic boundary conditions and the common initial condition for both models 
is given this time by 
$\rho=1$, $\mathbf{v}=(-\sin(y),\sin(x),0)$, $\mathbf{B}=(-\sin(y),\sin(2x), 0)$, $p=\frac{15}{4} + \frac{1}{4} \cos(4x) + \frac{4}{5} \cos(2x) \cos(y) - \cos(x) \cos(y) + \frac{1}{4} \cos(2y)$. 
The ratio of specific heats is set to $\gamma = \frac{5}{3}$. 
The other variables of the GPR model are set to $\AAA=\mathbf{I}$, $\mathbf{J}=0$ and $\mathbf{E}=-\mathbf{v} \times \mathbf{B}$. We choose 
$\mu = \eta = 10^{-2}$ and a Prandtl number of $Pr=1$ based on the heat capacity at constant volume of $c_v=1$, leading to a heat conduction coefficient of 
$\kappa = \gamma \mu$. 

\begin{figure}[!htbp]
  \begin{center}
	\begin{tabular}{cc} 
      \includegraphics[draft=false,width=0.45\textwidth]{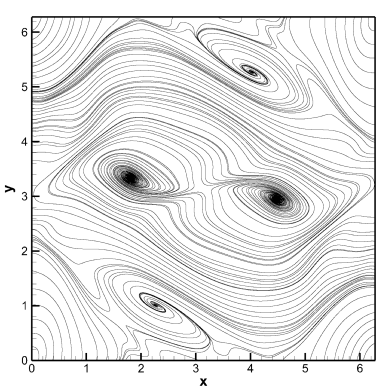}  
       &   
      \includegraphics[draft=false,width=0.45\textwidth]{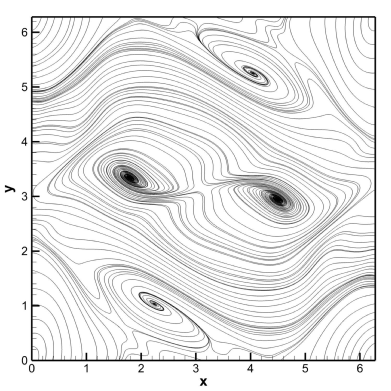}
       \\ 
      \includegraphics[draft=false,width=0.45\textwidth]{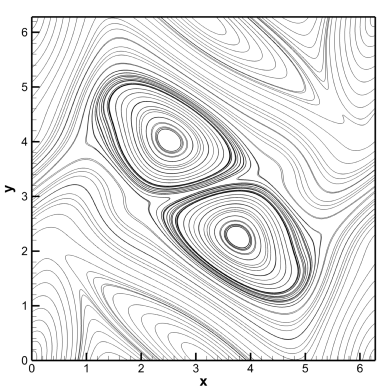}  
       & 
      \includegraphics[draft=false,width=0.45\textwidth]{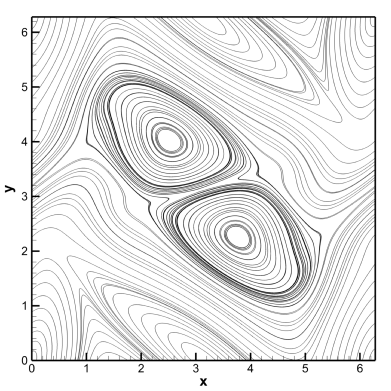}
         
	\end{tabular} 
    \caption{Viscous Orszag-Tang vortex problem ($\mu=\eta=10^{-2}$, $Pr=1$) at time $t=2.0$. First order hyperbolic GPR model (left) and
		classical VRMHD equations (right) for direct comparison. Velocity streamlines (top) and magnetic field lines (bottom).}
    \label{fig.vot}
	\end{center}
\end{figure}

For the GPR model, we run the test problem until $t=2$ with an ADER-DG $P_3$ scheme on a uniform Cartesian grid composed of $ 200 \times 200$ elements,   
while the numerical solution of the viscous and resistive MHD equations has been taken directly from \cite{ADERVRMHD}, where an eighth order $P_4P_7$ scheme 
has been used to solve the VRMHD equations on a very coarse unstructured triangular mesh composed of only 990 triangles. 
The direct comparison between the first order hyperbolic GPR model and the second order hyperbolic-parabolic 
VRMHD model is provided in Fig. \ref{fig.vot}, where the velocity streamlines as well as the magnetic field lines are plotted. Overall, we can observe an excellent  
agreement between the two computational results, which have been obtained by solving two completely different PDE systems and using two different mesh topologies (Cartesian
grid versus an unstructured simplex mesh). 

\subsection{Kelvin-Helmholtz instability in a viscous and resistive magnetized fluid} 

\begin{figure}[!htbp]
  \begin{center}
	\begin{tabular}{cc} 
      \includegraphics[draft=false,width=0.4\textwidth]{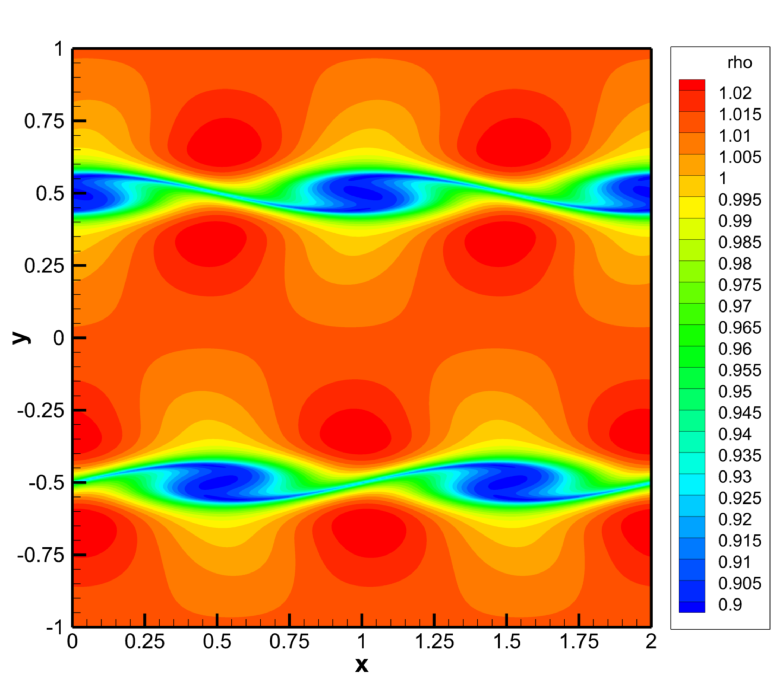} &  
      \includegraphics[draft=false,width=0.4\textwidth]{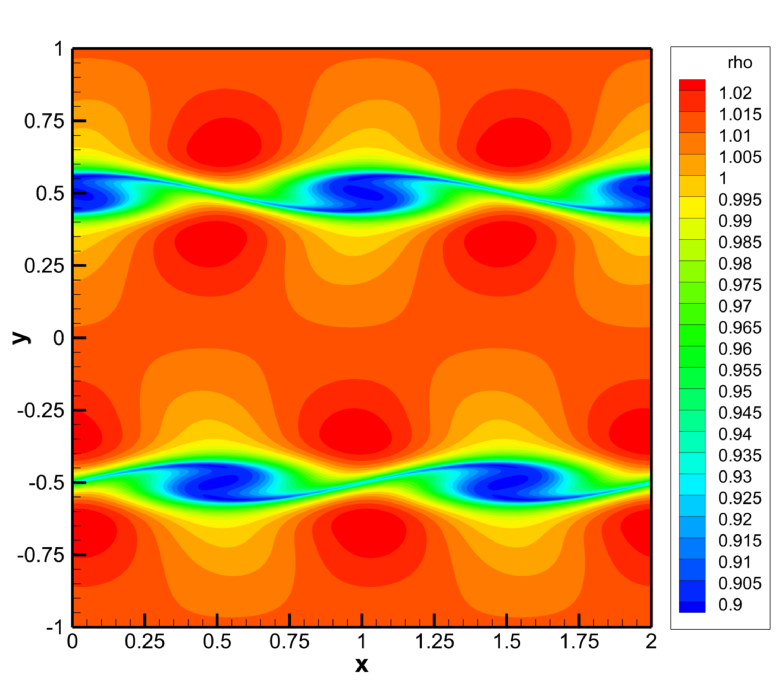}    
	\end{tabular} 
    \caption{Kelvin-Helmholtz instability in a viscous and resistive magnetized fluid ($\mu=\eta=10^{-3}$, $\kappa=0$) at time $t=4.0$. 
    Density contours obtained with an ADER-DG $P_3$ scheme for the first order hyperbolic GPR model (left) and for the VRMHD equations (right).} 
    \label{fig.kh}
	\end{center}
\end{figure}

\begin{figure}[!htbp]
  \begin{center}
	\begin{tabular}{cc} 
      \includegraphics[draft=false,width=0.4\textwidth]{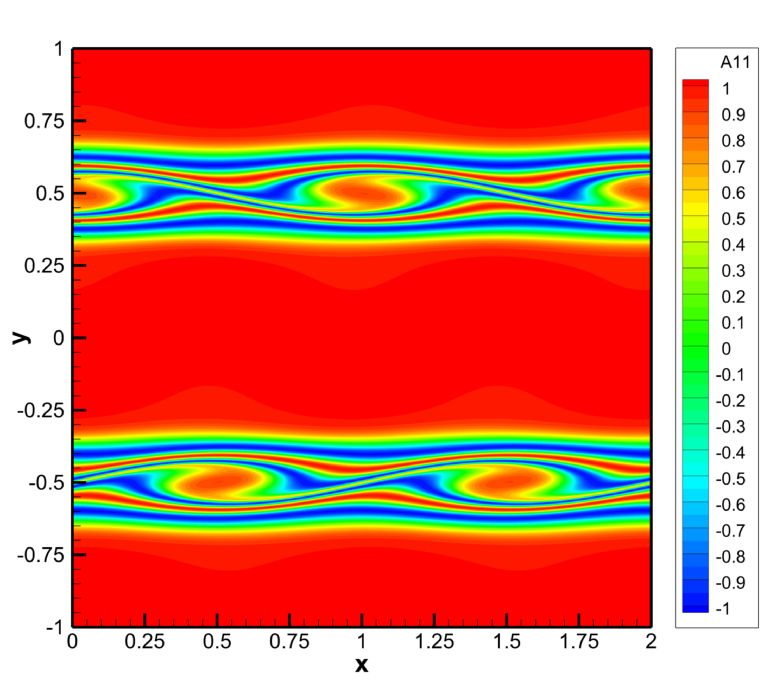} &  
      \includegraphics[draft=false,width=0.4\textwidth]{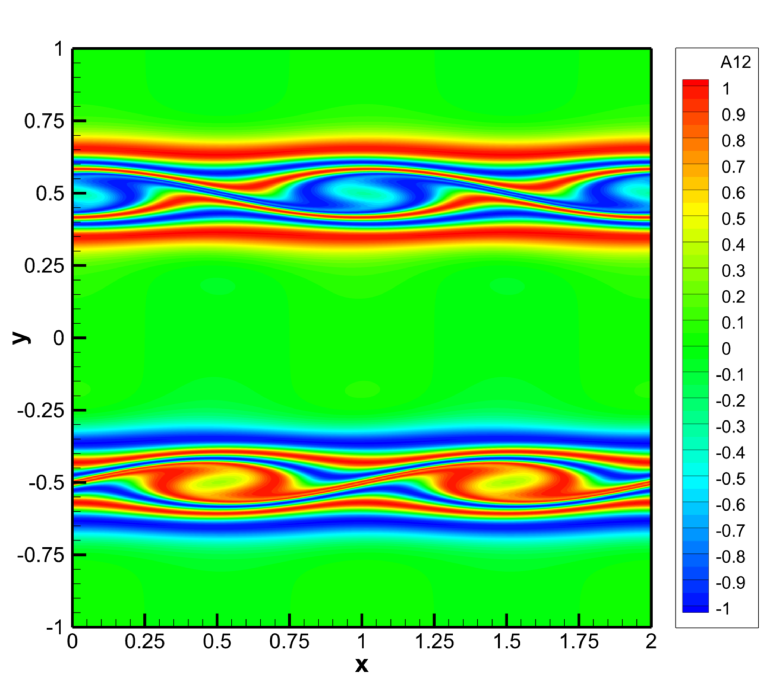}    
	\end{tabular} 
    \caption{Kelvin-Helmholtz instability in a viscous and resistive magnetized fluid ($\mu=\eta=10^{-3}$, $\kappa=0$) at time $t=4.0$. 
    Contours of the distortion components $A_{11}$ (left) and $A_{12}$ (right) for the first order hyperbolic GPR model.} 
    \label{fig.kha}
	\end{center}
\end{figure}

This test problem concerns the simulation of the Kelvin-Helmholtz instability in a magnetized fluid. As in the previous test, we solve the problem with the 
first order hyperbolic GPR model and with the viscous and resistive MHD equations (VRMHD).  The setup of the initial conditions of the problem is taken 
from \cite{ADERVRMHD} and references therein: $\rho=1$, $p=\frac{3}{5}$, $\gamma=\frac{5}{3}$, $\mathbf{v}=(u,v,0)$ with 
$u=-\halb U_0 \textnormal{tanh} \left( \frac{y-0.5}{a} \right)$ and $v=\delta v \sin(2\pi x) \sin(\pi y)$. The magnetic field is given by  
\begin{equation*}
	 \mathbf{B} = \left\{ 
	 \begin{array}{ccc}  
	 (B_0, 0, 0), & \textnormal{ if } & \halb + a < y < 1, \\ 
	 (B_0 \sin(\chi), 0, B_0 \cos(\chi)), & \textnormal{ if } & \halb - a < y < \halb + a, \\ 
	 (0, 0, B_0), & \textnormal{ if } & 0 < y < \halb - a,
	 \end{array}    	 
	   \right. 
\end{equation*}
with $ \chi = \frac{\pi}{2} \frac{y-0.5+a}{2a}$, $a=\frac{1}{25}$, $U_0=1$, $\delta v = 0.01$ and 
$B_0 = 0.07$. 
The computational domain is $\Omega=[0,1] \times [-1,1]$ with periodic boundary conditions in the $x$ direction and is discretized with an ADER-DG $P_2$ scheme using $200 \times 400$ elements for both PDE  models, i.e. for the first order hyperbolic GPR system and for the viscous and resistive MHD model  (VRMHD). The physical parameters are $\mu=\eta=10^{-3}$, $\kappa=0$, i.e. heat conduction is neglected in 
this test. The remaining parameters of the GPR model are set to $c=2$, $\alpha=0$, $c_s=0.8$, $c_h=2$ and $\rho_0=1$. 
The computational results are shown in Fig. \ref{fig.kh}, where an excellent agreement between 
the GPR results and the VRMHD results can be noted. One can clearly see the development of the so-called cat-eye vortices 
and the thin filaments connecting the individual vortices. For a detailed discussion of the MHD Kelvin-Helmholtz instability, 
see \cite{KeppensKH2D} and \cite{JeongRyuKH2D}. In Fig. \ref{fig.kha} we also show two components of the distortion
$\AAA$, which is the key quantity of the GPR model that allows the computation of the stress tensor in the case of
both, fluids and solids. As already emphasized in \cite{DPRZ2016}, the distortion $\AAA$ is very well suited for flow visualization. 


\subsection{High Lundquist number magnetic reconnection}

As next test problem we consider the case of a high Lundquist number magnetic reconnection. Reconnection occurs in unstable current sheets
due to the tearing instability that generates so-called plasmoid chains, see e.g. \cite{Biskamp1986,Loureiro2007,Samtaney2009,Landi2015}. 
For investigations of magnetic reconnection in the resistive relativistic case see \cite{Zanotti2011}, where high order ADER schemes 
similar to those employed in the present paper have been used \cite{DumbserZanotti}. 

\begin{figure}[!htbp]
  \begin{center}
	\begin{tabular}{cc} 
      \includegraphics[draft=false,width=0.45\textwidth]{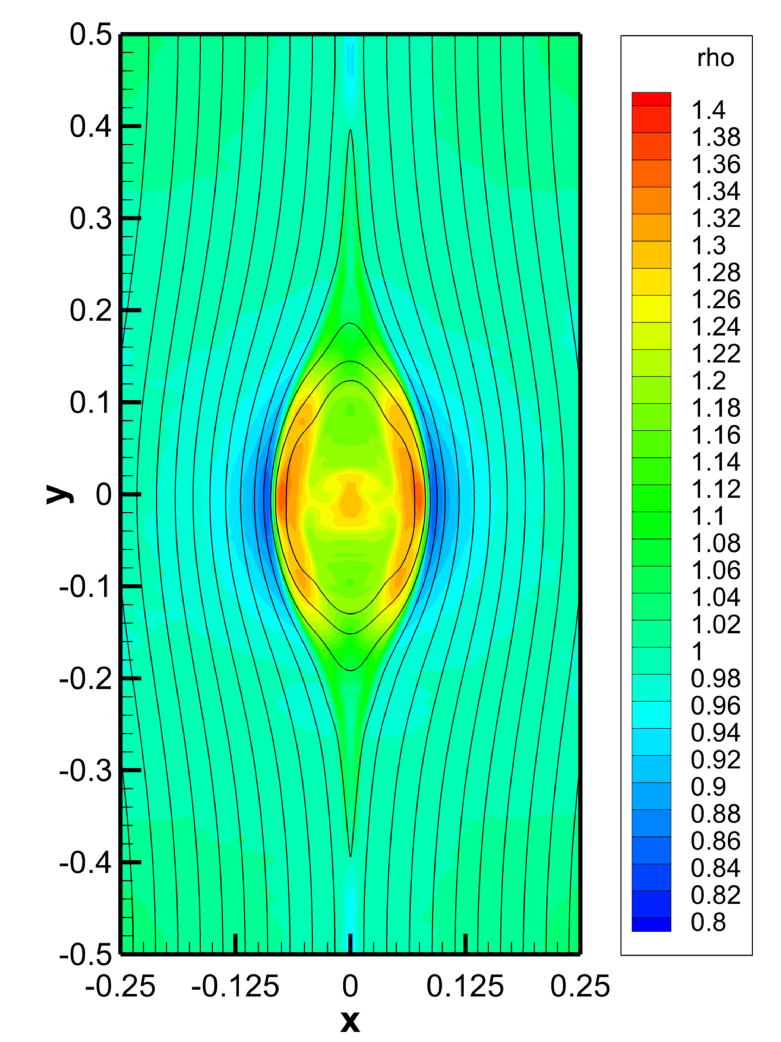}
       &  
      \includegraphics[draft=false,width=0.45\textwidth]{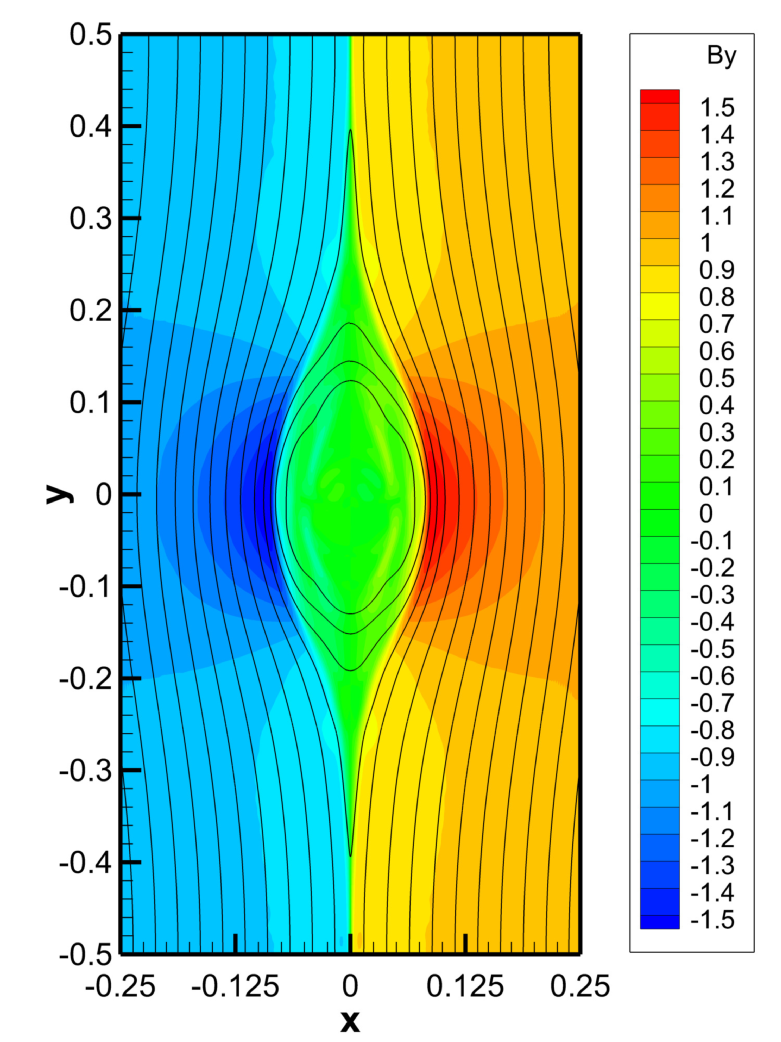}
        \\  
      \includegraphics[draft=false,width=0.45\textwidth]{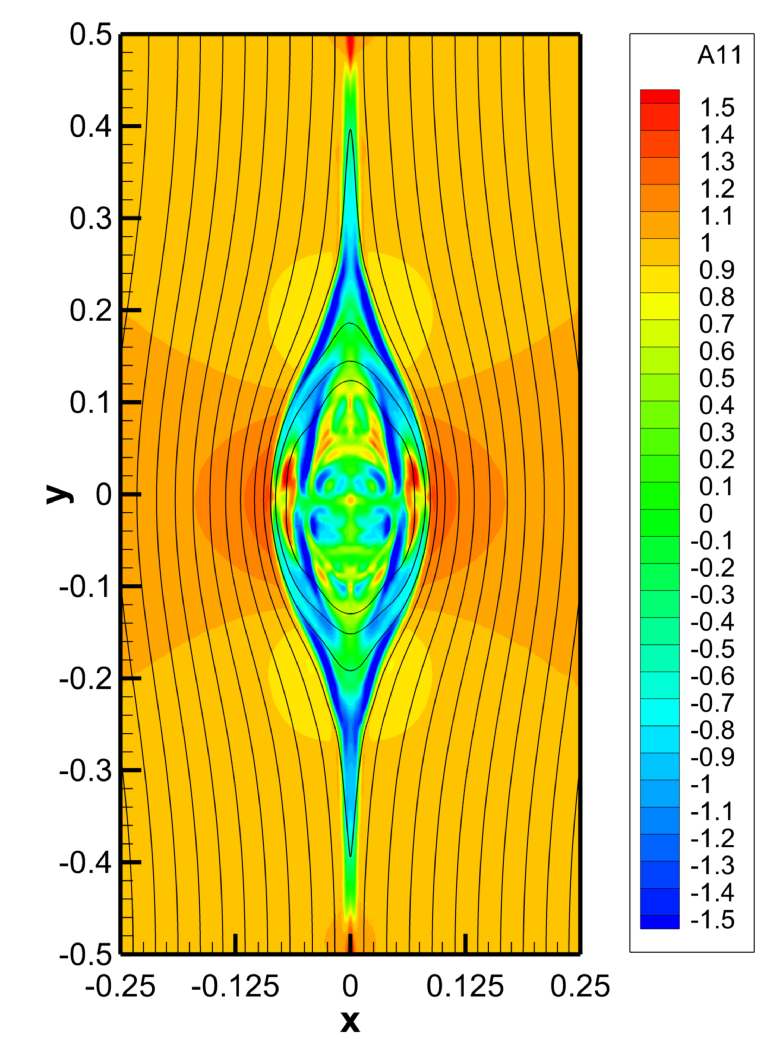}
        & 
      \includegraphics[draft=false,width=0.45\textwidth]{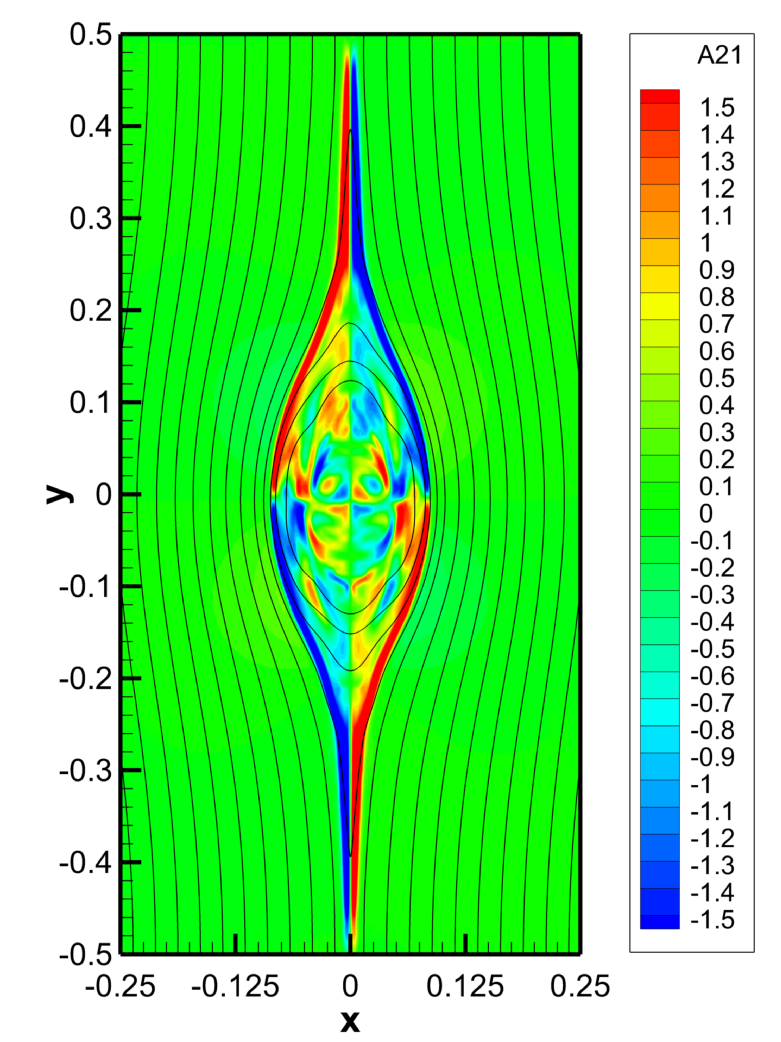}
        
	\end{tabular} 
    \caption{High Lundquist number magnetic reconnection ($\eta = 10^{-6}$, $\mu=\kappa=0$) at time $t=5.5$. Results obtained for the first order hyperbolic GPR model with an ADER-DG $P_2$ scheme.
		         Density (top left), magnetic field component $B_y$ (top right), distortion components $A_{11}$ (bottom left) and $A_{21}$ (bottom right). The magnetic field lines are also shown.} 
    \label{fig.reconnection}
	\end{center}
\end{figure}

The computational domain is given by $\Omega = [-25a,25a] \times [-L/2,L/2]$, where $L$ is the length of the domain, 
$a=L/S^{1/3}$ is the width of the current sheet and the Lundquist number is $S=L v_a/\eta$, with the Alfv\'en speed $v_a$. 
The initial condition for the magnetic field is  
\begin{equation}
\mathbf{B}= ( 0, B_0 \textnormal{tanh}(x/a), B_0 \textnormal{sech}(x/a) ), 
\end{equation}
with the relation between $B_0$ and $v_a$ given by $v_a^2=B_0^2/\rho$. 
The initial fluid pressure is set to $p=\rho/(\gamma M^2)$, where $M=v_a/c_0$ is the magnetic Mach number and 
$c_0$ is the sound speed. For our test we use $\rho=1$, $v_a=L=1$, $\gamma=5/3$, $M=0.7$ and $S=10^6$, hence 
the thickness of the current sheet is $a=0.01$, while the plasma parameter $\beta$ is given by $\beta=2.4$.  
The instability is triggered by adding a small perturbation to the velocity field of the form 
\begin{eqnarray}
u&=&\varepsilon \tanh\xi \exp(-\xi^2) \cos(ky) \\
v&=&\varepsilon (2\xi \textnormal{tanh}\xi - \textnormal{sech}^2\xi)\exp(-\xi^2)S^{1/2}\sin(ky)/k\,,
\end{eqnarray}
where $\varepsilon=10^{-3}$, $\xi = x S^{1/2}$ and the wave-number is computed from $k L=2\pi m$, with $m=10$. 
Free outflow and periodic boundary conditions are chosen along the $x$ and $y$ direction, respectively. 
The remaining variables and parameters of the GPR model are set to $\AAA=\sqrt[3]{\rho} \mathbf{I}$, 
$\mathbf{J}=0$, $\mathbf{E}=-\mathbf{v} \times \mathbf{B}$, $c=2$, $c_h=2$, $\rho_0=1$, $c_s=\alpha=0$, $\mu=\kappa=0$. 
Simulations have been carried out until a final time of $t=5.5$ using an ADER-DG $P_2$ scheme with \textit{a posteriori} subcell finite volume 
limiter \cite{Dumbser2014,Zanotti2015a,Zanotti2015b} on a uniform Cartesian grid composed of $200 \times 400$ elements. 
In Fig. \ref{fig.reconnection} the computational results for density, magnetic field component $B_y$ and for two components of 
the distortion $\AAA$ are shown. One can clearly see the formation of a main \emph{reconnection island} or \emph{major plasmoid}, 
which has the usual form similar to that observed also in other simulations reported in the literature \cite{Landi2015,Fambri2017}. 

\subsection{Scattering of a plane wave}

\begin{figure}[!htbp]
  \begin{center}
	\begin{tabular}{cc} 
      \includegraphics[draft=false,width=0.45\textwidth]{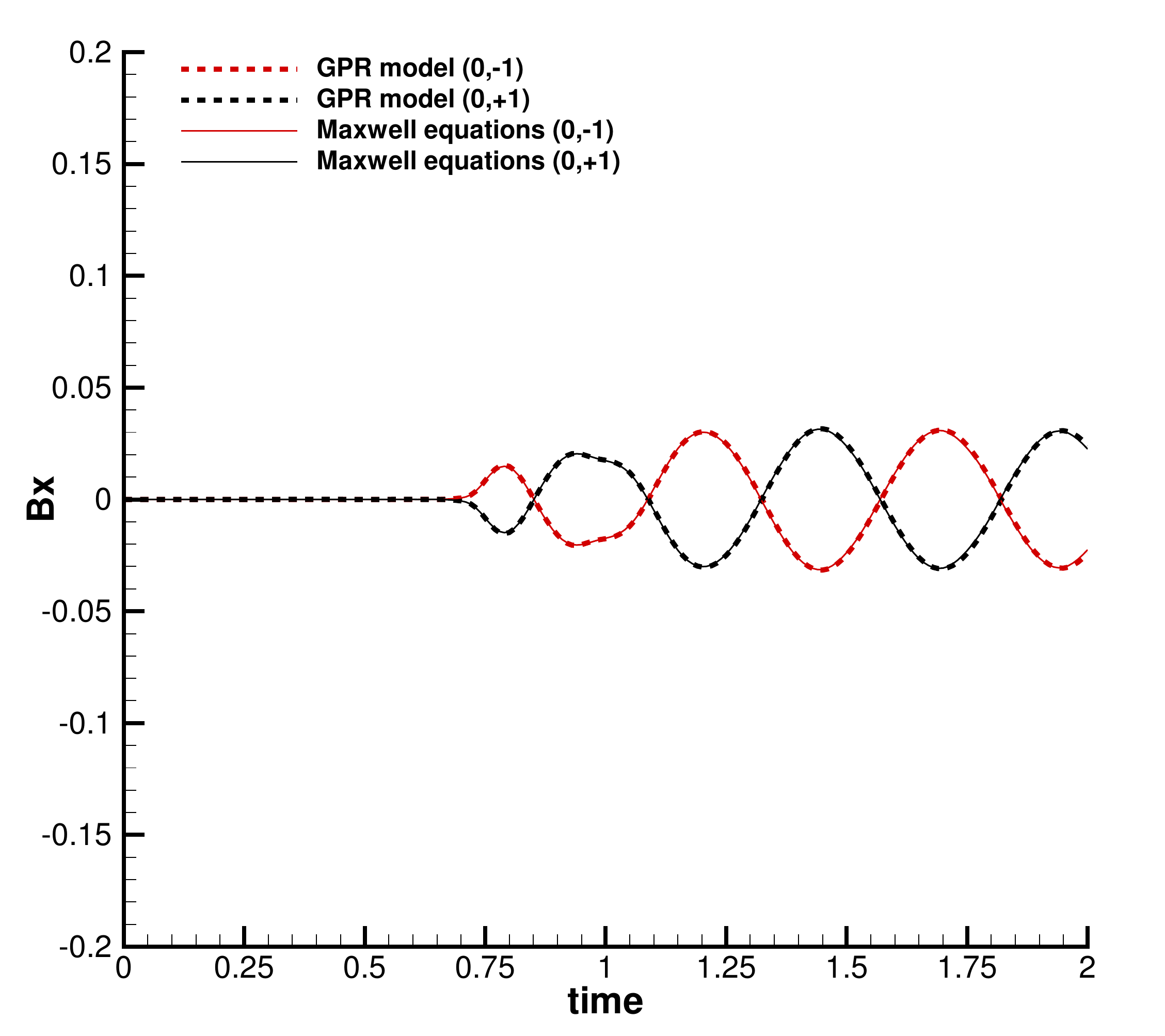}
       &  
      \includegraphics[draft=false,width=0.45\textwidth]{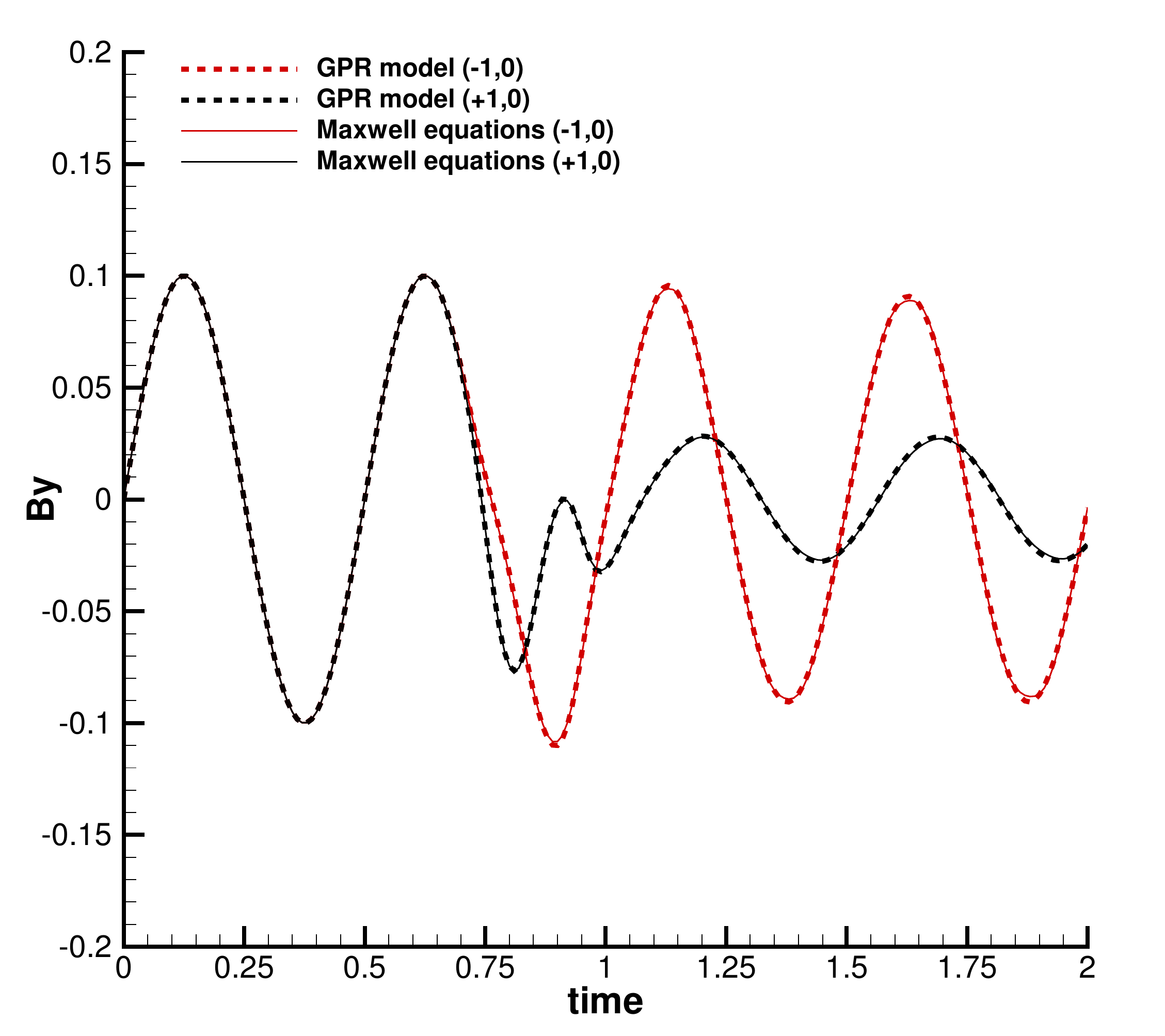}
        \\  
      \includegraphics[draft=false,width=0.45\textwidth]{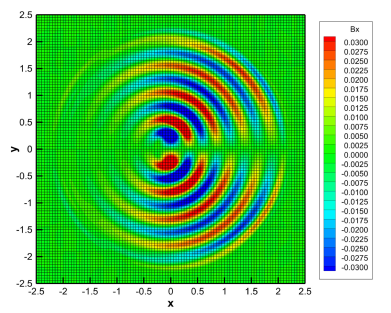}
        & 
      \includegraphics[draft=false,width=0.45\textwidth]{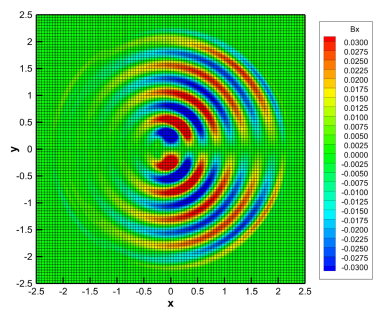}
        
	\end{tabular} 
    \caption{Scattering of a planar EM wave ($k=4\pi$) at a cylindrical inclusion of radius $R=0.25$. The light speed is $c=1$ in the ambient medium, while it is $c=2$ inside the inclusion. 
		         Comparison of the GPR model with the Maxwell equations using an ADER-DG $P_3$ scheme. Time series of the magnetic field components $B_x$ (top left) 
						 and $B_y$ (top right) registered in four observation points. Contour plot of the magnetic field component $B_x$ at time $t=2.0$ using the GPR model (bottom left) 
						and the Maxwell equations (bottom right).} 
    \label{fig.scatter}
	\end{center}
\end{figure}

Here, we run the GPR model inside a solid medium at rest in the limit $\tau_1 \to \infty$, $\tau_2 \to \infty$ and $\eta \to \infty$, i.e. without source terms. 
Hence, one expects to recover the behavior of the classical Maxwell equations concerning electromagnetic wave propagation. 
We therefore run the following test case twice, once with the full GPR model \eqref{eqn.conti}-\eqref{eqn.energy}, and once with the standard time domain Maxwell equations in a 
laboratory frame at rest. For high order ADER-DG schemes applied to the Maxwell equations, see \cite{TaubeMaxwell}. 
The computational domain for this test problem is $\Omega = [-2.5,+2.5]^2$ with four periodic boundary conditions. The initial electric and magnetic field vectors are set to 
$\mathbf{E}=(0,0,E_0 \sin (k x) )$ and $\mathbf{B}=(0, -B_0 \sin (k x), 0 )$, with $k=4\pi$ and $E_0=B_0=0.1$, while the remaining variables of the GPR model are initially set to 
$\rho=1$, $p=1$, $\mathbf{v}=\mathbf{J}=0$ and $\AAA=\mathbf{I}$. The light speed is set to $c=c_o=1$ everywhere in $\Omega$, apart from a small cylindrical inclusion 
of radius $R=0.25$, where it has been set to $c=c_i=2$. In order to avoid spurious oscillations, the transition has been smoothed by setting 
$c(r)=c_i (1-\xi) + c_o \xi$, with $\xi = \halb ( 1 + \textnormal{erf}((r-R)/\delta) )$ and $\delta = 0.05$. The remaining parameters in the GPR model are chosen as $\gamma=1.4$, 
$\tau_1 = \tau_2 = \eta = 10^{20}$, $c_h=2$, $c_s=0.8$, $\alpha^2=0.8$ and $\rho_0=1$. The problem is run with an unlimited ADER-DG $P_3$ scheme on a uniform Cartesian grid composed of $100 \times 100$
elements until a final time of $t=2.0$, so that the scattered waves have not yet reached the boundaries. 
In Fig. \ref{fig.scatter}, the computational results obtained with the GPR model are compared to those of the standard Maxwell equations. 
Overall, a very good agreement can be noted between the two different models, both, for the contour plot of $B_x$ that represents the scattered wave 
field, as well as for the time series recorded in the four observation points $\mathbf{x}_1=(-1,0)$, $\mathbf{x}_2=(+1,0)$, $\mathbf{x}_3=(0,-1)$ and $\mathbf{x}_4=(0,+1)$. 

\subsection{Comparison of different model regimes} 

In this last example we explore the behaviour of the GPR model in a large range of relaxation parameters, from the regime of a viscous and resistive magnetized fluid (case I, $\tau_1 \ll 1$, $\tau_2 \ll 1$, $\eta \ll 1$) 
over an electrically conducting elastic solid (case II, $\tau_1 \to \infty$, $\tau_2 \to \infty$, $\eta \ll 1$) to a non-conducting elastic solid (case III, $\tau_1 \to \infty$, $\tau_2 \to \infty$, $\eta \to \infty$). 
The initial data are essentially the same as for the MHD rotor problem solved in Section \ref{sec.mhdrotor}, i.e. density is set to $\rho=1$ in the ambient fluid and $\rho=10$ inside the rotor of radius $R=0.1$, 
while the initial pressure is constant everywhere $p=1$. The magnetic field vector is set to $\mathbf{B} = ( B_0, 0, 0)^T$ in the entire computational domain $\Omega$, with $B_0 = \frac{2.5}{\sqrt{4\pi}}$ and the velocity 
is zero in the ambient fluid and $\mathbf{v} = \boldsymbol{\omega} \times \mathbf{x}$ inside the rotor, with $\boldsymbol{\omega}=(0,0,10)$. 
In this test we use a light speed of $c=4$ and the stiffened gas equation of state \cite{DPRZ2016} with $p_0=1$. Furthermore, the reference density inside the rotor is chosen as $\rho_0=10$ so that the initial 
condition for the distortion is simply given by $\AAA=\mathbf{I}$. Furthermore, we initially set $\mathbf{J}=0$ and $\mathbf{E}=-\mathbf{v} \times \mathbf{B}$. The computational domain 
is $\Omega = [-1.25, +1.25]^2$, covered with a uniform Cartesian grid of $400 \times 400$ elements. All simulations are carried out with a third order ADER-WENO finite volume scheme \cite{AMR3DCL} and are 
run up to a final time of $t=0.25$. The parameters used for the three cases under consideration are summarized in Table \ref{tab.param}. 

\begin{table}  
\caption{Relaxation parameters of the GPR model used for the three different cases under consideration ($c_s=0.8$, $\alpha^2=0.8$, $c=4$).} 
\begin{center} 
\begin{small}
\renewcommand{\arraystretch}{1.0}
\begin{tabular}{cccc} 
\hline
         & Case I  & Case II  & Case III   \\ 
         & (viscous and resistive fluid) & (conducting elastic solid) &  (non-conducting elastic solid) \\ 
\hline
 $\tau_1$	& $10^{-3}$	& $10^{20}$	& $10^{20}$	     \\ 
 $\tau_2$	& $10^{-3}$	& $10^{20}$	& $10^{20}$	     \\ 
 $\eta$  	& $10^{-3}$	& $10^{-3}$	& $10^{20}$	     \\ 
\hline 
\end{tabular}
\end{small}
\end{center}
\label{tab.param}
\end{table} 

\begin{figure}[!htbp]
  \begin{center}
	\begin{tabular}{ccc} 
      \includegraphics[draft=false,width=0.3\textwidth]{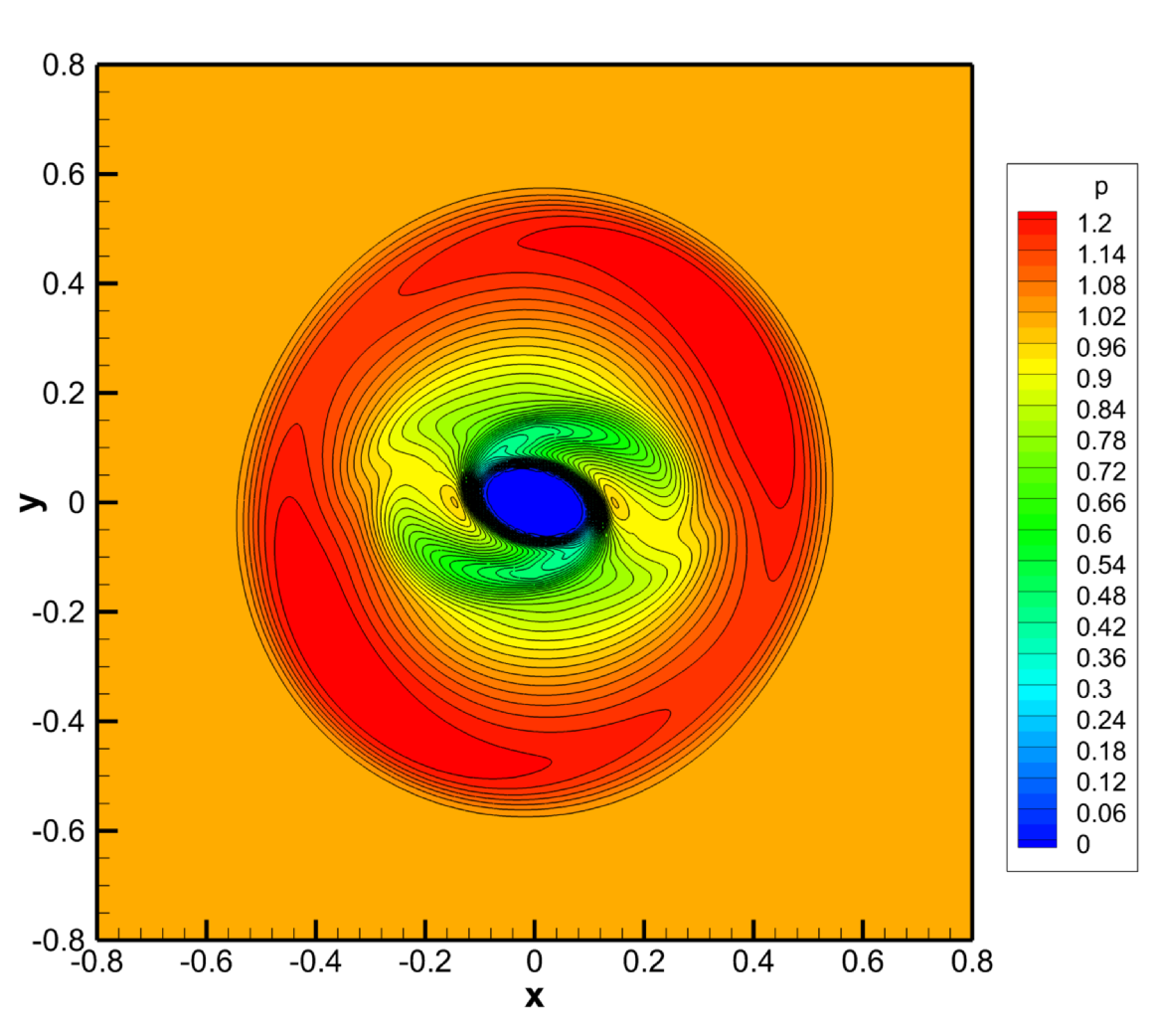}  
      &  
      \includegraphics[draft=false,width=0.3\textwidth]{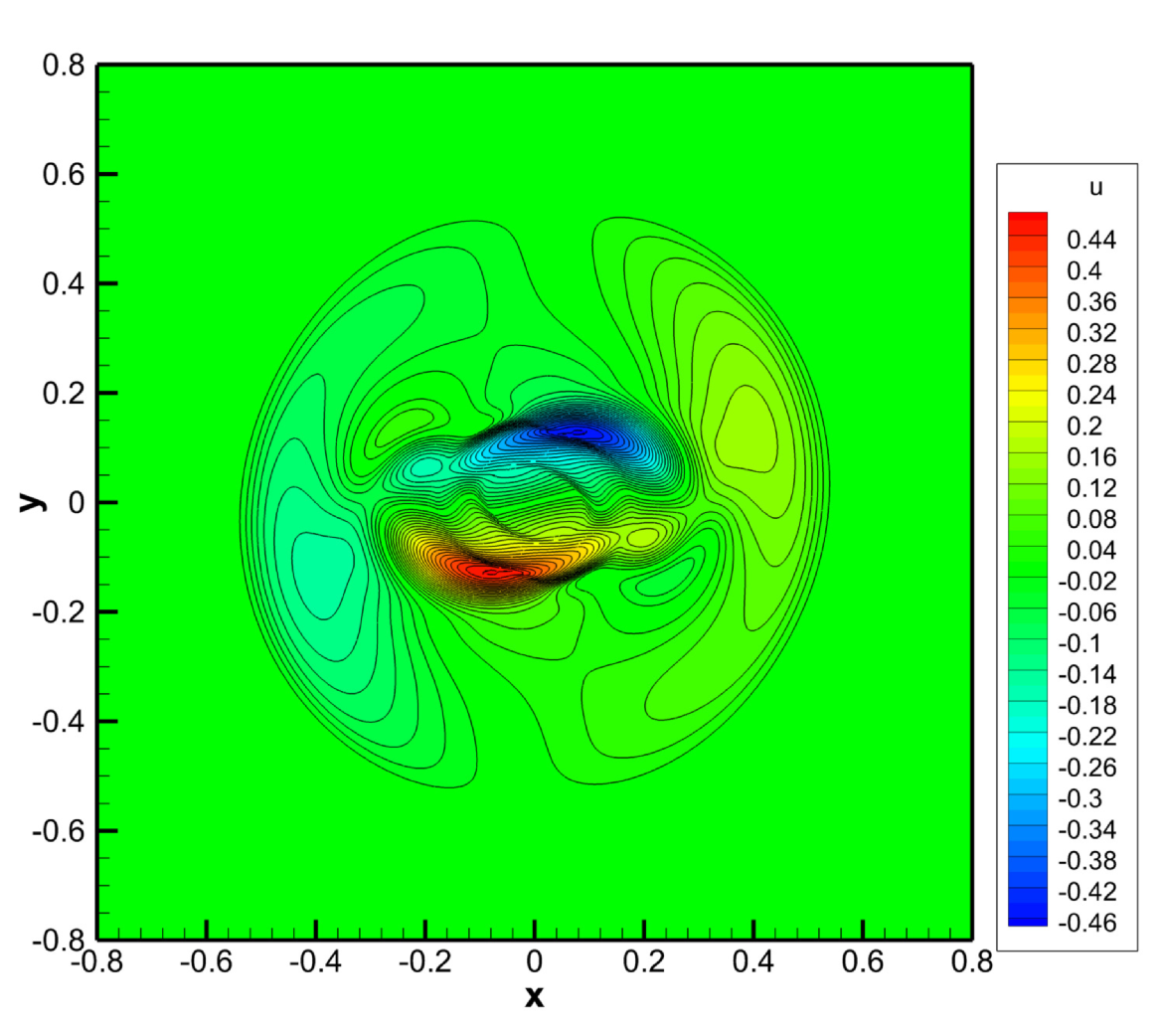}  
      &  
      \includegraphics[draft=false,width=0.3\textwidth]{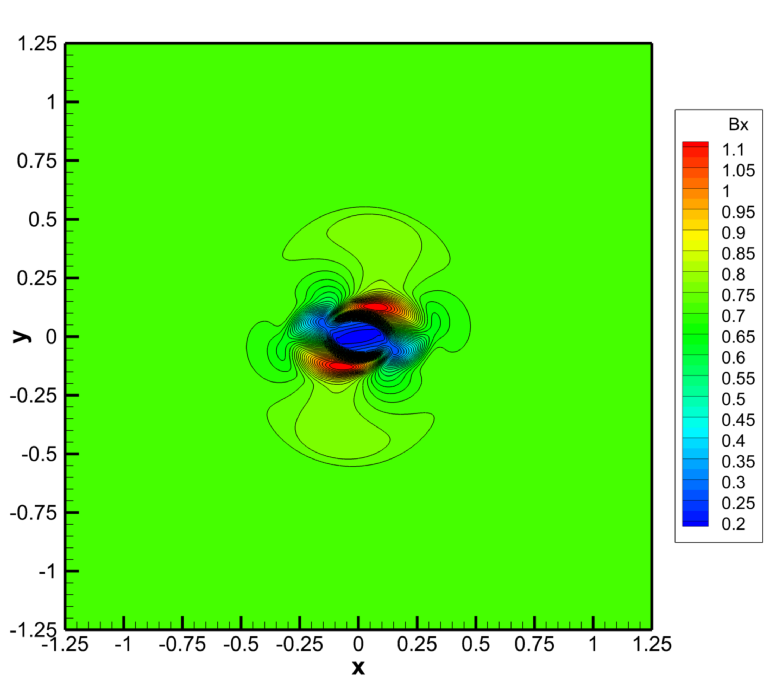} 
      \\   
      \includegraphics[draft=false,width=0.3\textwidth]{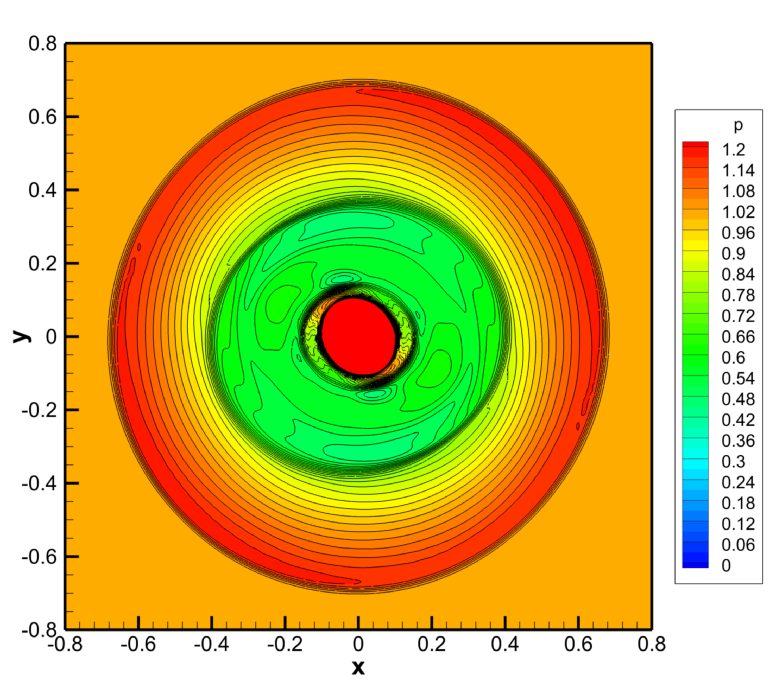}
        &  
      \includegraphics[draft=false,width=0.3\textwidth]{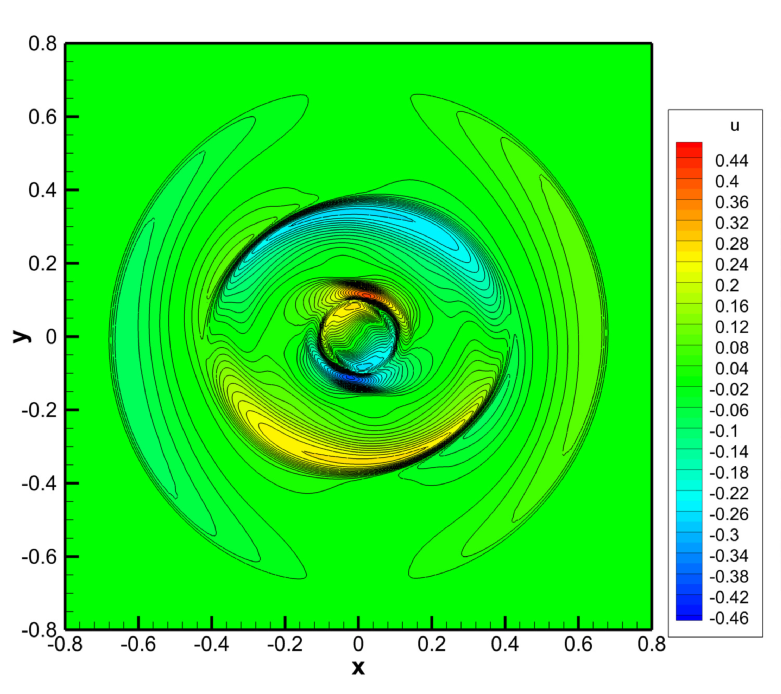}
        &  
      \includegraphics[draft=false,width=0.3\textwidth]{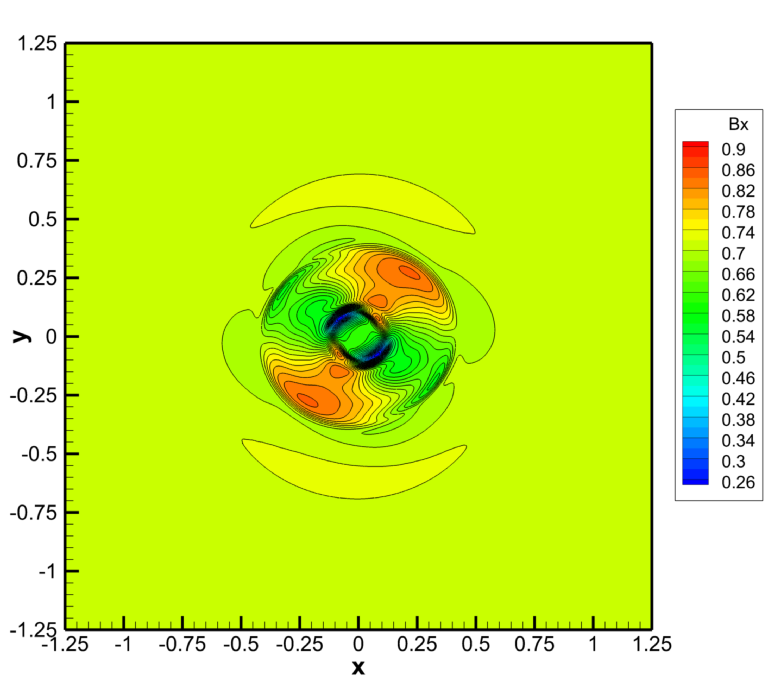}
       \\   
      \includegraphics[draft=false,width=0.3\textwidth]{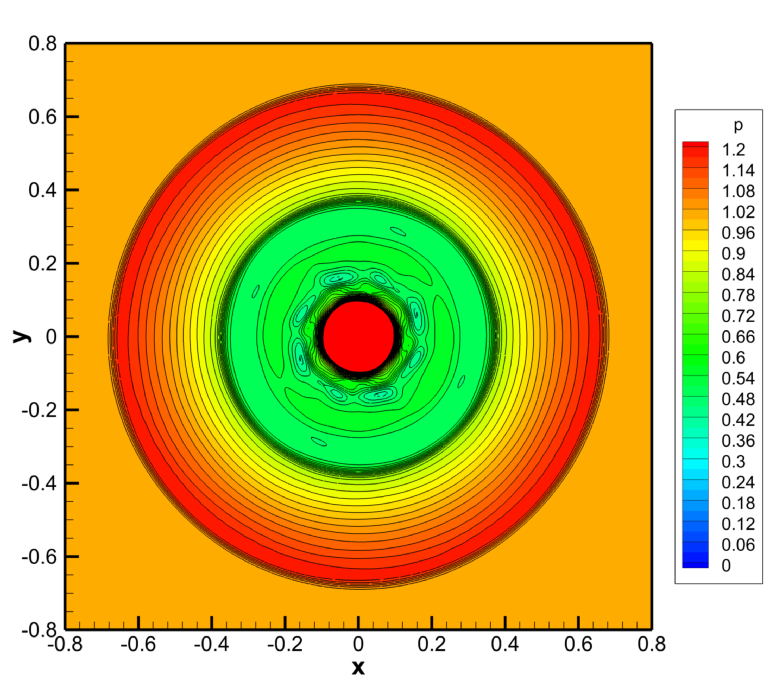}  
      &  
      \includegraphics[draft=false,width=0.3\textwidth]{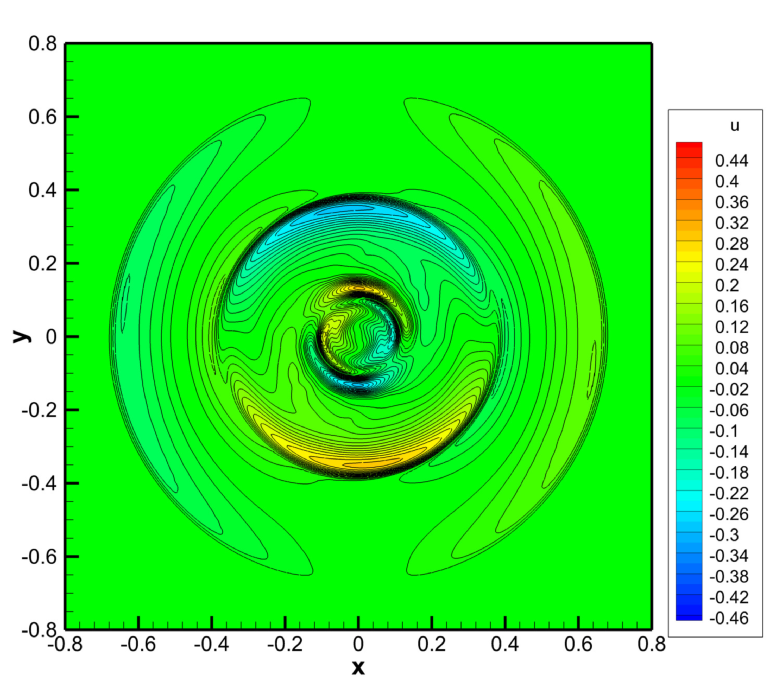}  
      &  
      \includegraphics[draft=false,width=0.3\textwidth]{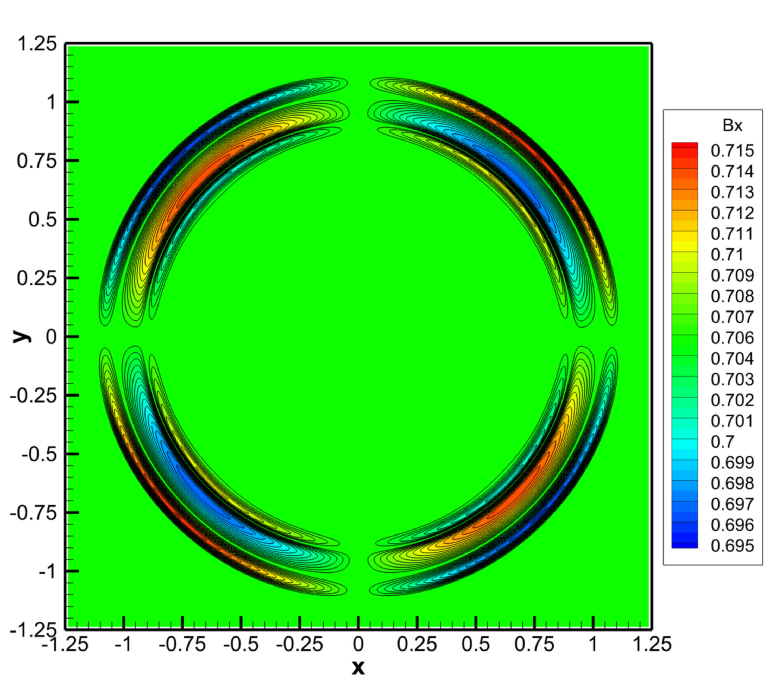}  
	\end{tabular} 
    \caption{Rotor problem in different regimes of the relaxation parameters. Case I: viscous and resistive magnetized fluid (top row). Case II: magnetized electrically conducting elastic solid (middle row). Case III: magnetized 
    electrically non-conducting elastic solid (bottom row). The contours of pressure (left column), velocity component $u$ (middle column) and magnetic field component $B_x$ are reported.} 
    \label{fig.solidrotor}
	\end{center}
\end{figure}

The computational results are depicted in Fig. \ref{fig.solidrotor}. We can see that in the case of electrically conducting material $\eta \ll 1$, the magnetic field is tight to the main pressure and shear waves 
arising in the medium, while in the case of an infinitely resistive or electrically non-conducting solid ($\eta \to \infty$), the electro-magnetic waves travel at the speed of light, independently of the other 
waves present in the medium. 
One can also clearly observe the effect of elasticity in the case of an elastic solid, since the rotor starts to oscillate and produce shear waves that are not visible in the case of the magnetized fluid,  
see the middle column of Fig. \ref{fig.solidrotor}. A similar behavior of the velocity field in elastic bodies has already been observed in \cite{Dumbser2008}.

\section{Conclusions} 

We have presented a new unified \textit{first order} symmetric \textit{hyperbolic} and thermodynamically compatible (HTC) model for the description of continuous media like fluids and solids interacting with electro-magnetic fields. 
The theoretical foundations of the model have been shown in great detail, in particular the connection of the final Eulerian form of model to the Euler-Lagrange differential equations associated with the 
underlying variational principle governed by the minimization of a Lagrangian. In the HTC framework, the dissipative terms are \textit{not} modeled by classical parabolic differential operators (as for example 
in the Navier-Stokes or in the viscous and resistive MHD equations), but via \textit{algebraic} relaxation source terms, like the Ohm law in the case of electro-magnetic wave propagation in conducting media. 
In the HTC framework, the same idea is also used to model dissipative momentum and heat transfer. 
The model satisfies the first and second principle of thermodynamics. It has also been shown via formal asymptotic analysis that in the stiff relaxation limit our first 
order hyperbolic system of PDEs reduces to the classical viscous and resistive MHD equations. A particular feature of the governing PDE system presented in this paper is the fact that all wave speeds remain \textit{finite}, 
even in the stiff relaxation limit when the relaxation parameters $\tau_1$, $\tau_2$ and $\eta$ tend to zero. This makes the model a potential candidate for a possible future extension to the more 
general case of special and general \textit{relativistic} continuum mechanics, in particular for the description of viscous and resistive relativistic fluids, where all propagation speeds in 
the medium must be necessarily bounded from above by the speed of light. In absence of relaxation source terms ($\tau_1 \to \infty$, $\tau_2 \to \infty$, $\eta \to \infty$), the PDE system proposed 
in this paper describes the propagation of electromagnetic waves in moving, electrically non-conducting dielectric media. 

Future investigations will concern a more detailed study of the coupling between the stress tensor and the electro-magnetic field, as it appears, for example, in piezoelectric actuators. 
For that purpose, the coupling between the distortion and the electro-magnetic field needs to be considered within the generating potential. Further work is also needed concerning the
development of new numerical methods which are able to preserve \textit{all} stationary compatibility conditions, i.e. those on the matrix $\AAA$ and on the electro-magnetic fields \textit{exactly} 
also at the discrete level. Finally, also the introduction of dispersive effects where the speed of propagation of electro-magnetic waves depends on the wave number will be subject to future research. 

\section*{Acknowledgments}

The research presented in this paper has been financed by the European Research Council (ERC) under the
European Union's Seventh Framework Programme (FP7/2007-2013) with the research project \textit{STiMulUs}, 
ERC Grant agreement no. 278267. 
M.D. and O.Z. have further received funding from the European Union's Horizon 2020 Research and 
Innovation Programme under the project \textit{ExaHyPE}, grant agreement number no. 671698 (call
FETHPC-1-2014). 
E.R. acknowledges a partial support by the Program N15 of the Presidium of RAS, project 121. 
I.P. acknowledges a partial support by ANR-11-LABX-0040-CIMI within the program ANR-11-IDEX-0002-02 and the Russian Foundation for Basic Research (grant number 16-31-00146).

The authors are grateful to the Leibniz Rechenzentrum (LRZ) for awarding access to the 
SuperMUC supercomputer based in Munich, Germany.

We also would like to thank Sergei Konstantinovich Godunov for his groundbreaking and very inspiring seminal ideas that are at the basis of both the 
theoretical as well as the numerical framework employed in this paper.

\appendix

\section{Connection of the HTC model with the Maxwell equations} 
\label{app.max.gpr} 

We start from the Maxwell equations in the \textbf{laboratory frame}, which are universally accepted to be true: 
\begin{subequations}\label{eqn.maxwell}
	\begin{align}
	& \frac{\partial \DD}{\partial t} - \nabla \times \HH = - \II \\ 
	& \frac{\partial \BB}{\partial t} + \nabla \times \EE = 0, \label{eqn.max.b} 
	\end{align}
\end{subequations}
with the constitutive relations
\begin{equation} \label{eqn.const.em}
	  \DD = \epsilon' \EE  \qquad \textnormal{ and } \qquad \BB = \mu' \HH. 
\end{equation} 
From \eqref{eqn.maxwell} and \eqref{eqn.const.em} one obtains 
\begin{subequations}\label{eqn.maxwell1b}
	\begin{align}
	&  \frac{\partial}{\partial t} \left(\epsilon' \EE \right)  - \nabla \times \HH = - \II \\ 
	&  \frac{\partial}{\partial t} \left(\mu' \HH \right) + \nabla \times \EE = 0,   
	\end{align}
\end{subequations}
Here, $\mu' = \mu_0 \mu_r$ denotes the magnetic permeability of the medium (not to be confounded with the fluid viscosity $\mu$), 
given as a product of the magnetic permeability of vacuum $\mu_0$ and the relative permeability $\mu_r$. 
The electric permittivity is denoted by $\epsilon' = \epsilon_0 \epsilon_r$, where $\epsilon_0$ is the electric permittivity 
of vacuum and $\epsilon_r$ is the relative permittivity of the medium. 
In \eqref{eqn.maxwell} the electric current $\II$ is given by the usual Ohm law
\begin{equation}
  \II = \rho_c \vv + \sigma \left( \EE + \vv \times \BB \right), 
\end{equation} 
with the conductivity $\sigma=1/\eta$ and the resistivity $\eta$. The charge density is $\rho_c = \epsilon' \nabla \cdot \EE$.  
We furthermore have the standard relation between the speed of light in the medium $c$, the magnetic permeability and the electric permittivity of the medium: 
\begin{equation}
  c^2 = \frac{1}{\epsilon' \mu'}. 
\end{equation} 
We now make the following change of variables 
\begin{subequations}\label{eqn.change}
	\begin{align}
	& \dd = \EE + \vv \times \BB,             \label{eqn.change.e} \\ 
	& \bb = \HH - \vv \times \DD = \HH - \frac{\vv \times \EE}{\mu' c^2} = \frac{1}{\mu'} \left( \BB - \frac{\vv \times \EE}{c^2} \right), \label{eqn.change.b}  
	\end{align}
\end{subequations}
where $\dd$ and $\bb$ denote respectively the electric and the magnetic field in the \textbf{comoving frame} $K^\prime$ that moves with velocity $\vv$. 
Substituting \eqref{eqn.change} into \eqref{eqn.maxwell} yields 
\begin{subequations}\label{eqn.maxwell.2}
	\begin{align}
	& \frac{\partial }{\partial t} \left[ \epsilon' \left( \dd - \vv \times \BB \right) \right]            - \nabla \times \left( \bb + \frac{\vv \times \EE}{\mu' c^2 } \right) = - \sigma  \dd 
	- \epsilon' \vv \nabla \cdot \EE, \label{eqn.max2.e} \\ 
	&  \frac{\partial }{\partial t} \left[ \mu' \left( \bb + \frac{\vv \times \EE}{\mu' c^2} \right) \right] +     \nabla \times \left( \dd - \vv \times \BB \right) = 0. \label{eqn.max2.b} 
	\end{align}
\end{subequations}
We now substitute \eqref{eqn.change} also into the remaining terms of \eqref{eqn.maxwell.2},
hence obtaining
\begin{subequations}\label{eqn.maxwell.3}
	\begin{align}
	&  \frac{\partial }{\partial t} \left[ \epsilon' \left( \dd - \vv \times \left( \mu' \bb + \frac{\vv \times \EE}{c^2} \right) \right) \right]  +  
	\nabla \times \left( - \bb - \frac{\vv \times (\dd - \vv \times \BB )}{\mu' c^2} \right) = - \sigma \dd - \epsilon' \vv \nabla \cdot \EE, \label{eqn.max3.e} \\ 
	&  \frac{\partial }{\partial t} \left[ \mu' \left( \bb + \frac{\vv \times \left(\dd - \vv \times \BB\right))}{ \mu' c^2 } \right)  \right]  +   
	\nabla \times \left(  \phantom{-} \dd - \vv \times \left( \mu' \bb + \frac{\vv \times \EE}{c^2}\right) \right) = 0. \label{eqn.max3.b} 
	\end{align}
\end{subequations}
The above equations are still fully consistent with the relativistic equations; \textit{no assumptions} have been made so far, only a change of variables. The formal structure
is still exactly the same as the one of the original Maxwell equations \eqref{eqn.maxwell}. We now insert
\eqref{eqn.change} once more into the fluxes of \eqref{eqn.maxwell.3}, which yields 
\begin{subequations}\label{eqn.maxwell.4}
	\begin{align}
	&  \frac{\partial }{\partial t} \left[ \epsilon' \left( \dd - \vv \times \left( \mu' \bb + \frac{\vv \times \EE}{c^2} \right) \right) \right] +  
	\nabla \times \left( - \bb - \frac{\vv \times \left(\dd - \vv \times \left( \mu' \bb + \frac{\vv \times \EE}{ c^2} \right) \right)}{\mu' c^2} \right) = - \sigma \dd -  \epsilon' \vv \nabla \cdot \EE, \label{eqn.max4.e} \\ 
	&  \frac{\partial }{\partial t} \left[ \mu' \left( \bb + \frac{\vv \times \left(\dd - \vv \times \BB\right))}{ \mu' c^2 } \right) \right]  +   
	\nabla \times \left(  \phantom{-} \dd - \vv \times \left( \mu' \bb + \frac{\vv \times \left(\dd - \vv \times \BB \right)}{c^2}\right) \right) = 0. \label{eqn.max4.b} 
	\end{align}
\end{subequations}
With the auxiliary variables 
\begin{subequations}\label{eqn.max.auxvar}
	\begin{align}
	& \ee':= \epsilon' \left( \dd - \vv \times \left( \mu' \bb + \frac{\vv \times \EE}{c^2} \right) \right) 
	       = \epsilon' \left( \dd - \mu' \vv \times \bb - \frac{1}{c^2} \vv \times \left( \vv \times \EE \right) \right), \label{eqn.aux.e} \\ 
	& \hh':= \mu' \bb + \frac{\vv \times \left(\dd - \vv \times \BB \right)}{c^2}  =  \mu' \bb + \frac{\vv \times \dd}{c^2} - \frac{1}{c^2} \vv \times \left( \vv \times \BB \right), \label{eqn.aux.b}  
	\end{align}
\end{subequations}
the system \eqref{eqn.maxwell.4} can be rewritten as 
\begin{subequations}\label{eqn.maxwell.5}
	\begin{align}
	& \frac{\partial \ee'}{\partial t} + \nabla \times \left( - \vv \times \ee' -  \bb \right) = - \sigma \dd - \epsilon' \vv \nabla \cdot \EE, \label{eqn.max5.e} \\[2mm] 
	& \frac{\partial \hh'}{\partial t} + \nabla \times \left( - \vv \times \hh' +  \dd \right) = 0. \label{eqn.max5.b} 
	\end{align}
\end{subequations}
We now assume that $\vv^2/c^2 \ll 1$, hence we neglect quadratic terms in $\vv/c$. 
This allows us to define the new simplified conserved variables 
\begin{subequations}\label{eqn.max.consvar}
	\begin{align}
	& \ee:= \epsilon' \dd - \epsilon' \mu' \vv \times \bb  \approx \ee' , \label{eqn.new.e} \\[3mm] 
	& \hh:= \mu'      \bb + \epsilon' \mu' \vv \times \dd  \approx \hh' , \label{eqn.new.b}  
	\end{align}
\end{subequations}
which yields the following \textit{simplified} system, where we have also used the identity $\nabla \cdot \BB=0$ and where again quadratic terms in $\vv/c$ 
have been neglected in the final expressions: 
\begin{subequations}\label{eqn.maxwell.gpr}
	\begin{align}
	& \frac{\partial \ee}{\partial t} + \nabla \times \left( - \vv \times \ee -  \bb \right) + \vv \, \nabla \cdot \ee = -\sigma  \dd, \label{eqn.maxg.e} \\[2mm]
	& \frac{\partial \hh}{\partial t} + \nabla \times \left( - \vv \times \hh +  \dd \right) + \vv \, \nabla \cdot \hh = 0. \label{eqn.maxg.b} 
	\end{align}
\end{subequations}
The above system is written in the form given in \cite{Rom1998}.  
Note that compared to \cite{Rom1998}, the consistency with the Maxwell equations in the laboratory frame requires the following definition of the generating potential
for electro-magnetic energy density: 
\begin{equation}
  L_{em} = 
	 \frac{1}{2} \left( \mu' \bb^2 + \epsilon' \dd^2 \right) +  \epsilon' \mu' \vv \cdot ( \dd \times \bb ) = 
   \frac{1}{2} \left( \mu' b_i b_i + \epsilon' d_i d_i \right) + \epsilon' \mu'  \, \varepsilon_{ijk} v_i d_j b_k,
									\label{eqn.gen.lem} 
\end{equation} 
since we must have that 
\begin{equation}
  e_i = \frac{\partial L_{em} }{\partial d_i} = \epsilon' d_i - \epsilon' \mu' \, \varepsilon_{ijk} v_j b_k, 
\end{equation} 
\begin{equation}
  h_i = \frac{\partial L_{em} }{\partial b_i} = \mu'      b_i + \epsilon' \mu' \, \varepsilon_{ijk} v_j d_k.
\end{equation} 
Using \eqref{eqn.max.consvar} and \eqref{eqn.change} the variables $\ee$ and $\hh$ can be expressed in terms of the 
electro-magnetic field quantities in the laboratory frame as follows: 
\begin{equation}
 \label{eqn.labframe}
	 \ee = \epsilon' \left( \EE + \frac{1}{c^2} \vv \times \vv \times \EE \right),  \qquad \textnormal{ and } \qquad 
	 \hh = \mu' \HH + \frac{1}{c^2} \vv \times \vv \times \EE,  
\end{equation}
which together with \eqref{eqn.const.em} reduces to the simple identities 
\begin{equation}
 \label{eqn.labframe.final}
	 \ee = \DD ,  \qquad \textnormal{ and } \qquad  	 \hh = \BB,  
\end{equation}
if quadratic terms in $\vv/c$ are again neglected in \eqref{eqn.labframe}.

\section{Euler-to-Lagrange field transformation}\label{app.EulertoLagr}

In this section, we demonstrate how to obtain Eulerian equations \eqref{eqn.HPR.Efield}--\eqref{eqn.HPR.Hfield} for electromagnetic fields $ e_i $ and $ h_i $ from their
Lagrangian counterparts~\eqref{eqn.MasterLagr.Electr}--\eqref{eqn.MasterLagr.Magn}. However, it is more convenient to chose the opposite strategy. Namely, we derive the
Lagrangian field equations from the Eulerian. Thus, if one likes to get the Lagrange-to-Euler derivation then the calculations should be repeated from the end to the
beginning of what follows. Recall that in \eqref{eqn.MasterLagr.Electr}--\eqref{eqn.MasterLagr.Magn} we use the same notations $ e_i $ and $ h_i $ for the fields however
they are different. As we mentioned earlier and as will be proven in what follows, the Eulerian and Lagrangian fields are related by~\eqref{eqn.eh_prime}. As discussed
in Section~\ref{sec.HTC.Lagr}, in the structure study, we can ignore the algebraic source terms as they are low order terms.

\subsection{Auxiliary relations}
Here, we summarize the definitions and formulas used in this section. The total deformation gradient $\FF= [F_{ij}] $, the distortion matrix $ \AAA=[A_{ij}] $ and the velocity are defined as
\begin{equation}\label{eq.gradient}
F_{ij}=\dfrac{\pd x_i}{\pd y_j},\qquad \AAA=\FF^{-1},\qquad w=\det (\FF)=\dfrac{\rho_0}{\rho},\qquad v_i=\frac{{\rm d} x_i}{{\rm d} t},
\end{equation}
where, as previously, $ y_j $ are the Lagrangian coordinates and $ x_i $ are the Eulerian ones, $ \rho $ and $ \rho_0 $ are the actual
and the reference mass densities, respectively. The time evolution equation for $ F_{ij} $ in the Lagrangian coordinates
\begin{equation}\label{eq.gradient.time}
\frac{\rmd F_{ij}}{\rmd t} - \frac{\pd v_i}{\pd y_j}=0
\end{equation}
is a trivial consequence of definitions \eqref{eq.gradient}.

The following standard definitions and formulas  are also introduced

\begin{equation}\label{eq.CofDefinition}
\boldsymbol{C}={\rm cof}(\FF)=w \AAA^\mathsf{T}=\left[C_{{ij}}\right], {\ \rm\ \ or\ \ }A_{{ij}}=w^{-1}C_{{ji}},
\end{equation}

\begin{equation}\label{eq.DivCof}
\frac{\partial C_{i j}}{\partial y_j}=0,\qquad \varepsilon _{{mjp}}\frac{\partial  F_{{mp}}}{\partial y_j}=0,
\end{equation}

%

\begin{equation}\label{eq.DetTimeEvol}
\frac{{\rm d} w}{{\rm d} t}-\frac{\partial w A_{j k}v_k}{\partial y_j}=0,{\rm or\ using\ (\ref{eq.CofDefinition})\ and\ (\ref{eq.DivCof}):\ \ \ }\frac{{\rm d} w}{{\rm d} t}-w A_{j k}\frac{\partial v_k}{\partial y_j}=0,
\end{equation}

\begin{equation}\label{eq.DetLeviCivita}
\varepsilon _{{ikl}}A_{{mi}}A_{{jk}}A_{{al}}=\varepsilon_{mja}w^{-1},
\end{equation}

\begin{equation}\label{eq.CofFormulaLeviCivita}
C_{{km}}=\frac{1}{2}\varepsilon _{{lnk}}\varepsilon _{{pqm}}F_{{lp}} F_{{nq}},
\end{equation}

\begin{equation}\label{eq.LeviCivitaKronecker}
\varepsilon _{{imn}}\varepsilon _{{jmn}}=2\delta _{{ij}}.
\end{equation}

\subsection{Transformation of \eqref{eqn.HPR.Efield}--\eqref{eqn.HPR.Hfield} to \eqref{eqn.MasterLagr.Electr}--\eqref{eqn.MasterLagr.Magn}}

We shall transform \eqref{eqn.HPR.Efield} into \eqref{eqn.MasterLagr.Electr} while \eqref{eqn.HPR.Hfield} transforms into \eqref{eqn.MasterLagr.Magn} analogously. Thus, \eqref{eqn.HPR.Efield} is equivalent to
\begin{equation}
\frac{\partial  e_i}{\partial  t}+v_k\frac{\partial e_i}{\partial  x_k}+\frac{\partial v_k}{\partial  x_k} e_i-\frac{\partial
v_i}{\partial  x_k}e_k-\varepsilon _{{ikl}}\frac{\partial \mcE_{h_l}}{\partial  
x_k}=0.
\end{equation}
Using \(\rmd/{\rmd t}=\partial /\partial t+v_k /\partial x_k\) we have

\begin{equation}
\frac{\rmd e_i}{{\rmd t}}+\frac{\partial v_k}{\partial  x_k} e_i-\frac{\partial v_i}{\partial  x_k}e_k-\varepsilon
_{{ikl}}\frac{\partial \mcE_{h_l}}{\partial  x_k}=0.
\end{equation}
From \eqref{eq.gradient} it follows that $\frac{\pd}{\pd x_k}=A_{{jk}}\frac{\pd}{\pd y_j}$ and thus we change the variables $ x_k $ on $ y_j $

\begin{equation}
\frac{\rmd e_i}{{\rmd t}}+A_{{jk}}\frac{\partial v_k}{\partial  y_j} 
e_i-A_{{jk}}\frac{\partial v_i}{\partial  
y_j}e_k-\varepsilon_{{ikl}}A_{{jk}}\frac{\partial \mcE_{h_l}}{\partial  y_j}=0.
\end{equation}
Applying (\ref{eq.DetTimeEvol})$_2$ to the second term and \eqref{eq.gradient.time} to the third term of the last equation, we have
\begin{equation}
\frac{\rmd e_i}{{\rmd t}}+\frac{1}{w}\frac{{\rmd w}}{{\rmd t}} e_i-A_{{jk}}\frac{\rmd F_{ij}}{\rmd t}e_k-\varepsilon_{{ikl}}A_{{jk}}\frac{\partial E_{h_l}}{\partial  y_j}=0,
\end{equation}

\begin{equation}
\frac{\rmd  w e_i}{{\rmd t}}-w e_k A_{{jk}}\frac{\rmd F_{ij}}{\rmd t}-w 
\varepsilon _{{ikl}}A_{{jk}}\frac{\partial \mcE_{h_l}}{\partial  y_j}=0,
\end{equation}

Now, we add \(0\equiv w e_k \dfrac{\rmd\delta_{ik}}{\rmd t}\) to the left hand side and then using that \(\delta _{ik}=F_{ij}A_{jk}\) one can obtain that
\begin{equation}
\frac{\rmd w e_i}{{\rmd t}}+w e_k\frac{\rmd \delta_{ik}}{ \rmd t}-w e_k A_{{jk}}\frac{{\rmd F}_{{ij}}}{{\rmd t}}-w
\varepsilon _{{ikl}}A_{{jk}}\frac{\partial \mcE_{h_l}}{\partial  y_j}=0,
\end{equation}
\begin{equation}
\frac{\rmd w e_i}{{\rmd t}}+w e_k\frac{\rmd F_{{ij}}A_{{jk}}}{{\rmd t}}-w e_k 
A_{{jk}}\frac{{\rmd F}_{{ij}}}{{\rmd t}}-w \varepsilon 
_{{ikl}}A_{{jk}}\frac{\partial \mcE_{h_l}}{\partial  y_j}=0,
\end{equation}
\begin{equation}
\frac{\rmd w e_i}{{\rmd t}}+w e_kF_{{ij}}\frac{\rmd A_{{jk}}}{{\rmd t}}+w 
e_k\frac{\rmd F_{{ij}}}{{\rmd t}}A_{{jk}}-w e_k A_{{jk}}\frac{{\rmd 
F}_{{ij}}}{{\rmd t}}-w \varepsilon _{{ikl}}A_{{jk}}\frac{\partial 
\mcE_{h_l}}{\partial  y_j}=0.
\end{equation}
After multiplying the last equation by $ A_{mi} $
\begin{equation}
A_{{mi}}\left(\frac{\rmd w e_i}{{\rmd t}}+w e_kF_{{ij}}\frac{\rmd 
A_{{jk}}}{{\rmd t}}-w \varepsilon _{{ikl}}A_{{jk}}\frac{\partial 
\mcE_{h_l}}{\partial  y_j}\right)=0,
\end{equation}
\begin{equation}
A_{{mi}}\frac{\rmd w e_i}{{\rmd t}}+w e_k\frac{\rmd A_{{mk}}}{{\rmd t}}-w 
\varepsilon _{{ikl}}A_{{mi}}A_{{jk}}\frac{\partial \mcE_{h_l}}{\partial  y_j}=0,
\end{equation}
we have an intermediate result:
\begin{equation}\label{eq.inter.result}
\frac{\rmd w A_{{mk}} e_k}{{\rmd t}}-w \varepsilon 
_{{ikl}}A_{{mi}}A_{{jk}}\frac{\partial \mcE_{h_l}}{\partial
 y_j}=0.
\end{equation}

Now, we introduce the change of unknowns~\eqref{eqn.eh_prime}: $w 
A_{{mk}}e_k=e'_m$, $w A_{{mk}}h_k=h'_m$, and we also change the energy 
potential $\mcE(e_i,h_i)= 
\mcE(w^{-1}F_{{ij}}e'_j,w^{-1}F_{{ij}}h'_j)=w^{-1}U(e'_j,h'_j)$. Hence, 
$ \mcE_{h_i} = A_{ji}U_{h'_j}$. After this, the intermediate 
result~\eqref{eq.inter.result} reads as
\begin{equation}
\frac{\rmd e'_m}{{\rmd t}}-w \varepsilon 
_{{ikl}}A_{{mi}}A_{{jk}}\frac{\partial   A_{{al}} U_{h'_a}}{\partial  y_j}=0,
\end{equation}
\begin{equation}
\frac{\rmd e'_m}{{\rmd t}}-w \varepsilon 
_{{ikl}}A_{{mi}}A_{{jk}}A_{{al}}\frac{\partial   U_{h'_a}}{\partial  y_j}-w 
\varepsilon
_{{ikl}}A_{{mi}}A_{{jk}}\frac{\partial  A_{{al}}}{\partial  y_j}   U_{h'_a}=0.
\end{equation}
Applying (\ref{eq.DetLeviCivita}) and \eqref{eq.CofDefinition}$_3 $ to the second term, we get
\begin{equation}
\frac{\rmd e'_m}{{\rmd t}} - \varepsilon _{{mja}}\frac{\partial   
U_{h'_a}}{\partial  y_j}-w \varepsilon _{{ikl}}A_{{mi}}A_{{jk}}\frac{\partial
A_{{al}}}{\partial  y_j}   U_{h'_a}=0,
\end{equation}
Now using the cofactor definition (\ref{eq.CofDefinition}) and then applying (\ref{eq.CofFormulaLeviCivita}) to the third term, we have
\begin{equation}
\frac{\rmd e'_m}{{\rmd t}} - \varepsilon _{{mja}}\frac{\partial   
U_{h'_a}}{\partial  y_j}-\varepsilon _{{ikl}}A_{{mi}}C_{{kj}}\frac{\partial
A_{{al}}}{\partial  y_j}   U_{h'_a}=0,
\end{equation}
\begin{equation}
\frac{\rmd e'_m}{{\rmd t}} - \varepsilon _{{mja}}\frac{\partial   
U_{h'_a}}{\partial  y_j}-\frac{1}{2}\varepsilon _{{ikl}}\varepsilon 
_{{lnk}}\varepsilon _{{pqj}}F_{{lp}} F_{{nq}}A_{{mi}}\frac{\partial 
A_{{al}}}{\partial  y_j}   U_{h'_a}=0.
\end{equation}
Using (\ref{eq.LeviCivitaKronecker}) in the third term gives us
\begin{equation}
\frac{\rmd e'_m}{{\rmd t}} - \varepsilon _{{mja}}\frac{\partial   
U_{h'_a}}{\partial  y_j}-\delta _{{in}}\varepsilon _{{pqj}}F_{{lp}} 
F_{{nq}}A_{{mi}}\frac{\partial A_{{al}}}{\partial  y_j}   U_{h'_a}=0,
\end{equation}
and subsequently,
\begin{equation}
\frac{\rmd e'_m}{{\rmd t}} - \varepsilon _{{mja}}\frac{\partial   
U_{h'_a}}{\partial  y_j}-\varepsilon _{{pqj}}F_{{lp}} 
F_{{iq}}A_{{mi}}\frac{\partial A_{{al}}}{\partial  y_j}   U_{h'_a}=0,
\end{equation}
\begin{equation}
\frac{\rmd e'_m}{{\rmd t}} - \varepsilon _{{mja}}\frac{\partial   
U_{h'_a}}{\partial  y_j}-\varepsilon _{{mjp}}F_{{lp}}\frac{\partial 
A_{{al}}}{\partial  y_j}   U_{h'_a}=0.
\end{equation}

Now, adding $0\equiv \varepsilon _{{mjp}}\frac{\partial  F_{{lp}}}{\partial 
y_j} A_{{al}}   U_{h'_a}$ (see \eqref{eq.DivCof}$_2$), we get
\begin{equation}
\frac{\rmd e'_m}{{\rmd t}} - \varepsilon _{{mja}}\frac{\partial   
U_{h'_a}}{\partial  y_j}-\varepsilon _{{mjp}}F_{{lp}}\frac{\partial 
A_{{al}}}{\partial  y_j} U_{h'_a}-\varepsilon _{{mjp}}\frac{\partial  
F_{{lp}}}{\partial y_j}A_{{al}}  U_{h'_a}=0,
\end{equation}
\begin{equation}
\frac{\rmd e'_m}{{\rmd t}} - \varepsilon _{{mja}}\frac{\partial   
U_{h'_a}}{\partial  y_j} -   U_{h'_a}\left(\varepsilon _{{mjp}}\frac{\partial
A_{{al}}F_{{lp}}}{\partial  y_j}\right)=0,
\end{equation}
\begin{equation}
\frac{\rmd e'_m}{{\rmd t}} - \varepsilon _{{mja}}\frac{\partial   
U_{h'_a}}{\partial  y_j} -   U_{h'_a}\left(\varepsilon _{{mjp}}\frac{\partial 
\delta _{{ap}}}{\partial  y_j}\right)=0.
\end{equation}
Eventually, we have 
\begin{equation}
\frac{\rmd e'_m}{{\rmd t}} - \varepsilon _{{mja}}\frac{\partial    
U_{h'_a}}{\partial  y_j}=0,
\end{equation}
which is identical to \eqref{eqn.MasterLagr.Electr}.

\section{Euler-to-Lagrange momentum transformation} 
\label{app.EulertoLagr.momentum}

In this section, we shall use $ m_i' $ to denote the Lagrangian momentum and $ 
e_i' $ and $ h_i' $ to denote Lagrangian electro-magnetic fields, i.e. 
exactly those vector fields appearing in~\eqref{eqn.MasterLagr.Momentum}, 
\eqref{eqn.MasterLagr.Electr} and \eqref{eqn.MasterLagr.Magn}. Now, 
\eqref{eqn.MasterLagr.Momentum} reads as
\begin{equation}\label{app.momentum}
\frac{{\rm d}m_i'}{{\rm d}t}-\frac{\pd U_{F_{ij}}}{\pd y_j}=0.
\end{equation}
Using that $ {\rm d}/{\rm d}t = \pd /\pd t + v_k\pd/\pd x_k $ and $ 
F_{ij}=\pd x_i/\pd y_j $, equation~\eqref{app.momentum} can be rewritten as
\begin{equation}\label{app.momentum2}
\frac{\pd m_i'}{\pd t} + v_k\frac{\pd m_i'}{\pd x_k} - F_{kj}\frac{\pd 
U_{F_{ij}}}{\pd x_k}=0.
\end{equation}
Subsequently, using the Eulerian stationary constraint for $ F_{ij} $ and time 
evolution of $ w=\det(\FF) $ (e.g. 
see~\cite{GodRom2003})
\begin{equation}
\frac{\pd w^{-1}F_{kj}}{\pd x_k}=0,\ \ \ \ \frac{\pd }{\pd t}\left ( 
\frac{1}{w}\right ) + \frac{\pd }{\pd x_k}\left( 
\frac{v_k}{w}\right) =0,
\end{equation}
\eqref{app.momentum2} can be rewritten as
\begin{equation}\label{app.momentum3}
\frac{\pd }{\pd t}\left(\frac{m_i'}{w}\right)  + \frac{\pd }{\pd 
x_k}\left(\frac{v_k m_i' - F_{kj}U_{F_{ij}}}{w}\right) =0.
\end{equation}
Finally, introducing the change of the variables $ 
m_i=w^{-1}m_i' $, $w 
A_{{mk}}e_k=e'_m$ and $w A_{{mk}}h_k=h'_m$ and the change of the potential $ 
\mcE(\rho,m_i,e_i,h_i,F_{ij})=w^{-1}U(m_i',e_i',h_i',F_{ij}) $, 
equation~\eqref{app.momentum3} transforms into 
equation~\eqref{eqn.HPR.momentum}.

\section*{References}
\bibliographystyle{plain}
\bibliography{biblio}

\end{document}